\documentclass[12pt]{article}
\usepackage{tikz}
\usetikzlibrary{decorations.pathreplacing,angles,quotes}
\usepackage{etex}
\usepackage[T1]{fontenc}
\usepackage{amssymb,amsmath,geometry}
\usepackage{mathtools}
\usepackage{graphicx}
\usepackage{fancyhdr}
\usepackage{fancybox}
\usepackage{shadow}
\usepackage{empheq}
\usepackage{titlesec}
\usepackage{titletoc}
\usepackage{soul, color}
\usepackage{float}
\usepackage{shorttoc}
\usepackage{xcolor}
\usepackage{fancybox}
\usepackage{natbib}
\usepackage{array}
\usepackage{tabularx}
\usepackage{booktabs}
\usepackage{rotating}
\usepackage{bigstrut}
\usepackage{natbib}
\usepackage{aeguill}

\usepackage{changepage}
\newenvironment{changemargin}[2]{\begin{list}{}{%
\setlength{\topsep}{0pt}%
\setlength{\leftmargin}{0pt}%
\setlength{\rightmargin}{0pt}%
\setlength{\listparindent}{\parindent}%
\setlength{\itemindent}{\parindent}%
\setlength{\parsep}{0pt plus 1pt}%
\addtolength{\leftmargin}{#1}%
\addtolength{\rightmargin}{#2}%
}\item }{\end{list}}
\usepackage{geometry}

\usepackage{hyperref}
\def\sym#1{\ifmmode^{#1}\else\(^{#1}\)\fi}
\usepackage{caption}
\usepackage{blindtext}

\usepackage{caption}
\usepackage{subcaption}
\usepackage{mathtools}

\usepackage{graphicx}

\usepackage{enumerate}

\usepackage{hyperref}
\hypersetup{backref,
colorlinks=true, citecolor=blue!50!black}
\usepackage{graphicx}

\renewcommand{\thetable}{\Roman{table}}
\renewcommand{\thefigure}{\Roman{figure}}

\usepackage{titlesec}
\usepackage{setspace}
\usepackage{indentfirst}

\makeatletter
  \newcommand\smalls{\@setfontsize\smalls{10.3pt}{6}}
\makeatother

\makeatletter
  \newcommand\footnotesizes{\@setfontsize\footnotesizes{9.6pt}{6}}
\makeatother

\newsavebox\tmpbox

\usepackage{caption}
\usepackage{subcaption}

\usepackage{tikz}
\usepackage{tikz-network}

\usepackage[export]{adjustbox}
\usepackage{graphicx}

\usepackage{fnpct}

\begin{document}

\title{Religious Competition, Cultural Change, and Domestic Violence: Evidence from Colombia\thanks{We thank seminar participants at various institutions, the Editor, and three anonymous referees for their valuable comments. All remaining errors are our own.}}
\author{Hector Galindo-Silva\thanks{%
Department of Economics, Pontificia Universidad Javeriana, E-mail: galindoh@javeriana.edu.co
} \\
Pontificia Universidad Javeriana\\
 \and Guy Tchuente\thanks{%
Department of Agricultural Economics, Purdue University, E-mail: gtchuent@purdue.edu} \\
Purdue University
}

 \maketitle
\begin{abstract}

We study how religious competition—defined as the entry of a religious organization with innovative worship practices into a predominantly Catholic municipality—affects domestic violence. Using municipality-level data from Colombia and a two-way fixed effects design, we find that the arrival of the first non-Catholic church leads to a significant reduction in reported cases of domestic violence. We argue that religious competition incentivizes churches to adopt and diffuse norms and practices that more effectively discourage such violence. Effects are largest in municipalities with smaller, younger, and more homogeneous populations—contexts that facilitate both intense competition and norm diffusion. Consistent with this mechanism, areas with more new non-Catholic churches exhibit greater rejection of domestic violence—particularly among the religiously observant—and higher female labor force participation. These findings contribute to the literature on the cultural determinants of domestic violence by identifying religious competition as a catalyst for cultural change.

\bigskip
\noindent \textbf{Keywords:} Marketplace for Religion, Domestic violence.\\
\noindent \textbf{JEL classification}:  H41, J13, O17.

\end{abstract}

\newpage

\section{Introduction}

Domestic violence is a pervasive global issue, affecting millions of individuals daily through various forms of abuse, including psychological and physical harm.  It persists across regions and income levels, occurring in both developing and developed countries \citep{UN2015}. The consequences of domestic violence are severe and long-lasting, with significant impacts on physical health, mental well-being, and overall quality of life.\footnote{According to the United Nations Office on Drugs and Crime (UNODC), approximately 78,000 people were killed by intimate partners or family members in 2017—64\% of whom were women \citep{UNODC2017}.}

\medskip

Among the various contributing factors, culture is increasingly recognized—by both scholars and policymakers—as a key determinant of domestic violence.\footnote{While there exists variation in scholarly views of culture \cite[see for instance][for a discussion of this concept]{Acemoglu2021}, we adopt a broad and conventional definition: culture refers to the customary beliefs, social structures, and material traits of particular social groups \citep{Fernandez2016}.}   For example, some cultural norms legitimize violence as a form of discipline or education within the household, or attribute abusive behavior to external factors such as alcohol or drug use rather than holding perpetrators accountable. To the extent that such beliefs are deeply embedded in broader social practices, the prevalence and persistence of domestic violence are shaped by the surrounding cultural environment.\footnote{\label{footnoteBIB1}Early research linked domestic violence to traditional masculinity, patriarchy, and social norms  \citep[see for instance][]{ONeilharway1997, Murnen2002, KoenigStephensonAhmedJejeebhoy2006}. Recent studies highlight the roles of family structure, gender roles, and cultural practices \cite[see][]{TurPrats2019, Bhalotraetalt2019, Gonzalezetalt2020, CalviKeskar2021, AlesinaBrioschiLaFerrara2021}. For broader reviews, see \cite{fernandez2011handbook} and  \cite{BauFernandez2023}.} 

\medskip

This cultural dimension helps explain the persistence of domestic violence across time and space  \cite[see e.g.,][]{TurPrats2019, Gonzalezetalt2020, AlesinaBrioschiLaFerrara2021}. Yet culture is not static. A growing body of research documents cultural shifts in domains such as fertility preferences \citep{LaFerraraetal2012}, attitudes toward same-sex relationships \citep{Fernandezetal2019}, women's labor force participation \citep{Galindoidarraga2023culture}, and domestic violence itself \citep{JensenOster2009}. Understanding the drivers of cultural change and their implications for domestic violence is therefore central to efforts to promote inclusive and sustainable development.

\medskip

This paper contributes to this understanding by studying how cultural change—specifically, shifts in the religious landscape—affects domestic violence. In recent decades, many countries have witnessed a decline in historically dominant religious institutions and the emergence of new religious organizations that promote alternative norms and practices \citep{MillerSargeantFlory2013}. If these new  practices discourage domestic violence and gain traction among the broader population, then religious competition may serve as a powerful channel of cultural transformation.

\medskip

We investigate this hypothesis in the context of Colombia—a deeply religious country that was overwhelmingly Catholic until the late 20th century. In recent decades, the religious landscape has changed markedly due to the expansion of non-Catholic, primarily Pentecostal, churches. We exploit this shift to examine whether, and through what mechanisms, the entry of non-Catholic churches affects domestic violence rates.

\medskip

Our empirical strategy employs a two-way fixed effects model that exploits variation across municipalities and over time. Specifically, we implement a difference-in-differences design, using municipalities that have not yet experienced the entry of a non-Catholic church as the control group. These municipalities serve as a valid counterfactual under the assumption that, in the absence of treatment, they would have followed similar trends in domestic violence as treated municipalities. This ``parallel trends" assumption implies that no unobserved, time-varying factors—such as shifts in conflict dynamics, economic shocks, or changes in reporting behavior—differentially affected treated and untreated areas around the time of church entry. To address potential violations, we include municipality and year fixed effects, control for observable covariates, and conduct pre-trend tests and robustness checks using alternative estimators. To assess the timing and persistence of effects, we also estimate a dynamic version of the model that examines how domestic violence evolves in the years before and after the arrival of a non-Catholic church.

\medskip

We find that the entry of a non-Catholic church into a municipality is associated with a significant decline in domestic violence. On average, we estimate a reduction of approximately 14 reported cases per 100,000 inhabitants per year following the church’s arrival.  These results are robust across datasets, specifications, and alternative estimators, including those of \cite{ChaisemartinDHaultfoeuille2020}, \cite{SunAbraham2021}, and \cite{CallawaySantAnna2021}.

\medskip

We propose the following mechanism to explain our findings. The initial establishment of a non-Catholic church in a Colombian municipality introduces new religious practices that actively involve both congregants and their families. In the context of religious competition, and due to the relative appeal of these practices, other religious organizations—most notably the Catholic Church—begin to adopt similar forms of engagement. Crucially, these practices promote behavioral norms and values associated with lower levels of domestic violence, such as increased household stability, reduced alcohol consumption, greater sexual restraint, and enhanced female leadership within families and congregations. As these norms diffuse more broadly through the community, they contribute to a general decline in domestic violence.

\medskip

This proposed mechanism is supported by anecdotal evidence, including interviews with members of non-Catholic churches, media coverage, and public statements by Catholic Church officials. It is also consistent with complementary quantitative evidence showing that a higher presence of non-Catholic churches is associated with stronger individual-level rejection of domestic violence—particularly among more religiously observant individuals—and with increased female participation in the local labor force.

\medskip

Our argument aligns with models of innovation diffusion in social networks, such as those developed by \cite{Morris2000} and \cite{AcemogluOzdaglarYildiz2011}, which suggest that innovations spread more effectively in networks with lower clustering and longer-range ties. In our context, the innovation consists of new cultural practices introduced by non-Catholic churches. If diffusion is the underlying mechanism, effects should be stronger in municipalities where diffusion is easier—i.e., those with lower ethnic fragmentation and fewer civil society organizations. This is precisely what we observe.

\medskip

We also consider several alternative explanations. One possibility is that the decline in domestic violence is driven by changes in conflict dynamics. Prior research suggests that religious competition can influence armed violence patterns \citep{galindo2021fighting}. However, within our sample, we find no evidence that the reduction in domestic violence is concentrated in municipalities that experienced larger changes in conflict-related violence. We also examine economic factors and changes in reporting behavior and find no evidence that they account for the observed effects.

\medskip

Although our data are limited to Colombia, we believe the mechanisms identified here may be relevant in other settings—such as Brazil, Guatemala, Bolivia, Kenya, and the Philippines—that share similar features of deep religiosity and growing religious pluralism \citep{MillerSargeantFlory2013}.

\medskip
This paper contributes to several strands of literature. First, we add to research examining how cultural factors shape social outcomes, with a particular focus on domestic violence and gender norms \citep{CoolsKotsadam2017, SvecAndic2018, TurPrats2019, Gonzalez2020, TurPrats2021}. Our specific contribution lies in identifying religious competition as a key mechanism driving cultural change.

\medskip

Second, we contribute to the literature on the role of religious institutions in shaping social, economic, and political outcomes  \citep{CampanteYanagizawaDrott2015, Cantoni2015, CantoniDittmarYuchtman2018, BeckerPascali2019, Schulzetalt2019, galindo2021fighting, Schulz2022}. We engage with the literature on religious competition by analyzing both the supply and demand sides of the religious marketplace \citep[see][]{brik2022club}. On the supply side, we examine how the entry of new religious organizations affects societal outcomes such as domestic violence. This connects to Church Competition theory \citep{brik2019church, gallego2010christian}, which argues that religious competition influences affiliation, attendance, and education. On the demand side, we assess how increased religious diversity shapes social behavior, aligning with broader work on the role of religious organizations in shaping individual and collective well-being \citep[e.g.,][]{woodberry2012missionary, becker2009weber, becker2016causes, akccomak2016did, waldinger2017long}.

\medskip

Third, our study contributes to the literature on cultural change and its implications for social outcomes beyond domestic violence \citep{LaFerraraetal2012, Fernandezetal2019, Galindoidarraga2023culture, JensenOster2009}. In particular, we build on \cite{BauFernandez2023} and related work showing that cultural change often requires an external ``shock” to institutions, technology, or information that alters the incentives for adhering to prevailing norms. In our context, we identify such a shock: the arrival of new religious organizations offering more dynamic and appealing worship practices in areas historically dominated by traditional churches. This constitutes both institutional change—via shifts in informal norms—and technological change, insofar as worship styles can be viewed as cultural technologies that enhance religious engagement. We investigate how these transformations influence domestic violence, a norm-driven and socially salient outcome. To our knowledge, this is the first study to analyze the effects of this type of cultural shock on domestic violence.

\medskip

Finally, within the broader framework of cultural change, we also contribute to the literature on the diffusion of innovations and the role of market conditions in shaping adoption \citep{Griliches1992, Aghionetalt2005, Aghionetalt2009, Aghionetalt2014}. Our preferred interpretation—supported by both anecdotal and empirical evidence—is that religious competition catalyzes the emergence and diffusion of new cultural technologies, thereby contributing to sustained reductions in domestic violence.  This perspective highlights the importance of competition—particularly among religious organizations—in driving cultural transformation, which we view as a central contribution to the field of cultural economics.

\medskip

The remainder of the paper is organized as follows. Section \ref{Background} provides background on the Colombian religious landscape and domestic violence. Section \ref{DataandEmpiricalStrategy} outlines the data and empirical strategy. Section \ref{MainResults} presents the main results. Section \ref{sec_mechanisms} explores potential mechanisms. Section \ref{Conclusion} concludes.


\section{Background} \label{Background}

In this section, we provide a brief overview of the Colombian context, focusing on the evolution of the religious landscape and patterns of domestic violence.

\subsection{Religious Landscape}

\begin{figure}[H]
\begin{center}
             \caption{Religious Adherence, 1996-2023}
        \label{fig_reladherencelatinobarometer}
        \vspace{-0.3cm}
\includegraphics[width=12cm,height=7cm]{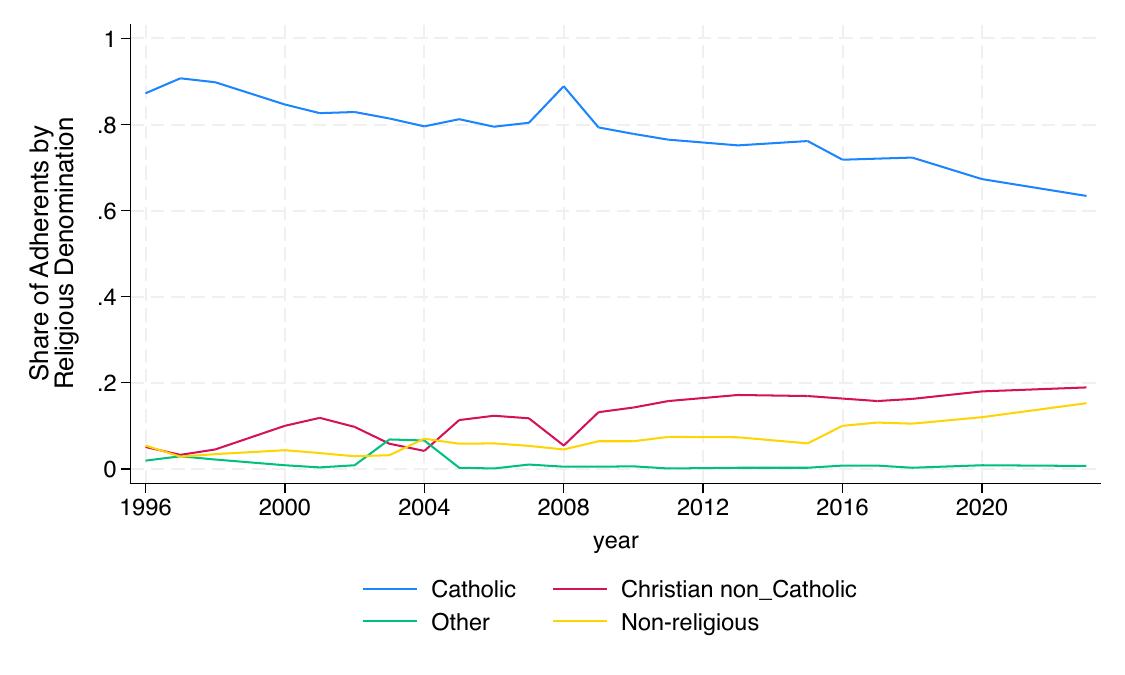}
     \begin{minipage}{12cm} \scriptsize \textbf{Note}: This figure uses data from the Latinobarómetro for Colombia, covering all available waves between 1996 and 2023. It plots the proportion of responses to the question, `What is your religion?' where ``Christian non-Catholic" includes the following denominations: Evangelical, Baptist, Methodist, Pentecostal, Adventist, Jehovah’s Witness, and Protestant. `Non-religious' includes responses such as Agnostic and Atheist.
\end{minipage}
\end{center}
\end{figure}

Over the past few decades, Colombia has experienced a major transformation in its religious composition. Historically, Catholicism dominated religious life for over four centuries, protected by state-imposed barriers that limited the growth of alternative affiliations. Between 1950 and 1990, the share of the population identifying as Catholic remained relatively stable, ranging between 91\% and 95\%. However, this began to change in the late 1990s. By the end of the 2010s, Catholic affiliation had declined to roughly 75\%—a drop of about 20 percentage points—while Protestantism had gained approximately 15 percentage points over the same period (see Figure \ref{fig_reladherencelatinobarometer}; also \citealt[][p. 27]{PewResearchCenter2014}).

\medskip

This trend reflects broader patterns across Latin America, although the timing varies by country (\citealp{PewResearchCenter2014}; \citealp{SommaBargstedValenzuela2017}). In Colombia, Pentecostalism has emerged as the fastest-growing Protestant denomination. By the late 2010s, at least 56\% of Protestants identified as Pentecostal (\citealt[][p. 62]{PewResearchCenter2014}; \citealt[][]{DANEECP2021}). Several factors explain this rapid growth: increased urbanization, the search for hope amid hardship, disillusionment with the Catholic Church, and the organizational agility of Pentecostal groups in addressing social needs \cite[][]{Bastian2005, Beltran2013}.

\medskip

Pentecostal churches are often perceived as effective due to their charismatic leadership, flexible structures, and adept use of marketing and media \cite[][]{Bastian2005, Beltran2013}. Women frequently lead this shift—often being the first to convert and subsequently encouraging the participation of husbands, children, and other relatives (\citealp[p. 5]{Brusco1995}; \citealp[p. 184]{Beltran2013}). The Pentecostal landscape in Colombia is also marked by ideological diversity, institutional fragmentation, and limited financial resources, with many congregations founded by independent leaders \citep[see][p. 19]{Tejeiro2010}.

\subsection{Domestic Violence}\label{sec_background_domesticviolence}

\begin{figure}[H]
\begin{center}
             \caption{Domestic Violence, 2005-2019}
        \label{domesticviolence20052019_fig}
        \vspace{-0.3cm}
\includegraphics[width=11.5cm,height=7.5cm]{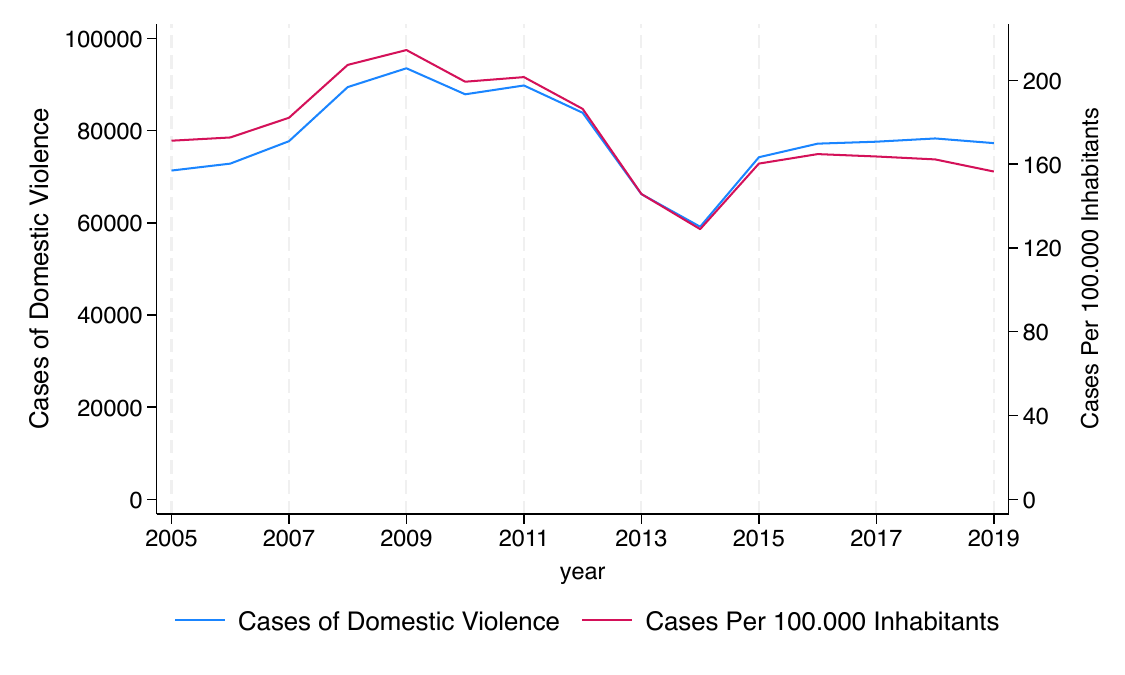}
     \begin{minipage}{12cm} \scriptsize\textbf{Source}: \emph{Forensis} (Colombian \emph{National Institute of Legal Medicine and Forensic Sciences})
\end{minipage}
\end{center}
\end{figure}

Domestic violence—defined as verbal, psychological, physical, or other forms of abuse by one family member against another—is a criminal offense in Colombia, punishable by a minimum of four years in prison.\footnote{For legislation defining this crime, see Law 599 of 2000 and Law 906 of 2004.} Over the past two decades, the incidence of reported domestic violence has fluctuated considerably. As shown in Figure  \ref{domesticviolence20052019_fig}, this variation from 2005 to 2019 partially reflects changing economic conditions (see Figure \ref{fig_domviolenceGDP}  in the Appendix). Despite short-term fluctuations, the overall trend appears slightly downward.\footnote{A key feature of domestic violence in Colombia is its high impunity rate: in 2006, over 70\% of cases went unpunished, largely due to underreporting and judicial inefficiencies \cite[][]{DefensoriaAUG2006, SemanaDEC2021}. In response, Law 1542 of 2012 allowed authorities to investigate cases without a formal complaint and enabled third-party reporting \cite[][]{AmbitojuridicoJUL2012}. Although its impact has not been rigorously assessed, the law’s implementation coincided with a rise in reported cases \cite[][]{eltiempoJUN252018}.  This increase may reflect improved reporting—possibly including duplicate or false claims—rather than a true rise in incidence. As a result, it is possible that actual cases declined, for example, due to the heightened threat of punishment.}
 
\medskip

Women constitute the majority of victims, particularly those aged 29 to 44. The main forms of domestic violence include intimate partner violence, family violence, abuse against minors, and abuse of the elderly \cite[][]{INMLCFforensis}. Empirical studies identify several risk factors: poverty, low household education, young motherhood, female household presence, and female labor force participation \cite[][]{RiberoSanchez2005, GafaroIbanez2012}. 

\medskip

Importantly, cultural factors are consistently cited by victims and researchers alike as core contributors to domestic violence. Surveys conducted by the INMLCF reveal that victims often attribute violence to alcohol consumption, machismo, and intolerance \cite[][]{INMLCFforensis}. These findings align with broader sociopolitical efforts in Colombia to address domestic violence through cultural transformation. For instance, national family policies have sought to replace hierarchical household norms with more egalitarian models  \cite[][]{Minsalud2016}.\footnote{In the Colombian main government's policy document regarding family matters, strategy 2A explicitly aims to confront and transcend cultural ideas rooted in authoritarianism while promoting the development of family and societal perspectives grounded in democratic relationships \cite[see][p. 21]{Minsalud2016}.} Yet, many of these state initiatives have been criticized for their limited reach, lack of coordination with civil society, and focus on victim assistance rather than prevention \cite[][p. 41-48]{Santanderetalt2020}.

\medskip

In this context, non-governmental actors—particularly religious institutions—have played a critical role in promoting behavioral norms that discourage violence. Both the Catholic Church and non-Catholic Christian churches have publicly condemned domestic violence and supported initiatives aimed at changing gender norms and improving family life (\citealp[][p. 220]{Beltran2013}; \citealp[][]{CEC2015, CEC2017, Caritas2017})


\section{Data and Empirical Strategy}\label{DataandEmpiricalStrategy}


\subsection{Data}\label{subdata}

The first key data source for this study documents the establishment of non-Catholic churches in Colombian municipalities. As discussed in the introduction, this is our primary variable for measuring the presence of a non-Catholic Christian religious organization in a given municipality.\footnote{The data source does not specify denominational affiliation. However, a preliminary analysis of church names indicates that the vast majority are Christian.} Given Colombia’s historically Catholic religious landscape, we interpret the establishment of these churches as indicative of increased religious competition. In the results section, we present evidence supporting this interpretation.

\medskip

Data on non-Catholic churches come from the \emph{Public Registry of Religious Organizations}, maintained by the Colombian Ministry of the Interior.\footnote{In Spanish, the \emph{Registro P\'ublico de Entidades Religiosas}. Publicly available at \url{https://www.mininterior.gov.co/registro-publico-de-entidades-religiosas/}.} The registry records the exact date and municipality in which a non-Catholic church obtains legal ``personhood.” While the Catholic Church is automatically granted legal status, non-Catholic churches must apply. The application process typically takes 60 working days and requires basic documentation, including the founding act, organizational bylaws, and identification of legal representatives. There are no specific requirements regarding membership size or duration of activity, and approval is generally granted \cite[see][]{Prieto2012}. Legal recognition allows churches to engage in a wide range of formal activities—such as opening bank accounts, hiring staff, paying leaders, acquiring property, and receiving tax benefits—and to collect member contributions under Law 133 of 1994. Our dataset covers the period from 1995 to 2021 and includes the date of legal recognition and the primary municipality of operation for each church.\footnote{There are two potential sources of measurement error in this dataset. First, a non-Catholic church may begin operating in a municipality well before applying for legal recognition—or may never apply at all. However, the process is simple, free, and legally guaranteed by Article 19 of the Colombian Constitution. It offers material benefits such as tax exemptions and requires only basic legal documentation. Thus, while we cannot rule out this issue entirely and must interpret our estimates with caution, it likely results in underestimation of religious competition and biases our results downward. Second, once a church receives legal recognition, it is permitted to operate nationwide. This means that the first legally recognized church in a municipality may not be the first to begin local activities—another church with legal status obtained elsewhere may already be present. Nonetheless, evidence suggests that non-Catholic churches in Colombia are highly fragmented (\citealt[][p. 140]{Beltran2013}; \citealt[][p. 19]{Tejeiro2010}), reducing the likelihood and severity of this issue.}

\medskip

\begin{figure}[H]
\begin{center}
             \caption{New non-Catholic Churches, 1995-2021}
             \vspace{-0.3cm}
        \label{nonCathoyears_fig}
\includegraphics[width=12cm,height=8cm]{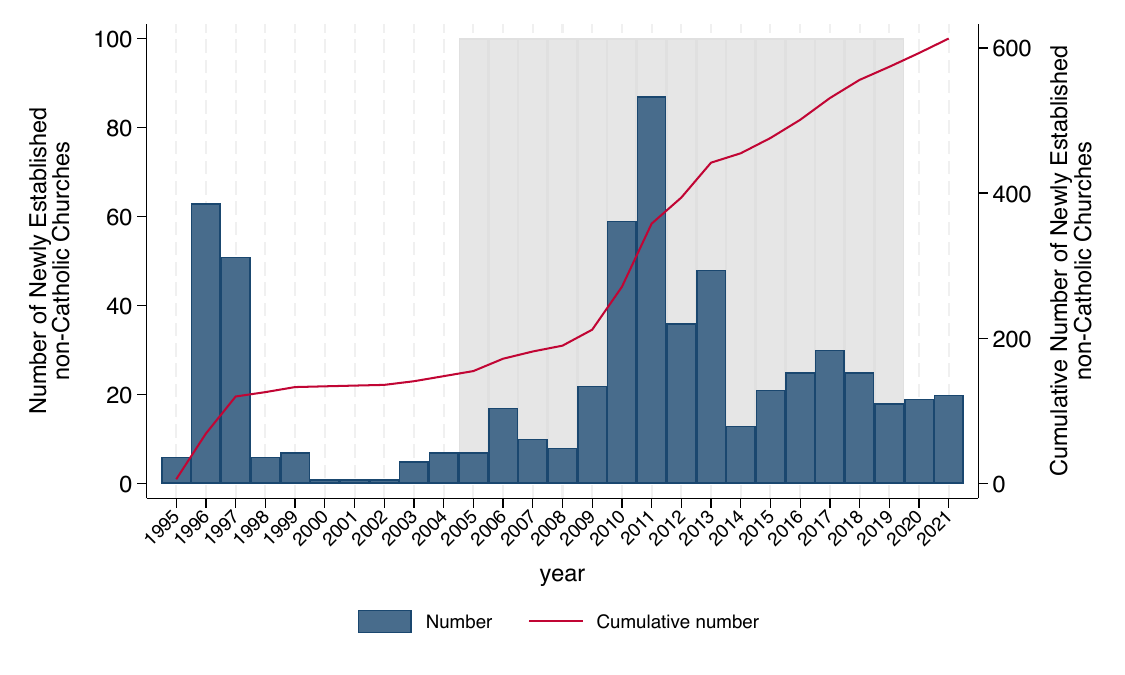}
     \begin{minipage}{12cm} \scriptsize \textbf{Source}: Public Registry of Religious Organizations (Office of Religious Affairs, Colombian \emph{Ministry of the Interior})
\end{minipage}
\end{center}
\end{figure}

Between 1995 and 2021, a total of 582 non-Catholic churches received legal recognition for the first time. Figure \ref{nonCathoyears_fig} shows the temporal distribution of these events, while Figures \ref{ncatchurchesmap_fig} and \ref{ncatchurchesmapchange_fig}  present their spatial distribution. These churches were established in 497 municipalities, with the vast majority (435 municipalities) recording only one new church in a given year. The phenomenon was geographically widespread and not confined to any specific region. During our primary period of analysis (2005–2019), an average of 28.4 new non-Catholic churches were established each year. This trend aligns with the broader religious transition discussed in Section \ref{Background}, which documents a decline in Catholic adherence and a concurrent rise in Protestantism. See Table \ref{nonCatho_tab} for descriptive statistics.

\medskip

Our second key data source provides information on domestic violence and is obtained from the \emph{National Institute of Legal Medicine and Forensic Sciences} (\emph{Instituto Nacional de Medicina Legal y Ciencias Forenses}, INMLCF). This institution, responsible for forensic investigations in Colombia, publishes annual statistical bulletins on violence through the \emph{Forensis} report.\footnote{Available at \url{https://www.medicinalegal.gov.co/cifras-estadisticas/forensis}.} Since 2005, the INMLCF has reported municipal-level data on domestic violence, including both total cases and those specifically involving intimate partner violence. It defines domestic violence as intentional acts committed by family members that result in physical, mental, or sexual harm, restriction of movement, or death. The data are based on reported injuries that undergo forensic medical evaluation.

\medskip

As a robustness check, we also use an alternative data source on domestic violence from the \emph{National Police of Colombia}.\footnote{Available at \url{https://www.policia.gov.co/grupo-informacion-criminalidad/estadistica-delictiva}.} While the definition of domestic violence is broadly consistent with that used by the INMLCF, the police data are less comprehensive, subject to fewer quality controls, and only available at the municipal level from 2010 onward. For these reasons, we use this dataset solely in robustness analyses.

\medskip
Finally, we incorporate several additional data sources for control variables. These include the number of civil society organizations and Catholic churches in each municipality (available for 1995), ethnic composition (available for 1993, 2005, and 2018), municipal population, an index of rurality, and the proportion of residents with unsatisfied basic needs (used as a proxy for poverty, also available for 1993, 2005, and 2018). Detailed information on the sources of these variables is provided in the note to Table \ref{precharacteristics_tab}.


\subsection{Empirical Strategy}

To estimate the impact of religious competition on domestic violence, we use a linear two-way fixed effects (FE) estimator with municipality and year fixed effects. Our main specification models a given outcome $y_{i,t}$ (e.g., the number of reported cases of domestic violence per 100{,}000 inhabitants) in municipality $i$ and year $t$ as a function of the arrival of the first non-Catholic church:

\begin{equation}
\label{didbaselines}
y_{i,t} = \alpha_i + \beta_t + \gamma D_{i,t} + \epsilon_{i,t}
\end{equation}

Here, $D_{i,t}$ is a treatment indicator equal to one if municipality $i$ is treated at time $t$ or earlier, and zero otherwise. We refer to this specification as \emph{static}, as it estimates the average long-term (non-transitory) effect of treatment across all post-treatment years. Untreated municipalities serve as the control group in this specification.

\medskip

To examine the time path of treatment effects and to assess the validity of the identifying assumptions, we also estimate a \emph{dynamic} specification:

\begin{equation}
\label{didbaseline}
v_{i,t} = \alpha_i + \beta_t + \sum_{\tau = -K}^{-2} \gamma_\tau D_{i,t}^\tau + \sum_{\tau = 0}^{L} \gamma_\tau D_{i,t}^\tau + \epsilon_{i,t}
\end{equation}

In this case, $D_{i,t}^\tau$ is an indicator equal to one if municipality $i$ is $\tau$ years away from treatment in year $t$ (with treatment occurring at $\tau = 0$).\footnote{Negative values of $\tau$ represent leads (i.e., years before the first non-Catholic church is established), while positive values represent lags (i.e., years after treatment). To avoid multicollinearity, we omit the indicator for $\tau = -1$ and normalize effects relative to that period, as discussed by \citet{laporte2005estimation}, \citet{BorusyakJaravel2017}, and \citet{SunAbraham2021}.} This model allows us to test for the presence of pre-treatment trends and to explore the evolution of effects over time.

\medskip

The key identifying assumption underlying both specifications is that municipalities not yet exposed to treatment provide a valid counterfactual for estimating what would have occurred in treated municipalities in the absence of treatment. In other words, conditional on observables, treated and not-yet-treated municipalities are assumed to follow parallel trends in domestic violence. This assumption could be violated if, for instance, the timing of church entry is systematically related to unobserved shocks that also influence domestic violence.

\medskip

To support the credibility of this conditional parallel trends assumption, we adopt several empirical strategies. First, we include municipality fixed effects ($\alpha_i$) to absorb time-invariant heterogeneity across municipalities—such as  historical factors and geographic features—and year fixed effects ($\beta_t$) to account for national trends or shocks. Second, we control for key time-varying municipal-level covariates, including population size, rurality, and poverty indicators, which may be correlated with both treatment timing and domestic violence. Third, we estimate the dynamic specification in Equation~(\ref{didbaseline}) to test for differential pre-treatment trends. Finally, we conduct robustness checks using alternative difference-in-differences estimators that allow for treatment effect heterogeneity and rely on weaker identification conditions.

\medskip

We also assume the absence of reverse causality—that is, the establishment of a non-Catholic church affects domestic violence, but not the other way around. This assumption is plausible given the staggered and plausibly exogenous timing of church entry, which we show is not systematically related to pre-existing trends in domestic violence.

\medskip

Additionally, we assume homogeneous treatment effects across municipalities and over time. While this is a strong assumption, particularly in settings with staggered adoption, we test the robustness of our findings using estimators explicitly designed to handle heterogeneous effects. Nonetheless, we focus primarily on the static specification, using the dynamic model mainly to assess identification validity and to explore treatment dynamics.

\medskip

We provide several pieces of evidence to support the validity of our empirical strategy. First, Tables~\ref{dfirstiglesiasnc_tab} and~\ref{dfirstiglesiasnc5lags_tab} show that pre-treatment levels of domestic violence and related factors—such as poverty, homicide, displacement, and political outcomes—are uncorrelated with the timing of church entry, including up to four years prior. This suggests that reverse causality is unlikely. Second, we show that our results are robust to the inclusion of various potentially confounding time-varying covariates. Third, we demonstrate that our estimates remain consistent when using heterogeneity-robust difference-in-differences estimators, including those proposed by \citet{ChaisemartinDHaultfoeuille2020}, \citet{SunAbraham2021}, and \citet{CallawaySantAnna2021}. Taken together, this evidence supports our interpretation of the estimates from Equations~(\ref{didbaselines}) and~(\ref{didbaseline}) as causal average treatment effects.

\section{Effects of Religious Competition on Domestic Violence}\label{MainResults}

Table \ref{twfe_domeviolenceMLrate_tab}  presents our main results.  The outcome of interest is the number of reported domestic violence cases per 100,000 inhabitants at the municipal level. We find that the establishment of the first non-Catholic church in a municipality is associated with a significant decline in domestic violence. Columns (1) and (2) of Panel A in Table \ref{twfe_domeviolenceMLrate_tab} show results from the static specification, indicating a reduction of 10.6 to 14.1 cases per 100,000 inhabitants. Panel B reports results from the dynamic specification, which, despite being less precise, displays a similar pattern over most post-treatment years.  The initial year of exposure shows a statistically significant reduction of 8.9 to 10.7 cases per 100,000. The effect persists in some of the subsequent years and peaks at 14.5 fewer cases by the fifth year (see columns (1) and (2) of Panel B).\footnote{An exception is the third post-treatment year, where the effect is smaller and not statistically significant. One possible explanation is that the initial decline was short-lived but later strengthened as religious competition intensified—potentially due to strategic responses by the Catholic Church. The effect appears to persist until year seven, vanish in year eight, and re-emerge in years nine and ten, possibly reflecting new adaptive strategies by both religious groups.} This overall decline corresponds to a decrease in domestic violence of 0.1 of its standard deviations.

\medskip

\begin{figure}[H]
             \caption{Effect of the First Non-Catholic Church on Domestic Violence (Against Intimate Partners and Other Household Members)}
        \label{twfe_domeviolenceMLratetypes_fig}
\vspace{-0.3cm}
\begin{subfigure}{0.5\textwidth}
\caption{Violence Against Intimate Partner} \label{}
\includegraphics[width=\linewidth]{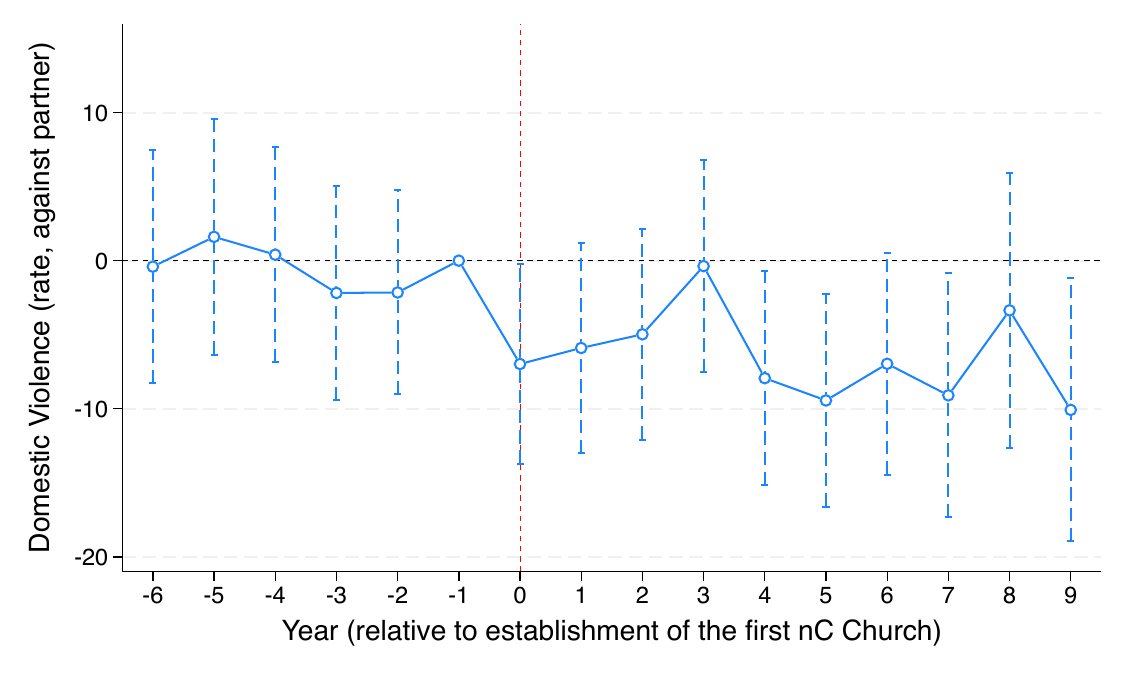}
\end{subfigure}\hspace*{\fill}
\begin{subfigure}{0.5\textwidth}
\caption{Violence Against Other Household Members} \label{}
\includegraphics[width=\linewidth]{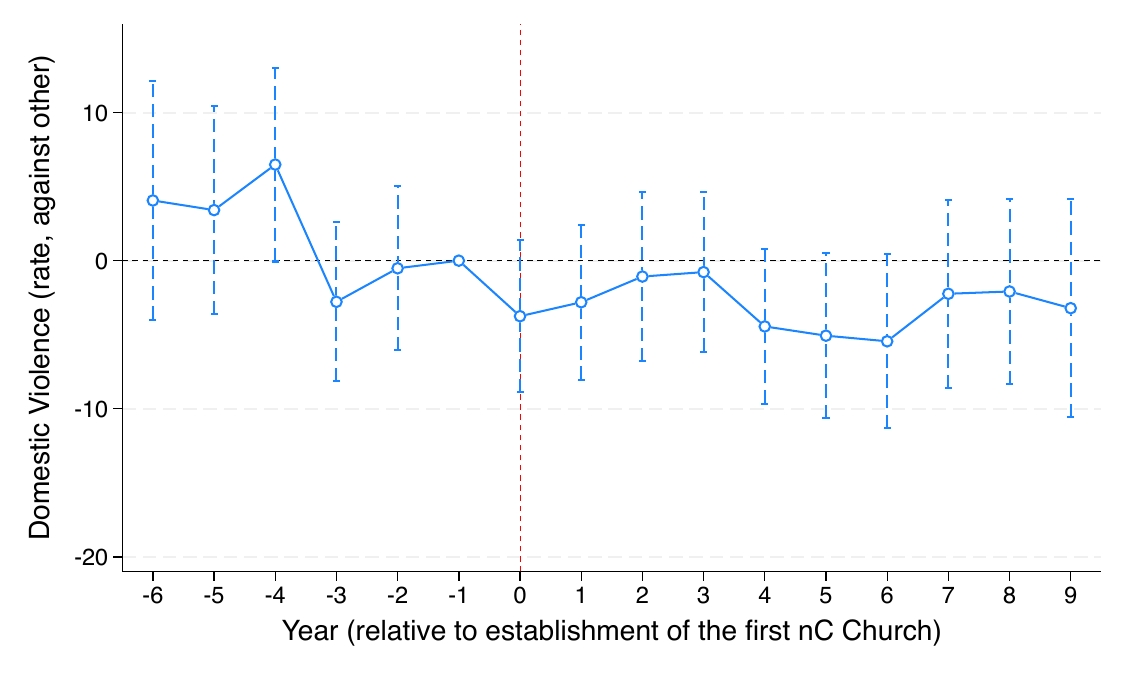}
\end{subfigure}\\
     \begin{minipage}{16.5cm} \footnotesizes \textbf{Note:}  These figures show the two-way fixed effects estimates from Eq. (\ref{didbaseline}), in a specification that includes municipality and  department $\times$ year fixed effects, and the following (lagged) covariates:  log of total population, rurality index, and proportion of the population with unsatisfied basic needs. Vertical lines indicate 90\% confidence intervals.
\end{minipage}
\end{figure}

Columns (3) to (6) of Table \ref{twfe_domeviolenceMLrate_tab} and Figure \ref{twfe_domeviolenceMLratetypes_fig} disaggregate the results by victim identity—distinguishing intimate partner violence from other intra-household violence. The estimates in these columns indicate that both types of violence exhibit similar but slightly smaller effects. Given that most victims of intimate partner violence in Colombia are women, these findings suggest that the observed effects may extend to broader household dynamics, not only violence against women.

\medskip
Tables \ref{twfe_domeviolenceMLrateALTESTM_tab} and  \ref{twfe_domeviolenceNPrate_tab}  present results that examine the robustness of our findings against alternative DID estimators and a different dataset on domestic violence. Table \ref{twfe_domeviolenceMLrateALTESTM_tab} shows that the results from Table \ref{twfe_domeviolenceMLrate_tab} remain qualitatively robust when using alternative DID estimators that do not assume homogeneous treatment effects. Specifically, we consider the estimators introduced by \cite{ChaisemartinDHaultfoeuille2020} (columns (1) and (2)), by \cite{CallawaySantAnna2021} (columns (3) and (4)), and by \cite{SunAbraham2021} (columns (5) and (6)). Importantly, for nearly all years and estimators, the estimates do not exhibit statistically significant differences compared to the results in Table \ref{twfe_domeviolenceMLrate_tab}.\footnote{We note some differences in standard errors, suggesting variation in statistical significance across specifications. Although we currently lack a definitive explanation for this divergence, we attribute it to the assumption of heterogeneous treatment effects underlying these estimators. To obtain these estimates, we used Stata commands, specifically \emph{twowayfeweights}, \emph{csdid}, and \emph{eventstudyinteract}.} As previously discussed, this supports the interpretation of our estimates as causal average treatment effects on the treated.

\medskip
Table \ref{twfe_domeviolenceNPrate_tab}, using domestic violence data from the Colombian National Police, yields similar results. As discussed in Section \ref{subdata}, this dataset is relatively independent from the INMLCF data used in our baseline analysis. The static model again shows a negative, statistically significant effect of comparable magnitude. Results from the dynamic specification are also consistent in sign.\footnote{These findings mitigate concerns about underreporting, which may affect domestic violence statistics. The INMLCF data likely has lower underreporting, as it is generated by medical professionals who document injuries without assigning legal culpability. Thus, if one dataset is more susceptible to underreporting and our results remain consistent across both datasets, we can reasonably infer that underreporting does not substantially affect our findings.}

\medskip
Finally, Table \ref{tab_twfe_em} explores the effect of religious competition intensity, measured by the number of non-Catholic churches in a municipality. We find small, statistically insignificant reductions—approximately 1 to 3 fewer reported cases per 100,000 inhabitants per additional church. This result suggests that the key factor driving the observed effects may be the novelty associated with the establishment of the first non-Catholic church and, in particular, we hypothesize, the new religious practices introduced by that initial church.


\section{Mechanisms}\label{sec_mechanisms}

The findings discussed in the previous section can be interpreted in various ways. Here, we present what we consider the most plausible explanation and offer both quantitative and qualitative evidence in support. We also evaluate several alternative mechanisms, which, while theoretically relevant, appear less consistent with the data.

\subsection{Religious Practices and Norm Change}

Our preferred explanation is that the arrival of a non-Catholic church introduces new social practices that are broadly perceived as beneficial and are closely associated with a reduction in domestic violence. In the context of religious competition, these practices are subsequently adopted by other religious institutions and disseminated more widely within the community. As a result, they gain traction among a substantial share of households, contributing to a meaningful decline in domestic violence.

\medskip

We support this interpretation with both quantitative and anecdotal evidence.

\medskip

First, numerous studies show that major religious organizations in Colombia actively promote values and behaviors that strengthen interpersonal relationships within households.  Drawing on extensive interviews with members of non-Catholic churches,  \cite{Beltran2013} identifies a set of practices consistently emphasized by these groups:
\begin{quotation} \small{``The new religious movements condemn all addictions, especially alcohol and cigarette consumption, and promote changes in consumption habits. [...] Additionally, thanks to their participatory dynamics—especially among Pentecostals—they promote female leadership, turning these churches into spaces of empowerment for women."} \cite[see][p. 220]{Beltran2013}.
\end{quotation}

These participatory dynamics often involve weekly workshops and seminars held on weekends, with strong expectations for member attendance. Although the topics vary, a consistent focus is the strengthening of couple and family relationships. These activities align with broader lifestyle changes observed among converts, including reductions in alcohol use, promotion of marriage and formal family structures, and increases in female leadership within congregations  (see \citealp[p. 184 and 379]{Beltran2013}).\footnote{Churches following this model include the Iglesia Adventista del Séptimo Día, Iglesia Integral Semillero de Fe y Esperanza, Iglesia Misión Familia, Iglesia Centro Bíblico de la Sabana, and the Iglesia Pentecostal Unida de Colombia. Leaders from several of these churches highlight the importance of weekend workshops \cite[see][]{CDLR2021youtube} and couples’ seminars \cite[see][]{CHIAsecretaria2022}. The Iglesia Pentecostal Unida de Colombia provides detailed information online, explaining how these sessions—part of Sunday School—guide participants through booklets that help them reflect on harmful behaviors experienced in childhood (e.g., domestic violence or coercion) and identify those they do not wish to repeat. Figure \ref{fig_booklet} provides an example of such materials.}

\medskip

A key feature of these practices is their direct relevance to domestic violence. Pastors interviewed by \citeauthor{Beltran2013} report that women constitute roughly two-thirds of their congregations and often derive tangible benefits from participation, including emotional support and protection from economic precarity and abuse \citep[see][p. 184]{Beltran2013}. As \cite{Brusco1995} observes: 
\begin{quotation} \small{“One outcome of conversion [to non-Catholic churches] is that the relative power positions of the spouses change. This is not to say that women now have power over their husbands. In evangelical households the husband may still occupy the position of head, but his relative aspirations have changed to coincide with those of his wife”} \cite[p. 137]{Brusco1995}.
\end{quotation}

Congregants themselves recognize the connection between these transformations and a reduction in domestic violence. As \citeauthor{Beltran2013} notes:  ``To the extent that new religious movements condemn alcohol consumption and sexual permissiveness, conversion brings benefits to the families of converts, which translate, for example, into a reduction in domestic violence” \cite[pp. 173, 185]{Beltran2013}. This view is echoed by  \cite{Reymartinez2008} and supported by testimonies collected by the national forensic institute \cite[see][]{INMLCFforensis}, which cite reduced alcohol consumption, increased female empowerment, and rejection of sexist norms as key protective factors. This association is consistent with testimonies by converts  \citep{Brusco1995, Reymartinez2008}. As \citeauthor{Brusco1995} summarizes: ``Aggression, violence, pride, self-indulgence, and an individualistic orientation in the public sphere are replaced by peace seeking, humility, self-restraint, and a collective orientation and identity with the church and the home” \cite[p. 137]{Brusco1995}.

\medskip

The Catholic Church has also promoted similar values and practices, though often through different strategies. In the past decade, violence against women has gained prominence in Catholic Church communications \citep[see][]{CEC2015, CEC2017, Caritas2017}. During the same period, the Colombian Catholic Church has supported lay associations that operate with varying levels of autonomy and play a key role in disseminating religious practices, particularly among women and families \citep[see][p. 292]{Beltran2013}. An illustrative example is captured in a testimony from a member of a lay Catholic organization in Bogotá:
\begin{quotation}\small{``To face all these challenges within the framework of social pastoral action, we founded the Fundehi Corporation 15 years ago, simultaneously committing to the promotion of women. [...] Before, I didn't have time for God or my children, as I left home at 4 a.m. and returned around 9 at night," she confesses. ``Now I share what I've learned, participate in Saturday Bible groups, and celebrate the Eucharist on Sundays, also with my children.'' \cite[see][]{eltiempoJUN242022}.}
\end{quotation}

\medskip
Tables \ref{tab_attitude_2019} and \ref{tab_attitude_wvs} provide further support for the hypothesis that Catholics and non-Catholic Christians in Colombia share their core values. Both groups place a high priority on religion and family, with non-Catholic Christians often expressing even greater emphasis. However, they do not consistently differ in their stated opposition to violence (see columns (1)–(2) in Table \ref{tab_attitude_2019} and columns (1), (2), (5), and (6) in Table \ref{tab_attitude_wvs}). This convergence in values contrasts with differences in behavior. High levels of church attendance are associated with similarly low tolerance for violence across both groups, but among low-attendance individuals, non-Catholic Christians express significantly less rejection of violence. In short, doctrinal similarity does not guarantee behavioral similarity; rather, it is religious practice—and not denomination—that plays a more decisive role. These findings support our broader argument: religious competition affects behavior not by altering official doctrine, but by shaping how norms are internalized through active community participation. This interpretation aligns with sociological evidence showing that infrequent or inactive believers often diverge from the normative stances of their religious communities \citep{ellison2009religious, bartkowski2000conservative}.

\medskip

Taken together, the evidence supports the view that major religious organizations in Colombia promote values and practices associated with reduced tolerance for domestic violence. This raises important questions about the underlying mechanisms. Our primary hypothesis centers on a cultural shift away from norms that have historically enabled domestic violence—such as rigid gender roles and alcohol use—toward norms emphasizing mutual respect, gender equity, and sexual restraint. As noted in the introduction, this mechanism is consistent with a growing body of research linking religious influence to reductions in domestic violence \citep[see][]{CoolsKotsadam2017, SvecAndic2018, TurPrats2019, Gonzalez2020, TurPrats2021}. In the Colombian context, we argue that newly established non-Catholic churches introduced alternative social practices and participatory dynamics that reinforced group norms and promoted behavioral accountability—particularly among individuals actively involved in religious life.

\medskip

To test this mechanism empirically, we draw on data from Waves 5 through 7 of the World Values Survey (WVS), which ask respondents to what extent they believe a man is justified in beating his wife (on a scale from 1 = never justified to 10 = always justified). For Colombia, data are available at the regional level. We examine whether regions with more new non-Catholic churches exhibit greater rejection of domestic violence. Table \ref{table_wvsnumbchurchers} presents the results, estimated using a strategy aligned with our main specification. The findings show that a higher number of new non-Catholic churches is significantly associated with lower justification of wife-beating, particularly among respondents who report above-median levels of religious attendance. These results support the view that religious competition influences social norms primarily through sustained engagement and norm reinforcement within religious communities. They are consistent with our broader argument that changes in religious dynamics in Colombia have contributed to a normative shift in attitudes toward gender roles and domestic violence.

\medskip

An additional mechanism that may complement our main hypothesis involves women’s participation in the labor market. While this channel also reflects broader cultural change, it operates specifically through economic empowerment. As gender norms evolve, women may become more likely to enter the labor force, thereby increasing their bargaining power within the household and reducing domestic violence by strengthening their outside options in marriage. The empirical literature offers mixed evidence on this channel: some studies find that increased female labor force participation reduces domestic violence \citep{StevensonWolfers2006, Aizer2010, PerovaReynoldsSchmutte2023}, while others highlight the potential for backlash or emphasize the importance of context-specific effects \citep{Chin2012, GuarnieriRainer2021, ErtenKeskin2021, ErtenKeskin2024, Bergvall2024}.

\medskip 

Table \ref{tab_labormarketdeptos} explores this mechanism by reporting correlations between the number of new non-Catholic churches and labor market outcomes at the departmental level—the most disaggregated level for which time-varying labor data are available in Colombia. The results indicate that departments with more new non-Catholic churches tend to have lower unemployment and higher employment rates for both men and women. Labor force participation also increases, although the effect is somewhat weaker for women. These patterns suggest that religious competition may foster environments more conducive to economic engagement—potentially by reshaping gender norms (for women) and by encouraging more disciplined behavior (for men). This evidence lends plausibility to the economic empowerment channel as an indirect pathway through which religious competition may reduce domestic violence.\footnote{Because this mechanism also entails a form of cultural transformation, we view it as consistent with—and complementary to—our broader argument. However, we do not present it as the sole specific channel, given the correlational (and not necessarily causal) nature of the results in Table \ref{tab_labormarketdeptos}.}

\subsection{Diffusion and Religious Competition Dynamics}

Having discussed what we consider the most plausible mechanism through which religious competition may lead to reductions in domestic violence, we now turn to a set of related questions concerning the diffusion of these practices. Specifically, we ask: Which religious organizations initially introduced and promoted these practices? And, if the timing of adoption differed across groups, did these practices eventually diffuse beyond the organizations that pioneered them—particularly through mechanisms of interdenominational competition?

\medskip 

With respect to the first question, anecdotal evidence suggests that while Catholic and non-Catholic Christian churches in Colombia have increasingly converged around similar agendas—emphasizing responsible consumption habits and stronger family relationships—non-Catholic churches likely introduced this agenda first, or at least promoted it on a broader scale. This diffusion appears to have been spurred by religious competition, which then prompted the Catholic Church to incorporate similar practices into its own activities \citep[see][]{OspinaSanabria2004, Cleary2011, Moreno2011, Beltran2013}.\footnote{Competition between Catholic and non-Catholic churches in Colombia parallels the dynamic between Catholic and Protestant missionaries analyzed in \cite{gallego2010christian}.}

\medskip

A clear example of this phenomenon is the so-called ``Pentecostalization of Catholicism,” referring to the expansion and revitalization of the Charismatic Renewal movement within the Colombian Catholic Church in recent decades \citep[see][pp. 54–77]{Cleary2011}; \citep[see also][p. 77]{Beltran2013paper}. One notable feature of this process was its timing, which coincided with a significant increase in the Catholic Church’s presence in mass media, including the creation of Catholic television channels and the rise of charismatic Catholic priests on widely viewed platforms.\footnote{Examples include Jesús Orjuela Pardo (``Chucho"), Alberto Linero, and Alirio López, all of whom gained prominence in Colombian mass media in the early 2000s \citep{eltiempoMARCH152006}.} Importantly, these strategies had long been employed by non-Catholic Christian churches since their entry into Colombia in the early 1990s \citep[p. 184]{Beltran2013}.\footnote{\citeauthor{Beltran2013} notes that non-Catholic churches reached wide audiences by broadcasting testimonies of households that had been “restored” through “the power of God” via radio and television \cite[see][p. 184]{Beltran2013}.} The subsequent adoption of similar practices by the Catholic Church has been interpreted as a response to this earlier and more expansive use of mass communication by non-Catholic denominations \citep[see][pp. 284, 292]{Beltran2013}.

\medskip 

Another important dimension of the Pentecostalization of Catholicism in Colombia is the growing endorsement by the Catholic Church of lay organizations centered on individual initiative and frequently led by women. This development is also seen as a strategic adaptation to the long-standing tradition within non-Catholic churches of fostering individual agency and elevating women’s roles within religious communities \citep[see][pp. 184, 291–292]{Beltran2013}.

\medskip

Turning to the second question—whether these practices diffused beyond members of non-Catholic churches—the available anecdotal evidence suggests that religious organizations have leveraged various media channels to reach a wide audience with the explicit goal of promoting values and behaviors conducive to healthier family relationships. This diffusion mechanism supports our broader interpretation of religious competition as a channel for cultural change, even among non-members.

\medskip

To complement the qualitative evidence discussed above, we now present additional quantitative results that reinforce the plausibility of our interpretation. The first set of results concerns the core assumption that the arrival of a non-Catholic church in a municipality increases interdenominational competition, which in turn drives reductions in domestic violence.

\medskip 

Table \ref{twfe_domeviolenceMLrate_heteroeffects_tab} provides empirical support for this interpretation by examining heterogeneous effects across municipalities where religious competition is likely to be more intense. Specifically, we focus on municipalities characterized by: (i) a stronger historical Catholic presence (which may prompt a more active response by the Catholic Church to retain followers), (ii) smaller population size (where new churches may reach a larger share of residents), and (iii) younger demographic composition (where individuals may be more receptive to religious alternatives).\footnote{We thank an anonymous referee for suggesting several of these hypotheses.} If religious competition is the underlying mechanism, we expect the effect of a new non-Catholic church to be larger in these settings.

\medskip

The estimates in Table \ref{twfe_domeviolenceMLrate_heteroeffects_tab} use the same empirical specification as Panel A of Table \ref{twfe_domeviolenceMLrate_tab}, with municipalities stratified by (i) the number of Catholic churches per 100,000 inhabitants in 1995 (Panel A),\footnote{\label{footnotecatholic} As previously noted, approximately 78.4\% of Colombia’s population identifies as Catholic, while about 10.1\% identifies as Protestant. Due to the lack of municipality-level data on religious affiliation, we use the number of Catholic churches in 1995 as a proxy.} (ii) population size in 2004 (Panel B), and (iii) the share of residents aged 25 to 44 (Panel C).  Tables \ref{twfe_domeviolenceMLrate_highlowcatholic_tab} through \ref{twfe_domeviolenceMLrate_highlowyoung_tab} report similar findings using a dynamic specification. Across models, the estimated effects are consistently larger in municipalities where we expect competition to be more pronounced.\footnote{Table \ref{twfe_domeviolenceMLrate_heteroeffects_tab_int} reports interaction terms between the treatment variable (first non-Catholic church arrival) and indicators for high Catholic presence, small population size, and large youth share. Table \ref{twfe_domeviolenceMLrate_highlowcivilethnicivil_tab_int} includes additional interactions with low ethnic fractionalization, few civil society organizations, and their joint effect. In both cases, the estimated interaction effects are negative, providing further support for our interpretation. The inclusion of these interaction terms also enables us to isolate and quantify the degree to which local social structures condition the impact of religious competition.}

\medskip 

The second set of evidence speaks to how the effects of a non-Catholic church’s arrival may diffuse within a community. Our hypothesis draws on theoretical insights from \cite{Morris2000} and \cite{AcemogluOzdaglarYildiz2011}, which suggest that the diffusion of cultural innovations—such as new norms around domestic violence—is more effective in social networks with low clustering and high connectivity. These dynamics are particularly relevant when the innovation originates from a small group of initial adopters, as is often the case in our context.

\medskip

Applied to the Colombian case, we hypothesize that norm diffusion is more likely in municipalities with greater population homogeneity. Specifically, we expect stronger diffusion effects—and consequently larger reductions in domestic violence—in municipalities with lower ethnic diversity and fewer civil society organizations. Such environments likely feature fewer social subgroups and greater cross-cutting interaction, thereby facilitating the spread of new norms.\footnote{This implication follows from \cite{AcemogluOzdaglarYildiz2011}, under the assumption that the seed set of adopters is small. In our case, most municipalities experience the arrival of only one non-Catholic church in the first year of exposure (see Table \ref{nonCatho_tab}).}

\medskip

Table \ref{twfe_domeviolenceMLrate_highlowcivilethnicivil_tab} presents empirical results consistent with this hypothesis. It compares municipalities with above- and below-median levels of ethnic diversity—measured using the 1993 ethnic fractionalization index—and civil society density—measured by the number of civil society organizations per 100,000 inhabitants in 1995.\footnote{Ethnic diversity is based on self-identification as Afro-Colombian, Indigenous, or Romani. Civil society organizations include NGOs, cooperatives, guilds, and advocacy groups, as recorded by the Social Foundation in 1995. See Table \ref{precharacteristics_tab} for details.} The results indicate that the effect of a new non-Catholic church on domestic violence is strongest in municipalities with both low ethnic diversity and a low density of civil society organizations (column 2).\footnote{Tables \ref{twfe_domeviolenceMLrate_highlowethnic_tab} and \ref{twfe_domeviolenceMLrate_highlowcivil_tab} report similar but somewhat less precise results when analyzing each factor separately.}


\subsection{Alternative Explanations}

The observed decline in domestic violence following the arrival of non-Catholic churches may plausibly result from mechanisms other than the cultural channel we propose. In this section, we assess several alternative explanations, which we regard as theoretically relevant but empirically less consistent with the data.

\medskip

A first possibility is that the decline in domestic violence is an indirect effect of reduced conflict-related violence. Armed conflict in Colombia is known to exacerbate domestic violence by normalizing physical force as a means of resolving disputes (\citealt{NoeRieckmann2013}; \citealt[][pp. 115–116]{GMH2011}).  If the arrival of a non-Catholic church were to reduce armed conflict—perhaps by reducing recruitment into armed groups—this could, in turn, lower domestic violence.

\medskip

However, evidence from \cite{galindo2021fighting} suggests the opposite: the establishment of a non-Catholic church is associated with an increase in conflict-related violence, particularly in municipalities with histories of coca cultivation or forced recruitment.  We also show that this increase in conflict-related violence is not accompanied by corresponding changes in domestic violence in these areas (Table \ref{twfe_domeviolenceMLrate_coca_tab}). Thus, our results suggest that changes in conflict dynamics are unlikely to mediate the relationship between religious competition and domestic violence

\medskip

A second hypothesis is that religious competition induces a broad moral awakening that reduces domestic violence—not necessarily through specific practices, but by promoting more ethical behavior overall. While this mechanism partially overlaps with our preferred explanation, it lacks specificity and is not strongly supported by the available evidence.  In particular, \cite{galindo2021fighting} find  no corresponding reductions in other types of violence—such as homicides or robberies—which would be expected to decline under a generalized moral improvement channel. Moreover, whereas non-Catholic churches are negatively correlated with domestic violence, civil society organizations are positively correlated (Table \ref{tab_corrorgcivsoc}), further suggesting that the effect is not simply driven by increased social interaction.

\medskip

A third alternative posits that economic development, coinciding with church arrival, may explain the reduction in domestic violence. This explanation could be complementary to ours, to the extent that cultural change induced by religious organizations improves household behavior—for instance, by reducing alcohol consumption or encouraging female labor force participation.

\medskip

However, this explanation could also suggest a more direct effect of churches on local economic development—for instance, if churches invest in public goods to attract adherents. The empirical evidence does not support this alternative mechanism. First, we examine whether department-level GDP per capita is associated with domestic violence. Figure \ref{fig_corrdomviolenceGDPdeptos} shows that it is not. Second, we assess whether the establishment of the first non-Catholic church in a municipality is followed by a reduction in poverty. Table \ref{poverty_tab} provides no evidence in support of this either. Taken together, these results suggest that, while economic channels are theoretically plausible, they are unlikely to be the primary drivers of the observed effects.\footnote{In the first analysis (Figure \ref{fig_corrdomviolenceGDPdeptos}), we use department-level GDP data, as this is the most disaggregated level for which such data are available for a substantial number of municipalities in Colombia. The specification controls for population size, rurality, homicide rates, and department and year fixed effects. In the second analysis (Table \ref{poverty_tab}), we use the share of the population with unsatisfied basic needs as a proxy for poverty. This variable is available for two years during our study period (2005 and 2011). The specification includes controls for rurality, homicide rates, and municipality and department $\times$ year fixed effects.}

\medskip

Another concern is that the observed reduction reflects changes in reporting behavior, not actual reductions in domestic violence. For example, if the arrival of a non-Catholic church increases stigma around violence, victims may be less likely to report abuse.\footnote{We thank an anonymous referee for suggesting this alternative explanation.}

\medskip

However, this explanation is inconsistent with two key findings. First, our results are robust across two independent data sources: INMLCF (medical reports) and National Police records. If stigma were driving underreporting, we would expect a larger decline in police reports, which are more likely to be influenced by legal consequences. In fact, the opposite is observed: the INMLCF data show a slightly larger reduction (Table \ref{twfe_domeviolenceNPrate_tab}). 

\medskip

Second, we exploit a 2012 legal reform (Law 1542) that empowered authorities to investigate domestic violence without a formal complaint. If stigma were the primary driver of reporting behavior, we would expect stronger effects after 2012. Instead, we find that the impact of religious competition on domestic violence is larger in the pre-2012 period (Table \ref{domv_before_after_2012}), when moral condemnation likely played a greater role than formal enforcement.\footnote{Note that this pattern is not a contradiction of our hypothesis, but rather a validation of its mechanism. Before 2012, legal accountability for domestic violence was limited, and enforcement depended more heavily on community and moral sanctions. In that context, increased religious competition likely influenced prevailing norms and attitudes, reinforcing informal deterrents to domestic violence. After 2012, however, the legal system assumed a stronger role in shaping behavior. We interpret the reduction of the religious competition effect as evidence that formal legal sanctions partially substituted for the community’s role in moral regulation. If this substitution story holds, then our findings across time — stronger effects when moral norms matter more, weaker effects when legal sanctions dominate — are consistent with our preferred interpretation of the mechanism.}


\section{Conclusion}\label{Conclusion}

Domestic violence is a widespread and persistent form of interpersonal harm with profound private and social consequences. This paper provides empirical evidence that cultural change—specifically, that induced by religious competition—can play a meaningful role in reducing domestic violence. Using administrative data from Colombia and a difference-in-differences framework, we find that the arrival of a non-Catholic church in a predominantly Catholic municipality leads to a significant and sustained decline in reported cases of domestic violence.

\medskip

Our interpretation of this finding is that religious competition fosters the diffusion of social norms and practices that discourage domestic violence. New religious organizations, particularly Pentecostal churches, introduce norms emphasizing household stability, reductions in alcohol use, sexual restraint, and female leadership. In a competitive religious environment, other religious institutions may adopt similar practices in an effort to retain or attract followers. These behavioral changes, we argue, are culturally transformative and contribute to the reduction in violence within households.

\medskip

We evaluate and rule out several alternative explanations—including shifts in conflict dynamics, economic development, and changes in reporting behavior—using a combination of robustness checks, alternative data sources, and institutional context. The consistency of our findings across multiple estimators and outcome measures reinforces the credibility of our interpretation.

\medskip

This study contributes to several strands of literature. First, it adds to research on the cultural determinants of domestic violence, particularly the role of gender norms. Second, it extends the literature on the social consequences of religious institutions, with a focus on how competition between them can influence behavior. Third, it contributes to the broader field of cultural economics by documenting how competition-induced innovations—whether institutional or technological—can shape informal norms and reduce harmful practices.

\medskip

Taken together, our findings highlight the potential for cultural mechanisms to generate positive social change. Understanding how these mechanisms operate—and under what conditions they are most effective—remains an important area for future research.


\hbox {} \newpage

\section*{Compliance with Ethical Standards}

\noindent The authors declare that they have no conflict of interest. 

\hbox {} \newpage
\section*{Figures and Tables}


 \begin{table}[H]
 \renewcommand{\arraystretch}{1}
\setlength{\tabcolsep}{2pt}
\hspace{-1cm}\begin{center}
\caption {Descriptive Statistics}  \label{precharacteristics_tab}
\vspace{-0.3cm}
\small
 \begin{tabular}{lcccccc}
\hline\hline \addlinespace[0.15cm]
& &&& \multicolumn{3}{c}{Municipalities where}\\
& &&& \multicolumn{3}{c}{a first non-Catholic}\\
& &&& \multicolumn{3}{c}{church was established}\\
& \multicolumn{3}{c}{All municipalities} & \multicolumn{3}{c}{ in 2005-2019}\\\cmidrule[0.2pt](l){2-4}\cmidrule[0.2pt](l){5-7}
    & \multicolumn{1}{c}{Obs.} & \multicolumn{1}{c}{Mean}  & \multicolumn{1}{c}{St. dev.} & \multicolumn{1}{c}{Obs.} & \multicolumn{1}{c}{Mean}  & \multicolumn{1}{c}{St. dev.} \\\cmidrule[0.2pt](l){2-2}\cmidrule[0.2pt](l){3-3} \cmidrule[0.2pt](l){4-4} \cmidrule[0.2pt](l){5-5} \cmidrule[0.2pt](l){6-6}  \cmidrule[0.2pt](l){7-7}
            &         (1)&         (2)&         (3)&         (4)&         (5)&         (6)\\
\midrule
\emph{\textbf{Domestic Violence}}&            &            &            &            &            &            \\
Domestic violence (cases)&       16830&      69.904&     625.802&        7530&     108.865&     353.813\\
Domestic violence (rate)&       16819&     104.737&     146.339&        7525&     124.907&     155.189\\
Intimate partner violence (cases)&       16830&      44.545&     408.012&        7530&      69.579&     228.013\\
Intimate partner violence  (rate)&       16819&      60.536&      86.484&        7525&      75.509&      93.354\\
Other domestic violence (cases)&       16830&      25.359&     219.191&        7530&      39.286&     128.531\\
Other domestic violence (rate)&       16819&      44.201&      69.460&        7525&      49.398&      68.868\\
Domestic violence (NP data, cases)&       11220&      58.299&     679.936&        5020&      90.506&     380.767\\
Domestic violence (NP data, rate)&       11220&      75.509&      92.048&        5020&      87.675&      95.539\\
\emph{\textbf{Sociodemographics and economy}}&            &            &            &            &            &            \\
Total population&       16830&   40190.309&  243082.412&        7530&   63685.983&  171820.532\\
Share of population aged 25 to 44&       16770&       0.535&       0.068&        7500&       0.556&       0.056\\
Rurality index&       14575&       0.572&       0.244&        6521&       0.441&       0.242\\
Index of ethnic fractionalization (1993)&       16800&       0.042&       0.119&        7530&       0.039&       0.112\\
Unsatisfied basic needs (1993)&        2236&      45.170&      21.043&        1001&      40.753&      20.905\\
Homicide (rate)&       12829&      36.441&      41.388&        6441&      35.235&      36.857\\
Internally displaced pop. (outflow, rate)&       14575&    1217.041&    3004.463&        6521&    1106.387&    2470.288\\
Internally displaced pop. (inflow, rate)&       14575&     656.547&    1650.789&        6521&     734.723&    1473.844\\
Catholic churches (1995)&       15045&       2.218&      10.994&        6915&       2.820&       7.873\\
Civil society organizations (1995)&       15780&     141.589&     341.569&        7185&     203.084&     324.619\\
Vote share for traditional parties&        3299&       0.336&       0.252&        1497&       0.305&       0.242\\

 \addlinespace[0.15cm]
\hline\hline
\multicolumn{7}{p{16.5cm}}{\footnotesizes{\textbf{Note:} The sample used in all columns is limited to data from the year 2005 onwards. Columns (1) through (3) encompass all municipalities, while columns (4) through (6) are restricted to municipalities in which non-Catholic churches acquired legal status within the specified time frame. `NP' denotes the Colombian \emph{National Police}, and the rates are presented per 100,000 inhabitants. Population data, information on the ethnic minority population, and data on people with unsatisfied basic needs (used as a proxy for poverty) are from the \emph{National Administrative Department of Statistics} (DANE). Data on internally displaced people and homicide rates are from the \emph{Center for Studies of Economic Development }(CEDE). Data on the number of civil society organizations and the count of Catholic churches in 1995 are from the \emph{Social Foundation}, a non-profit organization. Electoral data is from the from
the Colombian Electoral Agency.} }
\end{tabular}
\end{center}
\end{table}


\begin{table}[H]
\begin{center}
{
\renewcommand{\arraystretch}{0.7}
\setlength{\tabcolsep}{2pt}
\caption {Determinants of the Establishment of the First Non-Catholic Church}  \label{dfirstiglesiasnc_tab}
\vspace{-0.3cm}
\small
\centering  \begin{tabular}{lccccccc}
\hline\hline \addlinespace[0.15cm]
    & \multicolumn{4}{c}{Dep. Variable = 1 when first} \\\addlinespace[0.09cm]
    & \multicolumn{4}{c}{ non-Catholic church is established} \\\cmidrule[0.2pt](l){2-5}
& (1)& (2) & (3)  & (4)  \\    \addlinespace[0.15cm] \hline \addlinespace[0.15cm]
Domestic violence (rate) at t-1&       0.000         &                     &       0.000         &       0.000         \\
                    &     (0.000)         &                     &     (0.000)         &     (0.000)         \\
Log of total population at t-1&                     &      -0.091\sym{*}  &      -0.078         &      -0.068         \\
                    &                     &     (0.053)         &     (0.064)         &     (0.067)         \\
Rurality index at t-1&                     &      -0.147         &      -0.180         &      -0.246         \\
                    &                     &     (0.199)         &     (0.234)         &     (0.238)         \\
Unsatisfied Basic Needs at t-1&                     &       0.000         &       0.003         &       0.003         \\
                    &                     &     (0.000)         &     (0.004)         &     (0.004)         \\
Homicide (rate) at t-1&                     &                     &                     &      -0.000         \\
                    &                     &                     &                     &     (0.000)         \\
Internally displaced pop. (outflow, rate) at t-1&                     &                     &                     &      -0.000         \\
                    &                     &                     &                     &     (0.000)         \\
Internally displaced pop. (inflow, rate) at t-1&                     &                     &                     &       0.000         \\
                    &                     &                     &                     &     (0.000)         \\
Vote share for traditional parties at t-1&                     &                     &                     &      -0.027         \\
                    &                     &                     &                     &     (0.017)         \\
R-sq                &       0.097         &       0.103         &       0.103         &       0.117         \\
Observations        &        6967         &        6930         &        6456         &        5945         \\

\addlinespace[0.15cm]
\hline\hline
\multicolumn{5}{p{13.5cm}}{\footnotesizes{\textbf{Note:} All models include municipality and department $\times$ year fixed effects.  Samples for regression models include data from 2005 to 2019.   Robust standard errors (in parentheses) are clustered by municipality. * denotes statistically significant estimates at 10\%, ** denotes significant at 5\% and *** denotes significant at 1\%.} }
\end{tabular}
}
\end{center}
\end{table}


\begin{table}[H]
\begin{center}
{
\renewcommand{\arraystretch}{0.7}
\setlength{\tabcolsep}{8pt}
\caption {Effect of the First Non-Catholic Church on Domestic Violence}  \label{twfe_domeviolenceMLrate_tab}
\vspace{-0.3cm}
\small
\centering  \begin{tabular}{lcccccc}
\hline\hline \addlinespace[0.15cm]
  & \multicolumn{6}{c}{Dep. Var: Domestic Violence (per 100.000 inhabitants)} \\\addlinespace[0.1cm]\cmidrule[0.2pt](l){2-7}\addlinespace[0.05cm]
& (1)& (2) & (3)& (4)& (5)& (6) \\   \addlinespace[0.1cm] \hline \addlinespace[0.15cm]
& \multicolumn{2}{c}{Against any}& \multicolumn{2}{c}{Against}& \multicolumn{2}{c}{Against other} \\
& \multicolumn{2}{c}{household member}& \multicolumn{2}{c}{intimate partner}& \multicolumn{2}{c}{household members} \\
\cmidrule[0.2pt](l){2-3}\cmidrule[0.2pt](l){4-5}\cmidrule[0.2pt](l){6-7}\addlinespace[0.15cm]
            \multicolumn{1}{l}{\emph{\underline{Panel A}:}}            & \multicolumn{6}{c}{\emph{Static}  Specification}\\
            \addlinespace[0.3cm]
Effect at t$\geq 0$ &     -10.547\sym{*}  &     -14.089\sym{**} &      -5.069         &      -6.965\sym{*}  &      -5.478\sym{**} &      -7.124\sym{**} \\
                    &     (6.373)         &     (6.541)         &     (3.960)         &     (4.042)         &     (2.688)         &     (2.802)         \\
R-sq                &       0.715         &       0.722         &       0.722         &       0.729         &       0.626         &       0.631         \\
Observations        &        7465         &        6930         &        7465         &        6930         &        7465         &        6930         \\

\addlinespace[0.15cm]\hline\addlinespace[0.2cm]
            \multicolumn{1}{l}{\emph{\underline{Panel B}:}}            & \multicolumn{6}{c}{\emph{Dynamic} Specification }\\\addlinespace[0.3cm]
Effect at t=-3      &      -3.510         &      -4.962         &      -1.462         &      -2.182         &      -2.048         &      -2.780         \\
                    &     (6.976)         &     (7.138)         &     (4.283)         &     (4.397)         &     (3.167)         &     (3.260)         \\
Effect at t=-2      &      -0.192         &      -2.663         &      -0.609         &      -2.149         &       0.417         &      -0.515         \\
                    &     (6.988)         &     (6.837)         &     (4.196)         &     (4.178)         &     (3.418)         &     (3.363)         \\
Effect at t=-1      &       0.000         &       0.000         &       0.000         &       0.000         &       0.000         &       0.000         \\
                    &         (.)         &         (.)         &         (.)         &         (.)         &         (.)         &         (.)         \\
Effect at t=0       &      -8.930         &     -10.730         &      -6.002         &      -6.979\sym{*}  &      -2.928         &      -3.751         \\
                    &     (6.554)         &     (6.803)         &     (3.892)         &     (4.089)         &     (3.045)         &     (3.126)         \\
Effect at t=+1      &      -4.993         &      -8.704         &      -3.738         &      -5.896         &      -1.255         &      -2.808         \\
                    &     (6.555)         &     (6.953)         &     (4.028)         &     (4.315)         &     (3.034)         &     (3.182)         \\
Effect at t=+2      &      -2.950         &      -6.045         &      -3.559         &      -4.972         &       0.609         &      -1.072         \\
                    &     (6.948)         &     (7.257)         &     (4.137)         &     (4.313)         &     (3.293)         &     (3.476)         \\
Effect at t=+3      &      -0.288         &      -1.142         &       0.104         &      -0.368         &      -0.392         &      -0.773         \\
                    &     (7.043)         &     (7.150)         &     (4.296)         &     (4.347)         &     (3.146)         &     (3.265)         \\
Effect at t=+4      &      -7.390         &     -12.376\sym{*}  &      -5.073         &      -7.936\sym{*}  &      -2.317         &      -4.440         \\
                    &     (6.940)         &     (7.049)         &     (4.229)         &     (4.396)         &     (3.151)         &     (3.163)         \\
Effect at t=+5      &      -7.974         &     -14.504\sym{**} &      -6.012         &      -9.438\sym{**} &      -1.962         &      -5.066         \\
                    &     (6.796)         &     (7.229)         &     (4.047)         &     (4.350)         &     (3.189)         &     (3.383)         \\
Effect at t=+6      &      -7.945         &     -12.409\sym{*}  &      -4.966         &      -6.964         &      -2.979         &      -5.445         \\
                    &     (6.949)         &     (7.387)         &     (4.208)         &     (4.545)         &     (3.356)         &     (3.569)         \\
Effect at t=+7      &      -7.303         &     -11.330         &      -6.524         &      -9.094\sym{*}  &      -0.779         &      -2.236         \\
                    &     (7.079)         &     (8.286)         &     (4.217)         &     (5.000)         &     (3.333)         &     (3.862)         \\
Effect at t=+8      &     -10.488         &      -5.437         &      -5.328         &      -3.356         &      -5.160         &      -2.081         \\
                    &     (8.358)         &     (8.799)         &     (5.200)         &     (5.641)         &     (3.719)         &     (3.804)         \\
R-sq                &       0.716         &       0.722         &       0.723         &       0.729         &       0.627         &       0.632         \\
Observations        &        7465         &        6930         &        7465         &        6930         &        7465         &        6930         \\

 \addlinespace[0.15cm]\hline \addlinespace[0.15cm]
Baseline controls & No & Yes& No & Yes& No & Yes\\
\addlinespace[0.15cm]\hline\hline
\multicolumn{7}{p{14.3cm}}{\scriptsize{\textbf{Note:} All columns in Panel A report the estimates from Eq. (\ref{didbaselines}) and all columns in Panel B report the estimates from Eq. (\ref{didbaseline}). All models include municipality and department $\times$ year fixed effects.  The models in Panel B include 10 lags and 10 leads, normalized to the period prior to treatment. The models with baseline controls (columns (2), (4) and (6)) include the following (lagged) covariates:  log of total population, rurality index, and proportion with unsatisfied basic needs. Samples for regression models use data from 2005 to 2019. Robust standard errors (in parentheses) are clustered by municipality. * denotes statistically significant estimates at 10\%, ** denotes significant at 5\% and *** denotes significant at 1\%.} }
\end{tabular}
}
\end{center}
\end{table}


\begin{figure}[H]
\begin{center}
             \caption{Effect of the First Non-Catholic Church on Domestic Violence (Against Any Household Member)}
        \label{twfe_domeviolenceMLrate_fig}
        \vspace{-0.3cm}
\includegraphics[width=11cm,height=8cm]{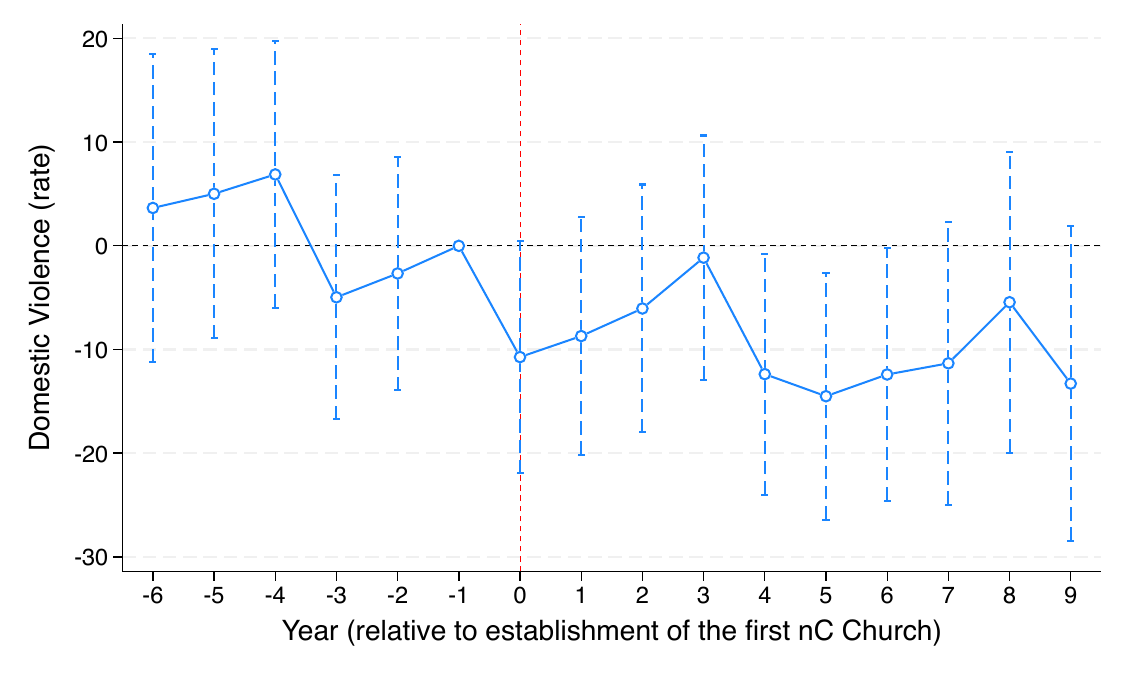}\\
     \begin{minipage}{11cm} \footnotesizes \textbf{Note:}  This figure shows the two-way fixed effects estimates from Eq. (\ref{didbaseline}), in a specification that includes municipality and  department $\times$ year fixed effects, and the following (lagged) covariates:  log of total population, rurality index, and proportion of the population with unsatisfied basic needs. Vertical lines indicate 90\% confidence intervals.
\end{minipage}
\end{center}
\end{figure}


\begin{table}[H]
\begin{center}
{
\renewcommand{\arraystretch}{0.8}
\setlength{\tabcolsep}{10pt}
\caption {Effect of the First Non-Catholic Church on Domestic Violence: Robustness to Using Alternative Heterogeneity-Robust DID Estimators}  \label{twfe_domeviolenceMLrateALTESTM_tab}
\vspace{-0.3cm}
\small
\centering  \begin{tabular}{lcccccc}
\hline\hline \addlinespace[0.15cm]
  & \multicolumn{6}{c}{Dep. Var: Domestic Violence (per 100.000 inhabitants)} \\\addlinespace[0.1cm]\cmidrule[0.2pt](l){2-7}\addlinespace[0.05cm]
& (1)& (2) & (3)& (4)& (5)& (6) \\   \addlinespace[0.1cm] \hline \addlinespace[0.15cm]
& \multicolumn{2}{c}{de Chaisemartin \& }& \multicolumn{2}{c}{Callaway \&}& \multicolumn{2}{c}{Sun \&} \\
& \multicolumn{2}{c}{D'Haultfoeuille}& \multicolumn{2}{c}{SantAnna}& \multicolumn{2}{c}{Abraham} \\
\cmidrule[0.2pt](l){2-3}\cmidrule[0.2pt](l){4-5}\cmidrule[0.2pt](l){6-7}
& \multicolumn{1}{c}{Estimate}& \multicolumn{1}{c}{SE}& \multicolumn{1}{c}{Estimate}& \multicolumn{1}{c}{SE} \\\cmidrule[0.2pt](l){2-2}\cmidrule[0.2pt](l){3-3}\cmidrule[0.2pt](l){4-4}\cmidrule[0.2pt](l){5-5}\cmidrule[0.2pt](l){6-6}\cmidrule[0.2pt](l){7-7}
Effect at t=-3&       13.638&       17.003&       -3.109&       10.915&           .&           .\\
Effect at t=-2&      10.765&      2.523&     -12.65&       8.186&           .&           .\\
Effect at t=-1&      -0.509&       1.720&       3.500 &       8.245&           .&           .\\
Effect at t=0&           .&           .&           .&           .&           .&           .\\
Effect at t=+1&      -8.532&       0.468&      -10.75&       9.461&     -10.269&       5.561\\
Effect at t=+2&     -10.815&       2.055&      -15.25&       11.568&      -6.328&       5.791\\
Effect at t=+3&      -7.775&       5.768&      -6.514&       11.664&      -4.410&       6.243\\
Effect at t=+4&      -4.272&        3.068&      -14.69&      13.787&      -2.540&       6.536\\
Effect at t=+5&        9.727&       6.931&       -7.319&      13.722&     -10.932&       6.886\\
Effect at t=+6&      -5.893&       3.671&       -18.71&      10.921&     -11.773&       6.936\\
Effect at t=+7&      -9.551&       3.589&       -15.89&      11.170&     -11.706&       7.403\\
Effect at t=+8&     -9.382&       6.858&      -25.92&      13.228&      -9.563&       7.149\\
Effect at t=+9&      -14.387&      5.751&      -36.37&      16.801&     -14.627&       7.232\\

\addlinespace[0.15cm]\hline\hline
\multicolumn{7}{p{14cm}}{\scriptsize{\textbf{Note:} All columns report the estimates of the effect of the first non-Catholic church on domestic violence, using three different heterogeneity-robust DID estimators. Columns (1) and (2) report  estimates using \citeauthor{ChaisemartinDHaultfoeuille2020}'s estimator, as implemented by the Stata command \emph{twowayfeweights} \cite[see][]{ChaisemartinDHaultfoeuille2020}. Columns (3) and (4) report estimates using \citeauthor{CallawaySantAnna2021}'s estimator as implemented by the Stata command \emph{csdid} \cite[see][]{CallawaySantAnna2021, RiosAvilaCallawaySantAnna2021}. Columns (5) and (6) report estimates using \citeauthor{SunAbraham2021}'s estimator as implemented by the Stata command \emph{eventstudyinteract}  \cite[see][]{SunAbraham2021}. All models include municipality and  department $\times$ year fixed effects.} }
\end{tabular}
}
\end{center}
\end{table}

\newpage


\begin{table}[H]
\begin{center}
{
\renewcommand{\arraystretch}{0.8}
\setlength{\tabcolsep}{11pt}
\caption {Effect of the First Non-Catholic Church on Domestic Violence: Heterogeneous Effects by Number of Catholic Churches, Population Size, and Share of Young Adults}  \label{twfe_domeviolenceMLrate_heteroeffects_tab}
\vspace{-0.3cm}
\small
\centering  \begin{tabular}{lcccc}
\hline\hline \addlinespace[0.15cm]
  & \multicolumn{4}{c}{Dep. Var: Domestic Violence (per 100.000 inhabitants)} \\\addlinespace[0.1cm]\cmidrule[0.2pt](l){2-5}\addlinespace[0.05cm]
& (1)& (2)& (3)& (4) \\   \addlinespace[0.1cm] \cmidrule[0.2pt](l){2-3}\cmidrule[0.2pt](l){4-5}
            & \multicolumn{2}{c}{Above median} & \multicolumn{2}{c}{Below median}\\
            \addlinespace[0.15cm]
\hline \addlinespace[0.15cm]
            \multicolumn{1}{l}{\emph{\underline{Panel A}:}}            & \multicolumn{4}{c}{Number of Catholic Churches in 1995}\\\cmidrule[0.2pt](l){2-5}
            \addlinespace[0.3cm]
Effect at t$\geq 0$ &     -25.963\sym{**} &     -31.757\sym{***}&       4.880         &       2.568         \\
                    &    (11.482)         &    (11.600)         &     (6.112)         &     (6.066)         \\
R-sq                &       0.676         &       0.687         &       0.817         &       0.823         \\
Observations        &        4000         &        3698         &        3420         &        3189         \\
\addlinespace[0.15cm]
\hline \addlinespace[0.15cm]
            \multicolumn{1}{l}{\emph{\underline{Panel B}:}}            & \multicolumn{4}{c}{Population Size  in 2004}\\\cmidrule[0.2pt](l){2-5}
            \addlinespace[0.3cm]
Effect at t$\geq 0$ &       8.596         &       6.946         &     -29.571\sym{**} &     -30.194\sym{***}\\
                    &     (5.385)         &     (5.466)         &    (11.656)         &    (11.575)         \\
R-sq                &       0.829         &       0.834         &       0.663         &       0.679         \\
Observations        &        3705         &        3453         &        3670         &        3394         \\
\addlinespace[0.15cm]
\hline \addlinespace[0.15cm]
            \multicolumn{1}{l}{\emph{\underline{Panel C}:}}            & \multicolumn{4}{c}{Population Share Aged 25 to 44}\\\cmidrule[0.2pt](l){2-5}
            \addlinespace[0.3cm]
Effect at t$\geq 0$ &     -20.296\sym{**} &     -22.454\sym{**} &       1.289         &      -2.189         \\
                    &    (10.107)         &    (10.330)         &     (7.439)         &     (7.664)         \\
R-sq                &       0.767         &       0.769         &       0.702         &       0.709         \\
Observations        &        3700         &        3418         &        3720         &        3470         \\

 \addlinespace[0.15cm]\hline \addlinespace[0.15cm]
Baseline Controls & No & Yes& No & Yes\\
\addlinespace[0.15cm]\hline\hline
\multicolumn{5}{p{13.1cm}}{\scriptsize{\textbf{Note:}  All columns in  report the estimates from Eq. (\ref{didbaselines}), in municipalities with an above median (columns (1) and (2)) and below median (columns (3) and (4)) of the respective variable.  All models include municipality and  department $\times$ year fixed effects. The models with baseline controls (columns (2) and (4)) include the following (lagged) covariates:  log of total population, rurality index, and proportion with unsatisfied basic needs. Samples for regression models use data from 2005 to 2019. Robust standard errors (in parentheses) are clustered by municipality. * denotes statistically significant estimates at 10\%, ** denotes significant at 5\% and *** denotes significant at 1\%} }
\end{tabular}
}
\end{center}
\end{table}

\newpage


\begin{table}[H]
\begin{center}
{
\renewcommand{\arraystretch}{0.8}
\setlength{\tabcolsep}{5pt}
\caption {Effect of the First Non-Catholic Church on Domestic Violence: Heterogeneous Effects by Ethnic Fractionalization and Number of Civil Society Organizations}  
\label{twfe_domeviolenceMLrate_highlowcivilethnicivil_tab}
\vspace{-0.3cm}
\small
\centering  \begin{tabular}{lcccc}
\hline\hline \addlinespace[0.15cm]
  & \multicolumn{4}{c}{Dep. Var: Domestic Violence (per 100.000 inhabitants)} \\\addlinespace[0.1cm]\cmidrule[0.2pt](l){2-5}\addlinespace[0.05cm]
& (1)& (2)& (3)& (4) \\   \addlinespace[0.1cm] \hline \addlinespace[0.1cm]
            & \multicolumn{2}{c}{ Below median} & \multicolumn{2}{c}{Above median}\\
                        & \multicolumn{2}{c}{ethnic fractionalization} & \multicolumn{2}{c}{ethnic fractionalization}\\\addlinespace[0.05cm] \cmidrule[0.2pt](l){2-3}\cmidrule[0.2pt](l){4-5}
            & \multicolumn{1}{c}{Above median \#} & \multicolumn{1}{c}{Below median \#} & \multicolumn{1}{c}{Above median \#} & \multicolumn{1}{c}{Below median \#}\\
                        & \multicolumn{1}{c}{of civil society} & \multicolumn{1}{c}{of civil society}& \multicolumn{1}{c}{of civil society} & \multicolumn{1}{c}{of civil society}\\
                      & \multicolumn{1}{c}{organizations} & \multicolumn{1}{c}{organizations}& \multicolumn{1}{c}{organizations} & \multicolumn{1}{c}{organizations}\\\addlinespace[0.05cm]
\cmidrule[0.2pt](l){2-2}\cmidrule[0.2pt](l){3-3}\cmidrule[0.2pt](l){4-4}\cmidrule[0.2pt](l){5-5} \addlinespace[0.15cm]
Effect at t$\geq 0$ &     -22.196         &     -32.515\sym{**} &      -8.833         &     -16.563         \\
                    &    (15.469)         &    (16.319)         &     (5.938)         &    (15.214)         \\
R-sq                &       0.695         &       0.762         &       0.674         &       0.790         \\
Observations        &        1670         &        1673         &        1805         &        1524         \\

Baseline Controls & Yes & Yes& Yes & Yes\\
\addlinespace[0.15cm]\hline\hline
\multicolumn{5}{p{15.5cm}}{\scriptsize{\textbf{Note:} All columns report the estimates from Eq. (\ref{didbaselines}),  in municipalities with an above or below median index of ethnic fractionalization in 1993, and, simultaneously, with an above or below median number of civil society organizations in 1995.  All models include municipality and department $\times$ year fixed effects, and the following (lagged) covariates:  log of total population, rurality index, and proportion with unsatisfied basic needs. Samples for regression models use data from 2005 to 2019. Robust standard errors (in parentheses) are clustered by municipality. * denotes statistically significant estimates at 10\%, ** denotes significant at 5\% and *** denotes significant at 1\%.} }
\end{tabular}
}
\end{center}
\end{table}


\hbox {} \newpage
\setcounter{table}{0}
\setcounter{figure}{0}
\setcounter{subsection}{0}
\renewcommand{\thefigure}{\Alph{section}\arabic{figure}}
\renewcommand{\thetable}{\Alph{section}\arabic{table}}

\appendix

\section{Web Appendix: Additional Figures and Tables}
\label{webapp1}


\begin{figure}[H]
\begin{center}
             \caption{Domestic Violence and GDP Growth}
        \label{fig_domviolenceGDP}
\includegraphics[width=12cm,height=8cm]{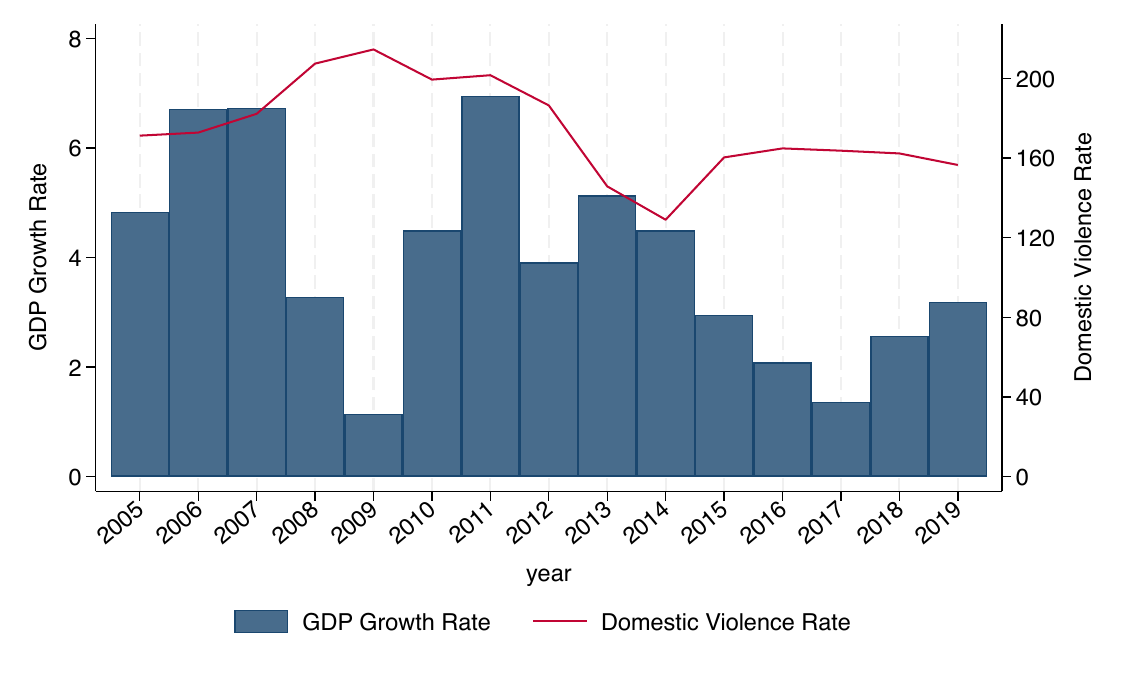}
     \begin{minipage}{12cm} \footnotesizes \textbf{Source:}  The data on GDP growth comes from the \emph{World Development Indicators}. The data on domestic violence is sourced from \emph{Forensis} (Colombian \emph{National Institute of Legal Medicine and Forensic Sciences}). 
\end{minipage}
\end{center}
\end{figure}


\begin{table}[H]
\begin{center}
  \centering \caption {Descriptive Statistics for Municipalities with New non-Catholic churches}  \label{nonCatho_tab}
\begin{tabular}{|c|c|c|}
\hline
 \# of New non- &  \# of  &  Average \\
Catholic Churches  & Municipalities  &  Population \\\hline
   \midrule
1           &         435&   45541.223\\
2           &          48&   51580.875\\
3           &           8&  123179.250\\
4           &           7&  155068.429\\
5           &           3&  480703.000\\

\hline\addlinespace[0.1cm]
\multicolumn{3}{p{11cm}}{\scriptsize{\textbf{Source}: Public Registry of Religious Organizations (Office of Religious Affairs, Colombian \emph{Ministry of the Interior})} }
\end{tabular}
\end{center}
\end{table}


\begin{table}[H]

\vspace{-1cm}
\begin{center}
{
\renewcommand{\arraystretch}{0.8}
\setlength{\tabcolsep}{15pt}
\caption {Determinants of the Establishment of the First Non-Catholic Church: 4 Lags}  \label{dfirstiglesiasnc5lags_tab}
\vspace{-0.3cm}
\footnotesize
\centering  \begin{tabular}{lccccccc}
\hline\hline \addlinespace[0.1cm]
    & \multicolumn{4}{c}{Dep. Variable = 1 when first} \\\addlinespace[0.0cm]
    & \multicolumn{4}{c}{ non-Catholic church is established} \\\cmidrule[0.2pt](l){2-5}
& (1)& (2) & (3)  & (4)  \\    \addlinespace[0.15cm] \hline \addlinespace[0.05cm]
Domestic violence (rate) at t-1&       0.000         &                     &       0.000         &      -0.000         \\
                    &     (0.000)         &                     &     (0.000)         &     (0.000)         \\
Domestic violence (rate) at t-2&      -0.000         &                     &      -0.000         &      -0.000         \\
                    &     (0.000)         &                     &     (0.000)         &     (0.000)         \\
Domestic violence (rate) at t-3&       0.000\sym{**} &                     &       0.000         &       0.000\sym{*}  \\
                    &     (0.000)         &                     &     (0.000)         &     (0.000)         \\
Domestic violence (rate) at t-4&       0.000         &                     &       0.000         &       0.000         \\
                    &     (0.000)         &                     &     (0.000)         &     (0.000)         \\
Log of total population at t-1&                     &       0.351         &       0.611         &       1.310\sym{*}  \\
                    &                     &     (0.298)         &     (0.403)         &     (0.751)         \\
Log of total population at t-2&                     &      -0.380         &      -1.183\sym{*}  &      -1.423         \\
                    &                     &     (0.456)         &     (0.679)         &     (1.081)         \\
Log of total population at t-3&                     &       0.065         &       0.331         &       0.282         \\
                    &                     &     (0.370)         &     (0.556)         &     (1.100)         \\
Log of total population at t-4&                     &      -0.098         &       0.258         &       0.005         \\
                    &                     &     (0.211)         &     (0.345)         &     (0.545)         \\
Rurality index at t-1&                     &      -0.041         &      -0.099\sym{*}  &      -0.196\sym{*}  \\
                    &                     &     (0.042)         &     (0.059)         &     (0.113)         \\
Rurality index at t-2&                     &       0.070         &       0.164\sym{*}  &       0.192         \\
                    &                     &     (0.062)         &     (0.094)         &     (0.156)         \\
Rurality index at t-3&                     &      -0.028         &      -0.062         &      -0.054         \\
                    &                     &     (0.048)         &     (0.075)         &     (0.158)         \\
Rurality index at t-4&                     &      -0.017         &      -0.074         &      -0.042         \\
                    &                     &     (0.030)         &     (0.050)         &     (0.079)         \\
Unsatisfied Basic Needs at t-1&                     &       0.000         &      -0.001         &       0.003         \\
                    &                     &     (0.000)         &     (0.007)         &     (0.008)         \\
Unsatisfied Basic Needs at t-2&                     &      -0.001         &      -0.001         &      -0.002\sym{*}  \\
                    &                     &     (0.001)         &     (0.001)         &     (0.001)         \\
Homicide (rate) at t-1&                     &                     &                     &      -0.000         \\
                    &                     &                     &                     &     (0.000)         \\
Homicide (rate) at t-2&                     &                     &                     &      -0.000         \\
                    &                     &                     &                     &     (0.000)         \\
Homicide (rate) at t-3&                     &                     &                     &       0.000         \\
                    &                     &                     &                     &     (0.000)         \\
Homicide (rate) at t-4&                     &                     &                     &      -0.000         \\
                    &                     &                     &                     &     (0.000)         \\
Internally displaced pop. (outflow, rate) at t-1&                     &                     &                     &      -0.000         \\
                    &                     &                     &                     &     (0.000)         \\
Internally displaced pop. (outflow, rate) at t-2&                     &                     &                     &      -0.000         \\
                    &                     &                     &                     &     (0.000)         \\
Internally displaced pop. (outflow, rate) at t-3&                     &                     &                     &       0.000         \\
                    &                     &                     &                     &     (0.000)         \\
Internally displaced pop. (outflow, rate) at t-4&                     &                     &                     &      -0.000         \\
                    &                     &                     &                     &     (0.000)         \\
Internally displaced pop. (inflow, rate) at t-1&                     &                     &                     &       0.000         \\
                    &                     &                     &                     &     (0.000)         \\
Internally displaced pop. (inflow, rate) at t-2&                     &                     &                     &       0.000         \\
                    &                     &                     &                     &     (0.000)         \\
Internally displaced pop. (inflow, rate) at t-3&                     &                     &                     &      -0.000         \\
                    &                     &                     &                     &     (0.000)         \\
Internally displaced pop. (inflow, rate) at t-4&                     &                     &                     &       0.000         \\
                    &                     &                     &                     &     (0.000)         \\
Vote share for traditional parties at t-1&                     &                     &                     &      -0.017         \\
                    &                     &                     &                     &     (0.044)         \\
Vote share for traditional parties at t-2&                     &                     &                     &       0.039         \\
                    &                     &                     &                     &     (0.042)         \\
R-sq                &       0.105         &       0.106         &       0.117         &       0.194         \\
Observations        &        5473         &        6782         &        4887         &        2862         \\

\addlinespace[0.05cm]
\hline\hline
\multicolumn{5}{p{16cm}}{\scriptsize{\textbf{Note:} All models include municipality and department $\times$ year fixed effects.  Samples for regression models include data from 2005 to 2019.   Robust standard errors (in parentheses) are clustered by municipality. * denotes statistically significant estimates at 10\%, ** denotes significant at 5\% and *** denotes significant at 1\%.} }
\end{tabular}
}
\end{center}
\end{table}


\begin{table}[H]
\begin{center}
{
\renewcommand{\arraystretch}{0.7}
\setlength{\tabcolsep}{15pt}
\caption {Effect of the First Non-Catholic Church on Domestic Violence: Robustness to using Colombian National Police Data}  \label{twfe_domeviolenceNPrate_tab}
\vspace{-0.3cm}
\small
\centering  \begin{tabular}{lcc}
\hline\hline \addlinespace[0.15cm]
  & \multicolumn{2}{c}{Dep. Var: Domestic} \\
    & \multicolumn{2}{c}{Violence (rate)} \\\addlinespace[0.1cm]\cmidrule[0.2pt](l){2-3}\addlinespace[0.05cm]
& (1)& (2) \\   \addlinespace[0.1cm] \hline \addlinespace[0.15cm]
            \multicolumn{1}{l}{\emph{\underline{Panel A}:}}            & \multicolumn{2}{c}{\emph{Static} Specification}\\
            \addlinespace[0.3cm]
Effect at t$\geq 0$ &     -10.043\sym{**} &     -10.361\sym{**} \\
                    &     (4.282)         &     (4.081)         \\
R-sq                &       0.725         &       0.736         \\
Observations        &        4980         &        4476         \\

\addlinespace[0.15cm]\hline\addlinespace[0.15cm]
            \multicolumn{1}{l}{\emph{\underline{Panel B}:}}            & \multicolumn{2}{c}{ \emph{Dynamic}  Specification}\\\addlinespace[0.3cm]
Effect at t=-3      &       2.341         &       2.341         \\
                    &     (4.119)         &     (4.119)         \\
Effect at t=-2      &      -4.097         &      -4.097         \\
                    &     (3.296)         &     (3.296)         \\
Effect at t=-1      &       0.000         &       0.000         \\
                    &         (.)         &         (.)         \\
Effect at t=0       &      -7.956\sym{**} &      -7.956\sym{**} \\
                    &     (3.249)         &     (3.249)         \\
Effect at t=+1      &     -11.115\sym{**} &     -11.115\sym{**} \\
                    &     (4.333)         &     (4.333)         \\
Effect at t=+2      &     -10.082\sym{*}  &     -10.082\sym{*}  \\
                    &     (5.306)         &     (5.306)         \\
Effect at t=+3      &     -15.067\sym{**} &     -15.067\sym{**} \\
                    &     (5.881)         &     (5.881)         \\
Effect at t=+4      &     -14.453\sym{**} &     -14.453\sym{**} \\
                    &     (6.182)         &     (6.182)         \\
Effect at t=+5      &     -15.262\sym{**} &     -15.262\sym{**} \\
                    &     (6.525)         &     (6.525)         \\
Effect at t=+6      &     -14.014\sym{*}  &     -14.014\sym{*}  \\
                    &     (7.232)         &     (7.232)         \\
R-sq                &       0.739         &       0.739         \\
Observations        &        4476         &        4476         \\

 \addlinespace[0.15cm]\hline \addlinespace[0.15cm]
Baseline controls & No & Yes\\
\addlinespace[0.15cm]\hline\hline
\multicolumn{3}{p{8.2cm}}{\scriptsize{\textbf{Note:} All columns in Panel A report the estimates from Eq. (\ref{didbaselines}) and all columns in Panel B report the estimates from Eq. (\ref{didbaseline}). All models include municipality and department $\times$ year fixed effects.  The models in Panel B include 10 lags and 10 leads, normalized to the period prior to treatment. The model with baseline controls (column (2)) includes the following (lagged) covariates:  log of total population, rurality index, and proportion with unsatisfied basic needs. Samples for regression models use data from 2005 to 2019. Robust standard errors (in parentheses) are clustered by municipality. * denotes statistically significant estimates at 10\%, ** denotes significant at 5\% and *** denotes significant at 1\%..} }
\end{tabular}
}
\end{center}
\end{table}

\newpage

\newgeometry{left=1.8cm,top=2cm, bottom=0.1cm}
\begin{figure}[H]
             \caption{Location of Treated and Control Group Municipalities}
        \label{ncatchurchesmap_fig}
        
\begin{subfigure}{0.4\textwidth}
\caption{2006} \label{}
\includegraphics[width=5.8cm, height=6.7cm]{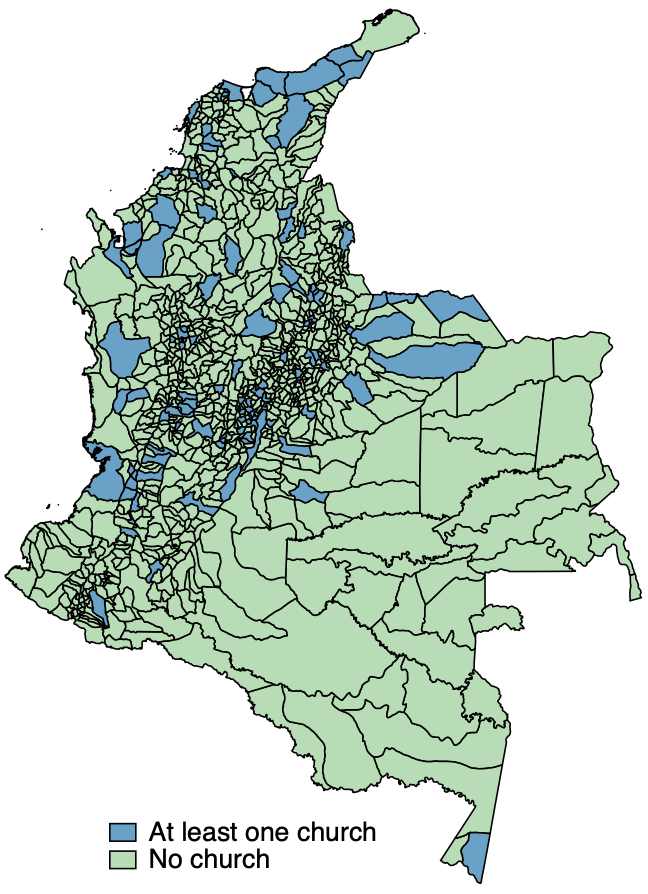}
\end{subfigure}\hspace*{\fill}
\begin{subfigure}{0.4\textwidth}
\caption{2008} \label{}
\includegraphics[width=5.8cm, height=6.7cm]{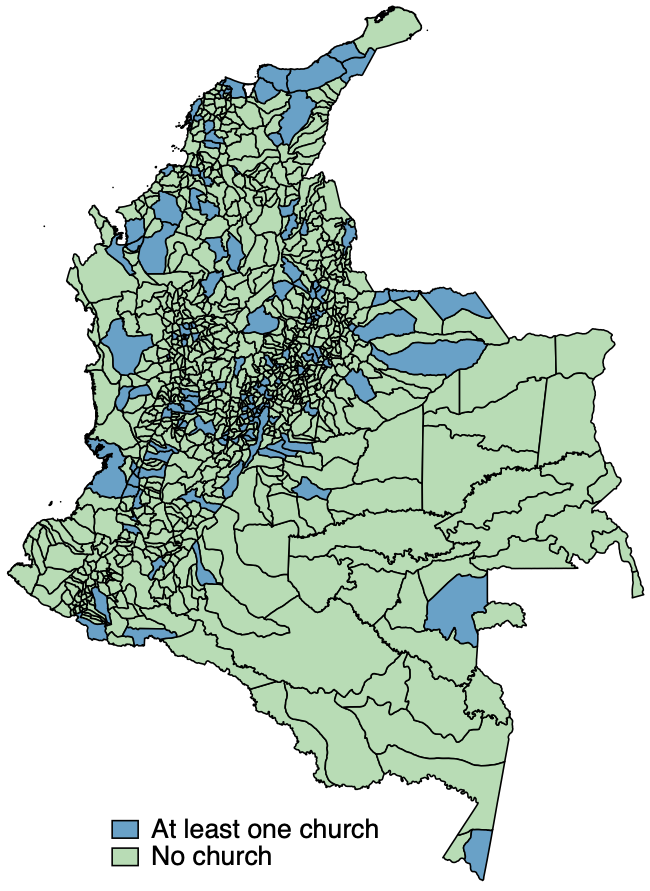}
\end{subfigure}\hspace*{\fill}
\begin{subfigure}{0.4\textwidth}
\caption{2010} \label{}
\includegraphics[width=5.8cm, height=6.7cm]{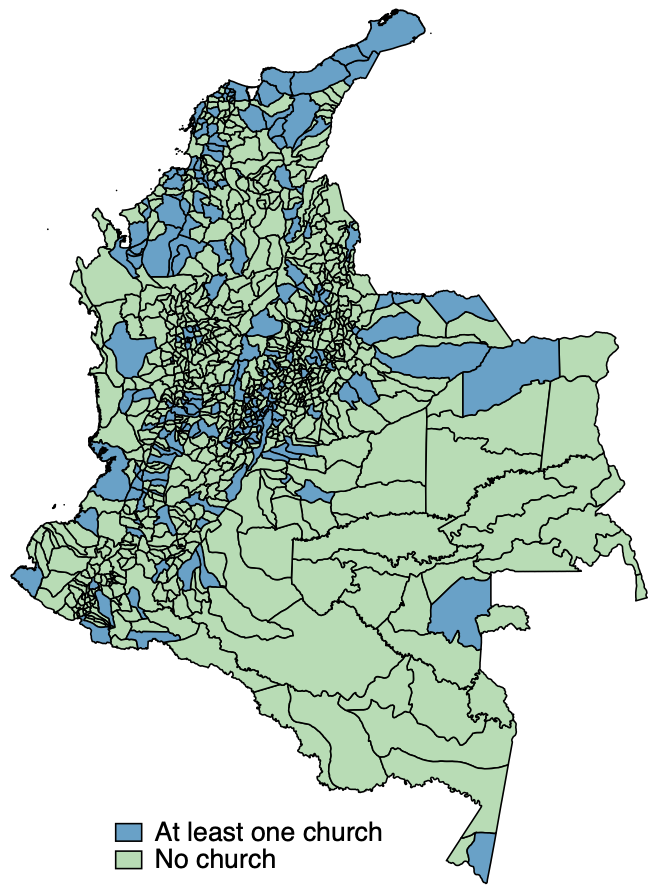}
\end{subfigure}\\
\begin{subfigure}{0.4\textwidth}
\caption{2012} \label{}
\includegraphics[width=5.8cm, height=6.7cm]{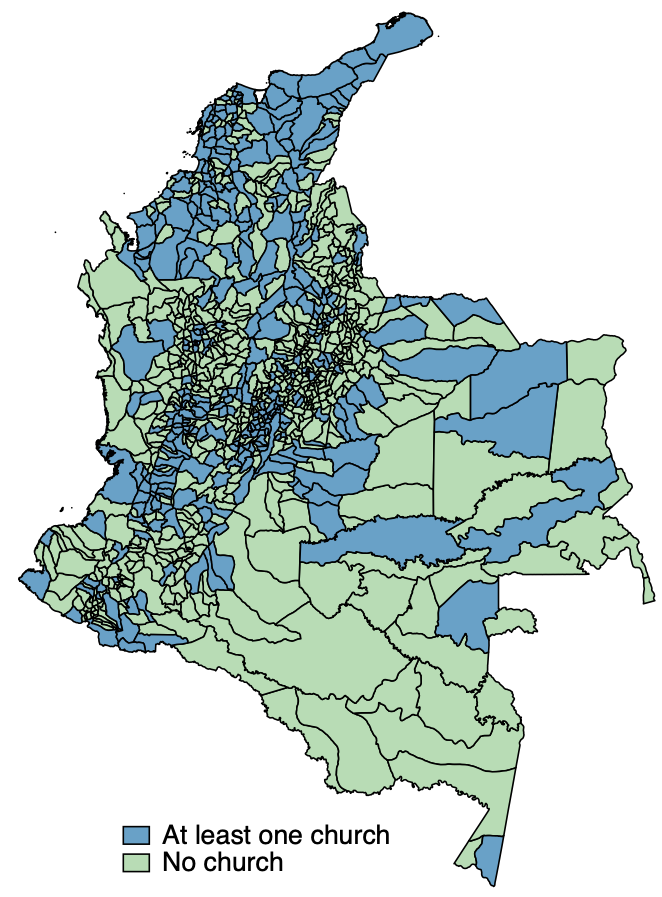}
\end{subfigure}\hspace*{\fill}
\begin{subfigure}{0.4\textwidth}
\caption{2014} \label{}
\includegraphics[width=5.8cm, height=6.7cm]{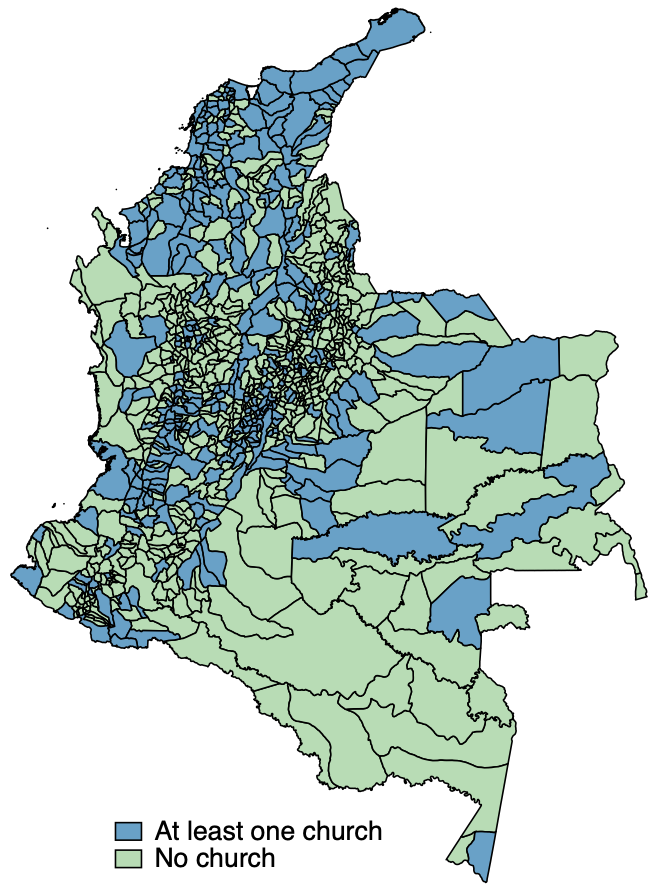}
\end{subfigure}\hspace*{\fill}
\begin{subfigure}{0.4\textwidth}
\caption{2016} \label{}
\includegraphics[width=5.8cm, height=6.7cm]{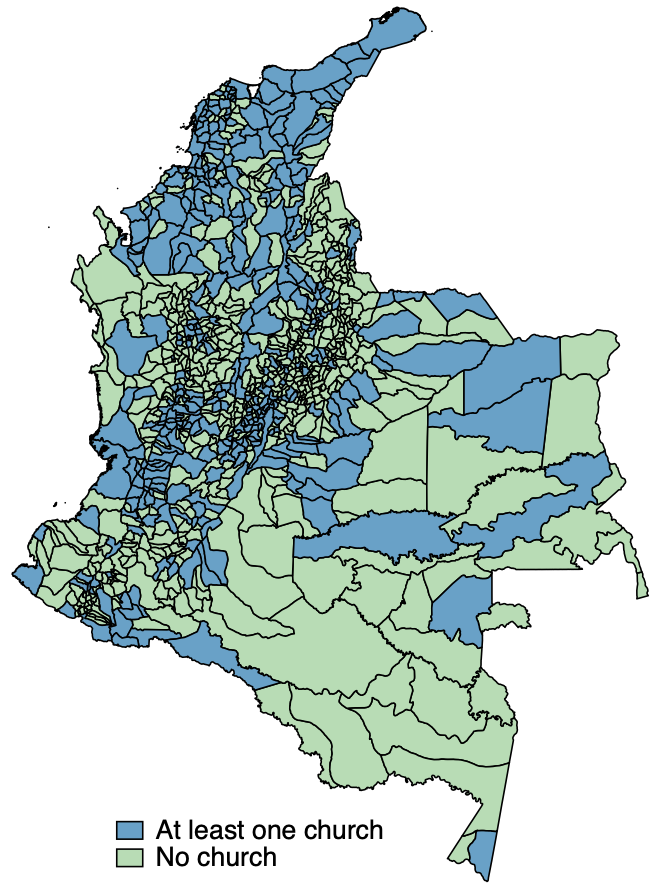}
\end{subfigure}\\
\begin{subfigure}{0.4\textwidth}
\caption{2018} \label{}
\includegraphics[width=5.8cm, height=6.7cm]{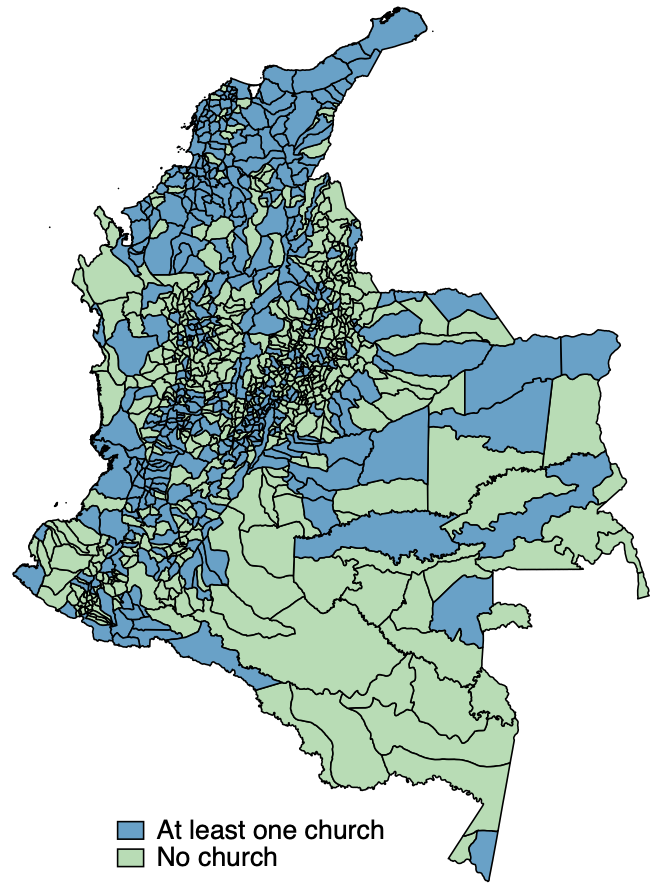}
\end{subfigure}\hspace*{\fill}
\begin{subfigure}{0.4\textwidth}
\caption{2020} \label{}
\includegraphics[width=5.8cm, height=6.7cm]{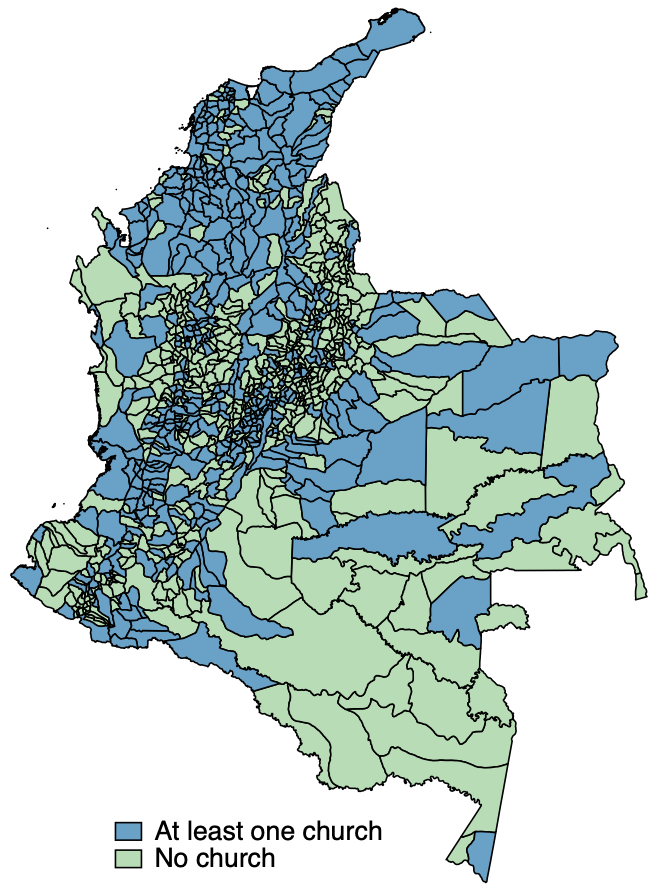}
\end{subfigure}\hspace*{\fill}

\vspace{0.3cm}
     \begin{minipage}{18cm} \footnotesizes \textbf{Note:}  This figure presents the spatial distribution of treated municipalities—defined as those with at least one non-Catholic church—for the years 2006, 2008, 2010, 2012, 2014, 2016, 2018, and 2020. The maps illustrate a steady increase in the number of treated municipalities over time, with expansion occurring across various regions of the country.
\end{minipage}
\end{figure}

\newpage 

\begin{figure}[H]
             \caption{Location of Treated and Control Group Municipalities}
        \label{ncatchurchesmapchange_fig}
        
\begin{subfigure}{0.4\textwidth}
\caption{2004-2006} \label{}
\includegraphics[width=5.8cm, height=6.7cm]{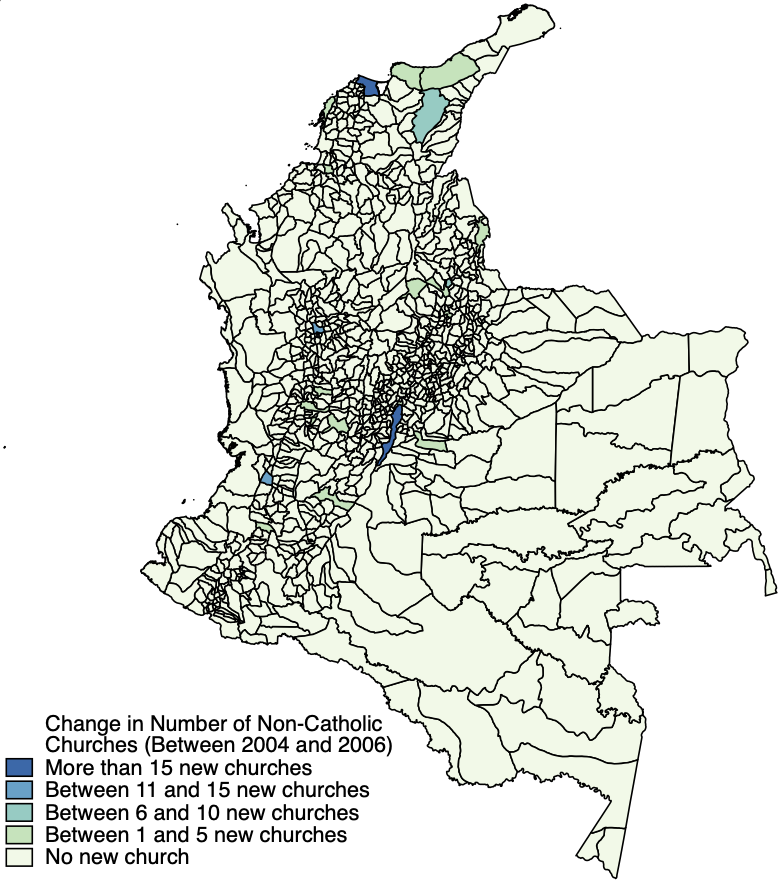}
\end{subfigure}\hspace*{\fill}
\begin{subfigure}{0.4\textwidth}
\caption{2006-2008} \label{}
\includegraphics[width=5.8cm, height=6.7cm]{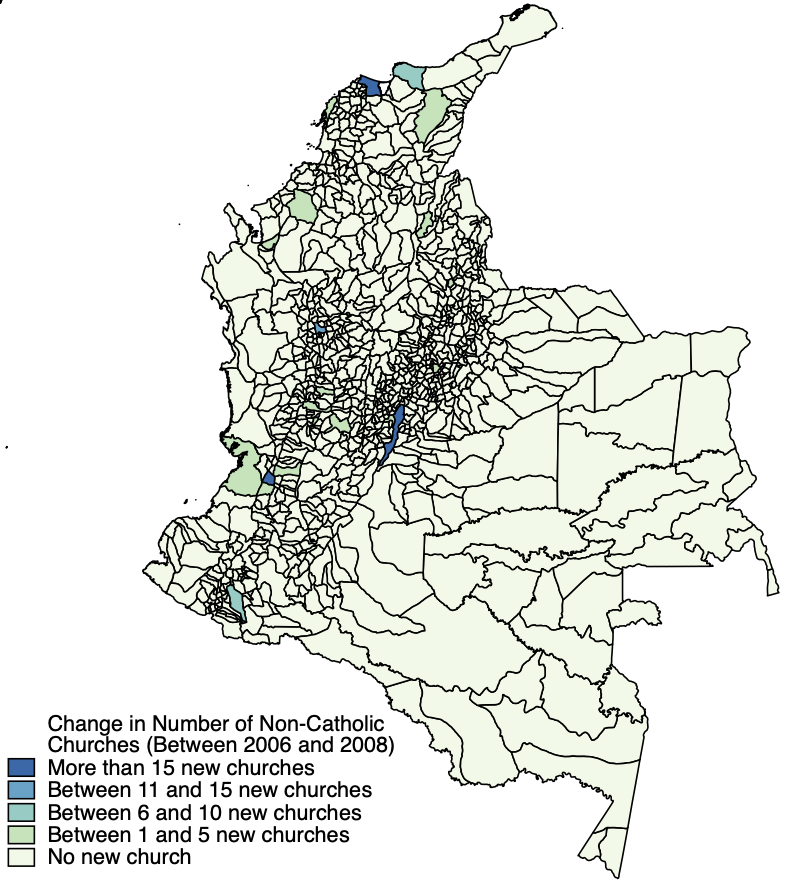}
\end{subfigure}\hspace*{\fill}
\begin{subfigure}{0.4\textwidth}
\caption{2008-2010} \label{}
\includegraphics[width=5.8cm, height=6.7cm]{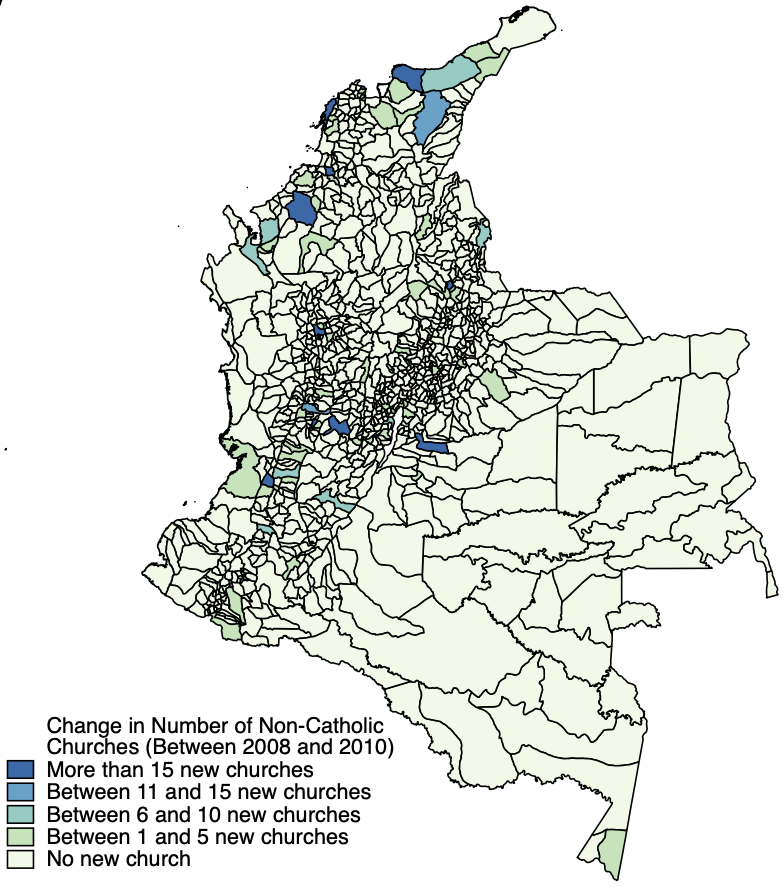}
\end{subfigure}\\
\begin{subfigure}{0.4\textwidth}
\caption{2010-2012} \label{}
\includegraphics[width=5.8cm, height=6.7cm]{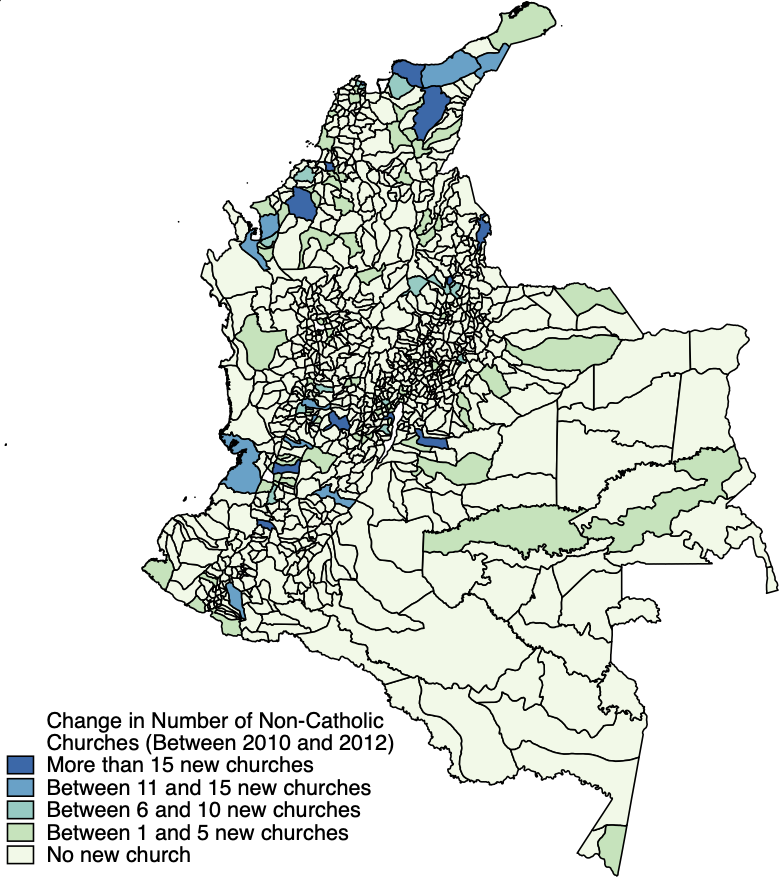}
\end{subfigure}\hspace*{\fill}
\begin{subfigure}{0.4\textwidth}
\caption{2012-2014} \label{}
\includegraphics[width=5.8cm, height=6.7cm]{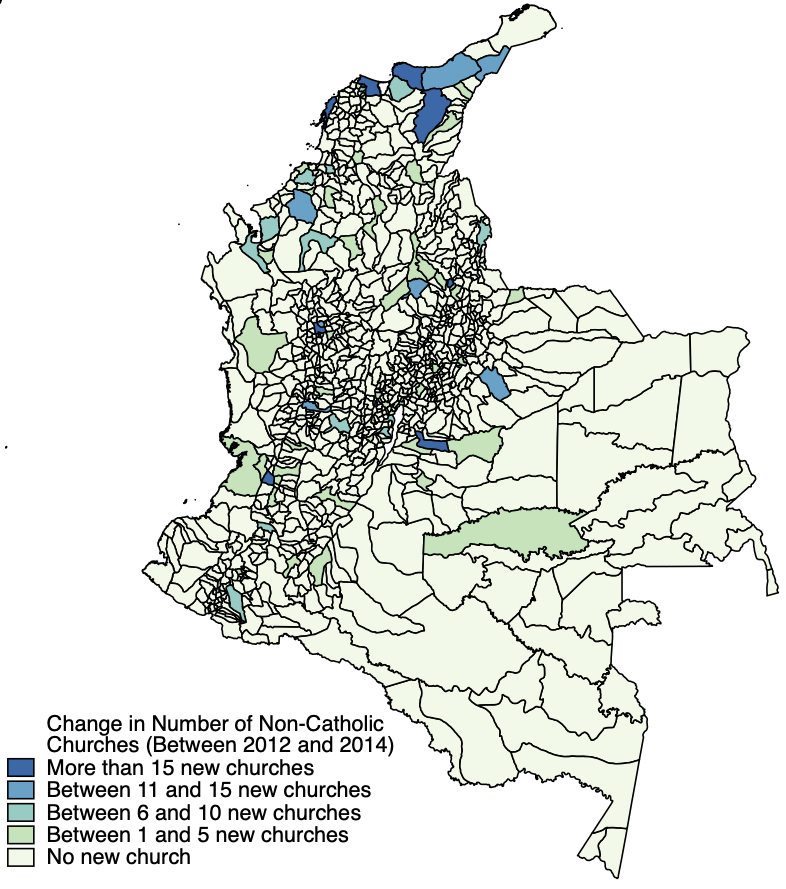}
\end{subfigure}\hspace*{\fill}
\begin{subfigure}{0.4\textwidth}
\caption{2014-2016} \label{}
\includegraphics[width=5.8cm, height=6.7cm]{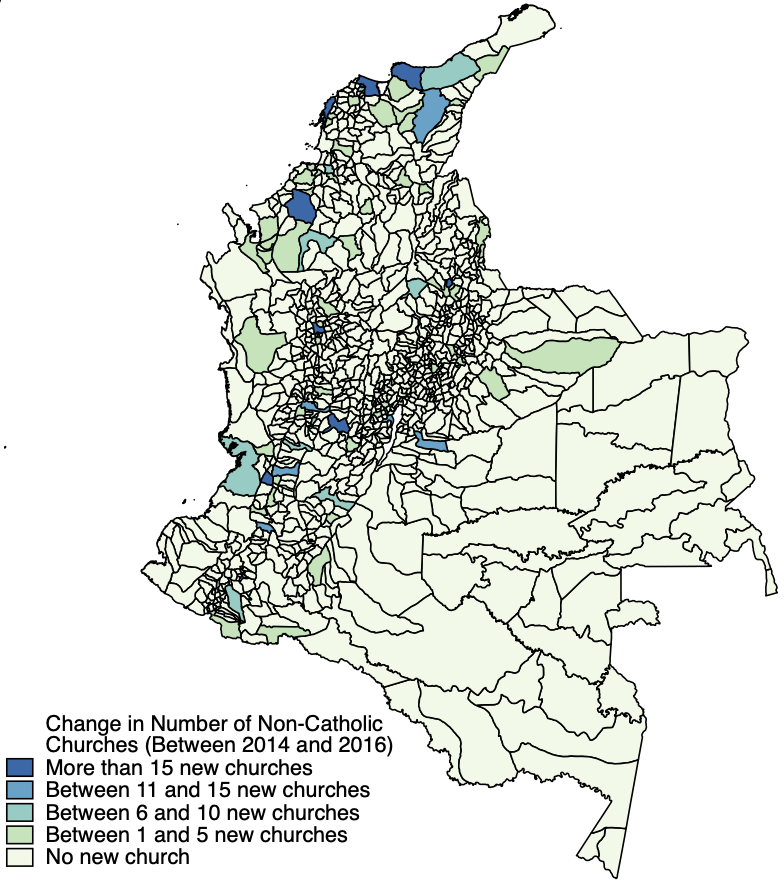}
\end{subfigure}\\
\begin{subfigure}{0.4\textwidth}
\caption{2016-2018} \label{}
\includegraphics[width=5.8cm, height=6.7cm]{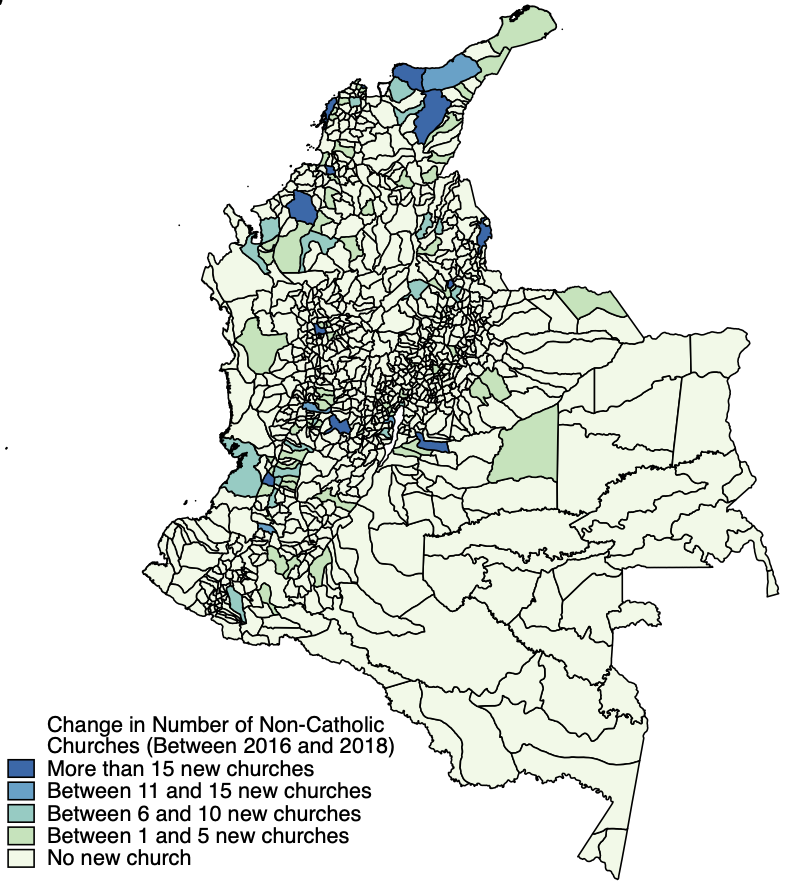}
\end{subfigure}\hspace*{\fill}
\begin{subfigure}{0.4\textwidth}
\caption{2018-2020} \label{}
\includegraphics[width=5.8cm, height=6.7cm]{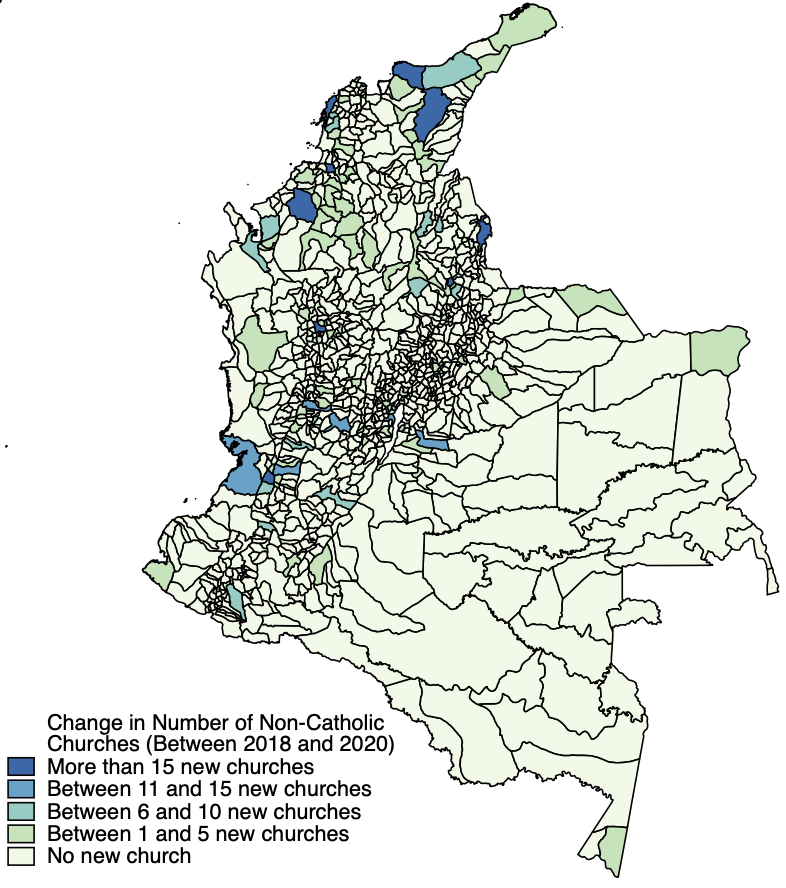}
\end{subfigure}\hspace*{\fill}

\vspace{0.3cm}
     \begin{minipage}{18cm} \footnotesizes \textbf{Note:} This figure depicts the spatial distribution of changes in the number of non-Catholic churches across all Colombian municipalities between 2004 and 2020. The change is measured as the difference between the number of non-Catholic churches in a given year and the number recorded in the same municipality two years earlier. The maps reveal substantial spatial heterogeneity in these changes: municipalities experiencing the largest increases in non-Catholic churches vary over time and are not concentrated in any single region of the country.
\end{minipage}
\end{figure}

\restoregeometry


\begin{table}[H]
\begin{center}
{
\renewcommand{\arraystretch}{0.7}
\setlength{\tabcolsep}{8pt}
\caption {Effect of the Number Non-Catholic Church on Domestic Violence}  \label{tab_twfe_em}
\vspace{-0.3cm}
\small
\centering  \begin{tabular}{lcccccc}
\hline\hline \addlinespace[0.15cm]
  & \multicolumn{6}{c}{Dep. Var: Domestic Violence (per 100.000 inhabitants)} \\\addlinespace[0.1cm]\cmidrule[0.2pt](l){2-7}\addlinespace[0.05cm]
& (1)& (2) & (3)& (4)& (5)& (6) \\   \addlinespace[0.1cm] \hline \addlinespace[0.15cm]
& \multicolumn{2}{c}{Against any}& \multicolumn{2}{c}{Against}& \multicolumn{2}{c}{Against other} \\
& \multicolumn{2}{c}{household member}& \multicolumn{2}{c}{intimate partner}& \multicolumn{2}{c}{household members} \\
\cmidrule[0.2pt](l){2-3}\cmidrule[0.2pt](l){4-5}\cmidrule[0.2pt](l){6-7}\addlinespace[0.15cm]
            \addlinespace[0.3cm]
Effect of number of churches&      -3.540         &      -2.670         &      -0.705         &      -0.082         &      -2.835\sym{**} &      -2.588\sym{**} \\
                    &     (2.461)         &     (2.591)         &     (1.524)         &     (1.623)         &     (1.136)         &     (1.192)         \\
R-sq                &       0.704         &       0.711         &       0.709         &       0.716         &       0.617         &       0.622         \\
Observations        &        7011         &        6519         &        7011         &        6519         &        7011         &        6519         \\

 \addlinespace[0.15cm]\hline \addlinespace[0.15cm]
Baseline controls & No & Yes& No & Yes& No & Yes\\
\addlinespace[0.15cm]\hline\hline
\multicolumn{7}{p{15.8cm}}{\scriptsize{\textbf{Note:}  All columns report the estimates of a variation of Eq. (\ref{didbaselines}), in which the dummy variable capturing the presence of at least one non-Catholic church in each municipality and year has been replaced by an ordinal variable equal to the number of non-Catholic churches in the municipality and year.  All models include municipality and department $\times$ year fixed effects.  The models with baseline controls (columns (2), (4) and (6)) include the following (lagged) covariates:  log of total population, rurality index, and proportion with unsatisfied basic needs. Samples for regression models use data from 2005 to 2019. Robust standard errors (in parentheses) are clustered by municipality. * denotes statistically significant estimates at 10\%, ** denotes significant at 5\% and *** denotes significant at 1\%.} }
\end{tabular}
}
\end{center}
\end{table}


\begin{table}[H]
\begin{center}
{
\renewcommand{\arraystretch}{0.7}
\setlength{\tabcolsep}{2pt}
\caption {Non-Catholic Christianity and Attitudes Towards Religion, Family and Violence}  
\label{tab_attitude_2019}
\vspace{-0.3cm}
\small
  \begin{tabular}{lcccccc}
\hline\hline \addlinespace[0.15cm]
& (1)& (2)& (3)& (4) & (5)& (6)\\   \addlinespace[0.1cm] \hline \addlinespace[0.1cm]
&  \multicolumn{3}{c}{Religious} & \multicolumn{3}{c}{Actively religious} \\\addlinespace[0.1cm]\cmidrule[0.2pt](l){2-4}\cmidrule[0.2pt](l){5-7}\addlinespace[0.05cm]
            & \multicolumn{1}{c}{Importance} & \multicolumn{1}{c}{Importance}& \multicolumn{1}{c}{Anti-violence}& \multicolumn{1}{c}{Importance}& \multicolumn{1}{c}{Importance}& \multicolumn{1}{c}{Anti-violence}\\
                        & \multicolumn{1}{c}{of religion} & \multicolumn{1}{c}{of the family}& \multicolumn{1}{c}{stance}& \multicolumn{1}{c}{of religion}& \multicolumn{1}{c}{of the family}& \multicolumn{1}{c}{stance}\\\addlinespace[0.05cm] \cmidrule[0.2pt](l){2-2}\cmidrule[0.2pt](l){3-3}\cmidrule[0.2pt](l){4-4}\cmidrule[0.2pt](l){5-5}\cmidrule[0.2pt](l){6-6}\cmidrule[0.2pt](l){7-7}
\cmidrule[0.2pt](l){2-2}\cmidrule[0.2pt](l){3-3}\cmidrule[0.2pt](l){4-4}\cmidrule[0.2pt](l){5-5} \addlinespace[0.15cm]
Non Catholic Christian&       0.319\sym{***}&       0.013\sym{*}  &       0.006         &       0.302\sym{***}&       0.035\sym{**} &       0.014         \\
                    &     (0.032)         &     (0.005)         &     (0.003)         &     (0.046)         &     (0.012)         &     (0.008)         \\
R-sq                &       0.060         &       0.004         &       0.002         &       0.048         &       0.008         &       0.003         \\
Observations        &       83308         &       83055         &       81328         &        8832         &        8776         &        8698         \\

\addlinespace[0.15cm]\hline\hline
\multicolumn{7}{p{17.5cm}}{\scriptsize{\textbf{Note:} All columns report the correlation coefficients between the corresponding outcome and a dummy variable indicating whether the individual identifies as a non-Catholic Christian (i.e., Protestant non-Evangelicals, Protestant Evangelicals, or Pentecostals). The data come from the Political Culture Survey conducted by Colombia’s National Administrative Department of Statistics (DANE). The sample is restricted to Catholic and non-Catholic Christians and to the years 2019 and 2021—the only years for which this information is available.  Columns (1) to (3) include those individuals who reported that they identified as Catholics or Protestants (non-evangelicals, evangelicals, and pentecostals). Columns (4) to (6) include individuals who, in addition to identifying as Catholics or Protestants, reported that in the last year they attended meetings or had contact with a religious group or organization. The question in columns (1) and (4) is ``On a scale of 1 to 5, where 1 means not important at all and 5 very important, how important is religion in your life?''. The question in columns (2) and (5) is ``On a scale of 1 to 5, where 1 means not important at all and 5 very important, how important is family in your life?''. The question in columns (3) and (6) is ``Do you agree with people using violence to solve problems?''. Robust standard errors (in parentheses) are clustered by municipality. * denotes statistically significant estimates at 10\%, ** denotes significant at 5\% and *** denotes significant at 1\%.} }
\end{tabular}
}
\end{center}
\end{table}


\begin{table}[H]
\begin{center}
{
\renewcommand{\arraystretch}{0.7}
\setlength{\tabcolsep}{2pt}
\caption {Religious Adherence, Church Attendance, and Attitudes Towards Violence}  
\label{tab_attitude_wvs}
\vspace{-0.3cm}
\small
  \begin{tabular}{lcccccccc}
\hline\hline \addlinespace[0.15cm]
& (1)& (2)& (3)& (4) & (5)& (6)& (7)& (8)\\   
\addlinespace[0.1cm] \hline \addlinespace[0.1cm]
\multicolumn{1}{l}{\emph{\underline{Panel A}:}}        & \multicolumn{8}{c}{Dependent Variable: Degree to which respondent } \\\addlinespace[0.05cm]
    & \multicolumn{8}{c}{considers physical violence to be justified} \\\cmidrule[0.2pt](l){2-9}\addlinespace[0.10cm]

                        & \multicolumn{4}{c}{Ordinal Variable} & \multicolumn{4}{c}{Dummy Variable (1 if  justified)}\\
\addlinespace[0.05cm] \cmidrule[0.2pt](l){2-5}\cmidrule[0.2pt](l){6-9}
 \addlinespace[0.15cm]
non-Catholic Christian&       0.025         &      -0.015         &       0.529\sym{*}  &       0.474\sym{*}  &       0.017         &       0.005         &       0.174\sym{**} &       0.174\sym{**} \\
                    &     (0.064)         &     (0.064)         &     (0.270)         &     (0.263)         &     (0.017)         &     (0.016)         &     (0.066)         &     (0.067)         \\
non-Christian       &      -0.306\sym{*}  &      -0.280         &      -0.604\sym{**} &      -0.540\sym{*}  &      -0.039         &      -0.031         &      -0.108         &      -0.109         \\
                    &     (0.179)         &     (0.198)         &     (0.234)         &     (0.269)         &     (0.075)         &     (0.083)         &     (0.139)         &     (0.147)         \\
non-religious       &       0.071         &       0.060         &       0.125         &       0.121         &       0.007         &      -0.008         &      -0.023         &      -0.042         \\
                    &     (0.073)         &     (0.078)         &     (0.101)         &     (0.088)         &     (0.018)         &     (0.018)         &     (0.035)         &     (0.033)         \\
Church attendance   &      -0.029\sym{***}&      -0.017\sym{*}  &      -0.013         &      -0.001         &      -0.012\sym{***}&      -0.008\sym{**} &      -0.013\sym{***}&      -0.009         \\
                    &     (0.009)         &     (0.009)         &     (0.017)         &     (0.017)         &     (0.003)         &     (0.003)         &     (0.005)         &     (0.005)         \\
non-Catholic $\times$ attendance&                     &                     &      -0.095\sym{*}  &      -0.092\sym{*}  &                     &                     &      -0.029\sym{**} &      -0.031\sym{**} \\
                    &                     &                     &     (0.052)         &     (0.049)         &                     &                     &     (0.012)         &     (0.012)         \\
non-Christian $\times$ attendance&                     &                     &       0.051         &       0.045         &                     &                     &       0.013         &       0.014         \\
                    &                     &                     &     (0.039)         &     (0.043)         &                     &                     &     (0.017)         &     (0.017)         \\
non-religious $\times$ attendance&                     &                     &      -0.008         &      -0.011         &                     &                     &       0.011         &       0.012         \\
                    &                     &                     &     (0.030)         &     (0.029)         &                     &                     &     (0.008)         &     (0.008)         \\
R-sq                &       0.023         &       0.032         &       0.025         &       0.034         &       0.043         &       0.052         &       0.047         &       0.057         \\
Observations        &        3022         &        2958         &        3022         &        2958         &        3022         &        2958         &        3022         &        2958         \\

\addlinespace[0.15cm]\hline
\addlinespace[0.1cm]  \addlinespace[0.1cm]
\multicolumn{1}{l}{\emph{\underline{Panel B}:}}        & \multicolumn{8}{c}{Dependent Variable: Degree to which respondent } \\\addlinespace[0.05cm]
    & \multicolumn{8}{c}{considers physical violence against wives to be justified} \\\cmidrule[0.2pt](l){2-9}\addlinespace[0.10cm]
                        & \multicolumn{4}{c}{Ordinal Variable} & \multicolumn{4}{c}{Dummy Variable (1 if  justified)}\\
\addlinespace[0.05cm] \cmidrule[0.2pt](l){2-5}\cmidrule[0.2pt](l){6-9}
 \addlinespace[0.15cm]
non-Catholic Christian&      -0.063         &      -0.041         &      -0.028         &       0.010         &      -0.018\sym{*}  &      -0.014         &      -0.006         &       0.001         \\
                    &     (0.047)         &     (0.050)         &     (0.147)         &     (0.154)         &     (0.010)         &     (0.011)         &     (0.027)         &     (0.029)         \\
non-Christian       &      -0.084         &      -0.060         &      -0.175         &      -0.172         &      -0.022         &      -0.016         &      -0.068\sym{**} &      -0.068\sym{**} \\
                    &     (0.096)         &     (0.098)         &     (0.140)         &     (0.160)         &     (0.023)         &     (0.024)         &     (0.028)         &     (0.032)         \\
non-religious       &       0.036         &       0.062         &      -0.042         &      -0.003         &       0.008         &       0.010         &      -0.025         &      -0.020         \\
                    &     (0.047)         &     (0.047)         &     (0.067)         &     (0.076)         &     (0.012)         &     (0.012)         &     (0.017)         &     (0.018)         \\
Church attendance   &      -0.009         &      -0.003         &      -0.016\sym{*}  &      -0.008         &      -0.004         &      -0.001         &      -0.007\sym{**} &      -0.004         \\
                    &     (0.007)         &     (0.007)         &     (0.009)         &     (0.006)         &     (0.002)         &     (0.002)         &     (0.003)         &     (0.003)         \\
non-Catholic $\times$ attendance&                     &                     &      -0.005         &      -0.009         &                     &                     &      -0.002         &      -0.002         \\
                    &                     &                     &     (0.024)         &     (0.024)         &                     &                     &     (0.004)         &     (0.004)         \\
non-Christian $\times$ attendance&                     &                     &       0.018         &       0.022         &                     &                     &       0.009\sym{*}  &       0.010\sym{*}  \\
                    &                     &                     &     (0.021)         &     (0.025)         &                     &                     &     (0.005)         &     (0.006)         \\
non-religious $\times$ attendance&                     &                     &       0.026         &       0.022         &                     &                     &       0.011\sym{*}  &       0.010         \\
                    &                     &                     &     (0.024)         &     (0.025)         &                     &                     &     (0.006)         &     (0.006)         \\
R-sq                &       0.023         &       0.029         &       0.023         &       0.029         &       0.031         &       0.040         &       0.032         &       0.041         \\
Observations        &        6023         &        5714         &        6023         &        5714         &        6023         &        5714         &        6023         &        5714         \\

\hline \addlinespace[0.15cm]
Baseline controls & No & Yes& No & Yes& No & Yes& No & Yes\\ \addlinespace[0.15cm]
\hline\hline
\multicolumn{9}{p{17.5cm}}{\scriptsize{\textbf{Note:} All columns report the correlation coefficients between the corresponding outcome variables and dummies for the type of religious denomination reported as the respondent's own religion. The omitted category is Catholic. The data come from the World Values Survey—specifically, waves 6 and 7 for the estimates in Panel A, and waves 5, 6, and 7 for those in Panel B (the only waves in which these outcome variables are available). All models include region and year fixed effects, and control for age, gender, income, education, and marital status. In Panel A, the question asked is: "Please tell me for each of the following actions whether you think it can always be justified, never be justified, or something in between – Any form of violence." In Panel B, the question is: "Please tell me for each of the following statements whether you think it can always be justified, never be justified, or something in between, using this card – For a man to beat his wife." For both questions, the scale ranges from 1 to 10, where 1 means "Never justifiable" and 10 means "Always justifiable." In columns (5) to (8), the dependent variable is a dummy equal to 1 if the response is greater than 1.. Robust standard errors (in parentheses) are clustered by municipality. * denotes statistically significant estimates at 10\%, ** denotes significant at 5\% and *** denotes significant at 1\%.} }
\end{tabular}
}
\end{center}
\end{table}


\begin{figure}[H]
\begin{center}
             \caption{Booklet from the Iglesia Pentecostal Unida de Colombia}
        \label{fig_booklet}
\includegraphics[width=5cm,height=7.5cm]{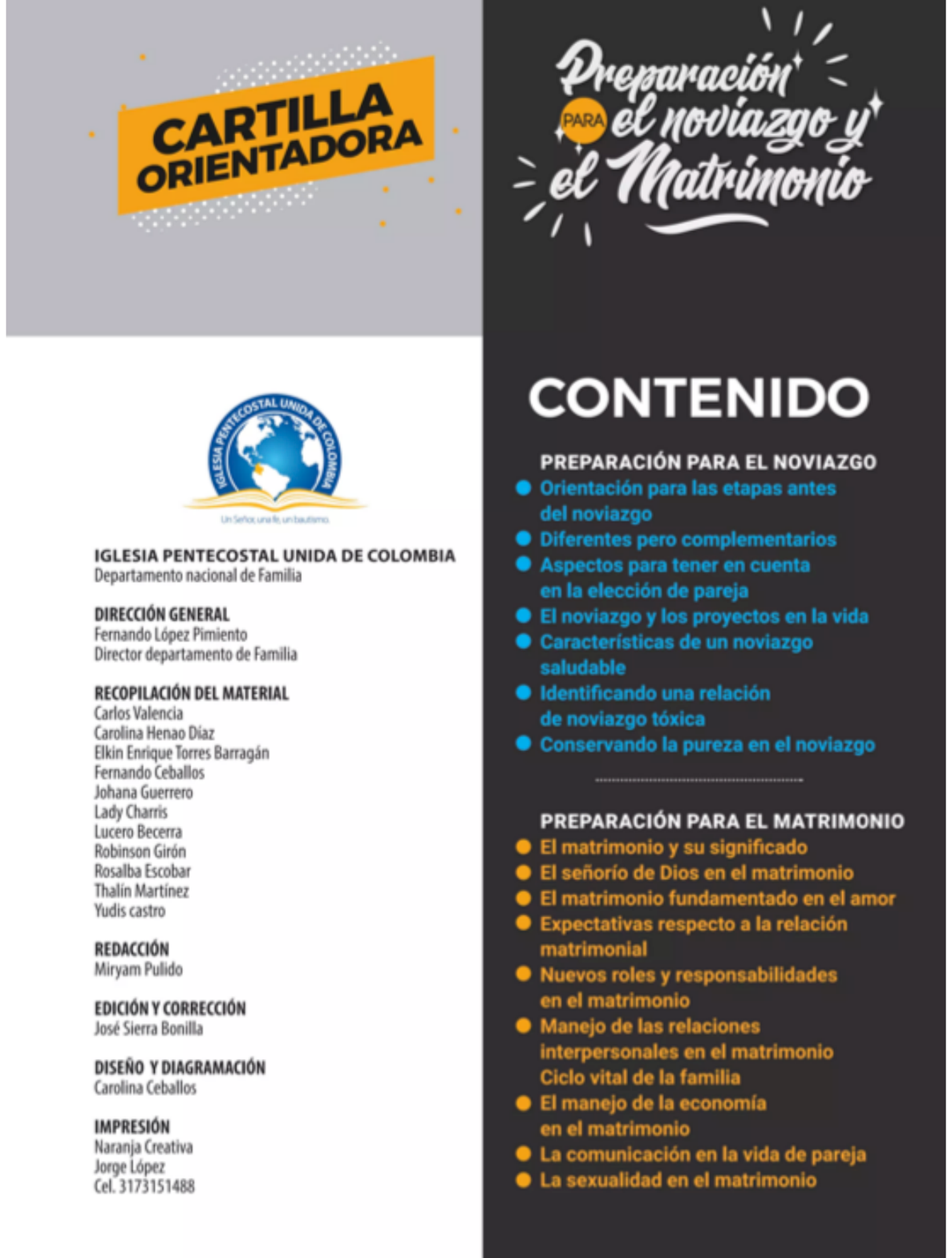}
\includegraphics[width=5cm,height=7.5cm]{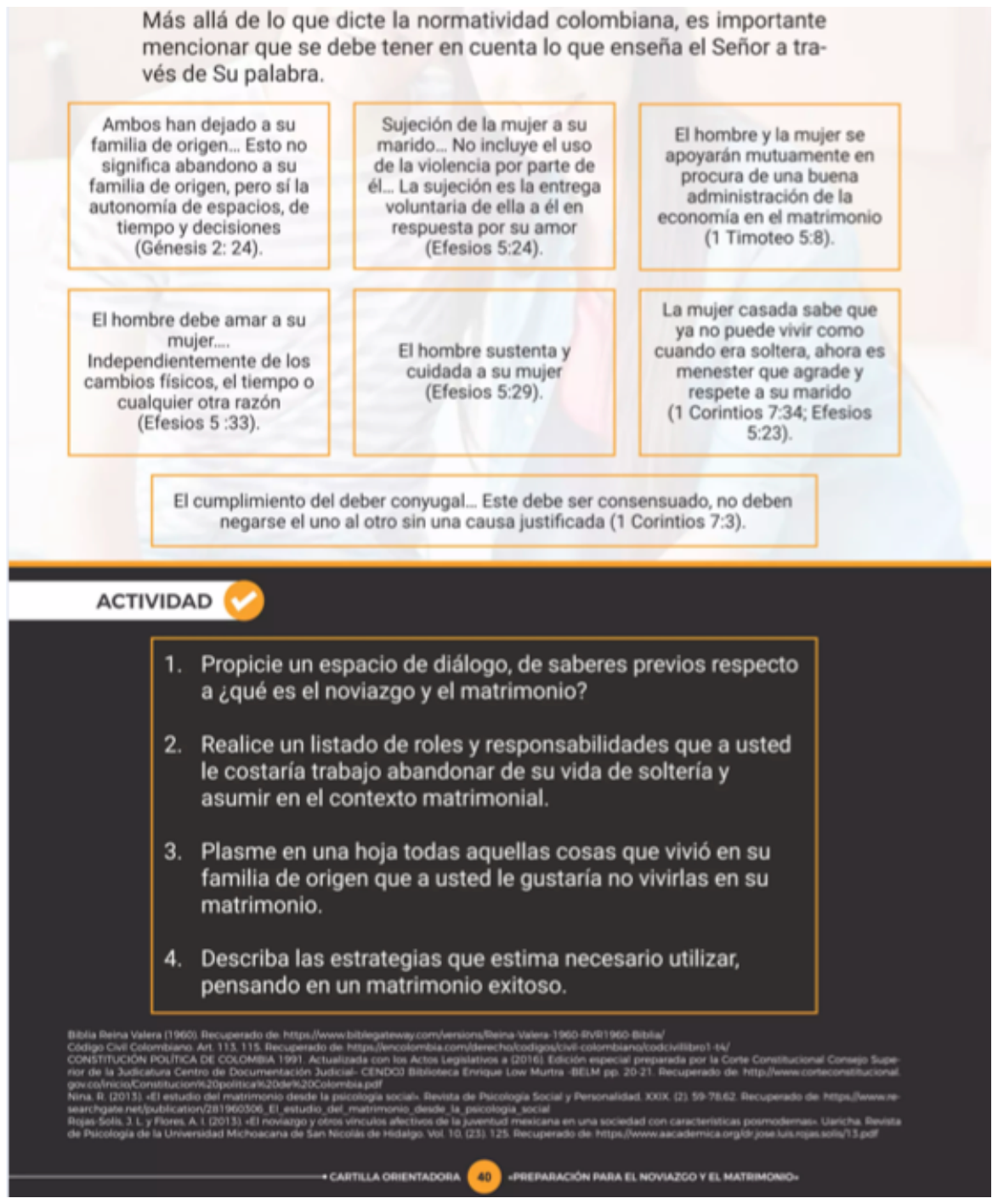}
\end{center}
\end{figure}


\begin{table}[H]

\vspace{-0.5cm}
\begin{center}
{
\renewcommand{\arraystretch}{1}
\setlength{\tabcolsep}{5pt}
\caption {Individual Attitudes Toward Wife-Beating and Non-Catholic Church Expansion: Regional-Level
Correlations}  \label{table_wvsnumbchurchers}
\vspace{-0.3cm}
\footnotesize
\centering  \begin{tabular}{lccccccc}
\hline\hline \addlinespace[0.1cm]
    & \multicolumn{4}{c}{Dependent Variable: Degree to which respondent } \\\addlinespace[0.05cm]
    & \multicolumn{4}{c}{considers physical violence against wives to be justified} \\\cmidrule[0.2pt](l){2-5}
        & \multicolumn{2}{c}{} & \multicolumn{2}{c}{Above-median}  \\
               & \multicolumn{2}{c}{Full sample} & \multicolumn{2}{c}{religious attendance}  \\\cmidrule[0.2pt](l){2-3}\cmidrule[0.2pt](l){4-5}
& (1)& (2) & (3)  & (4)  \\    \addlinespace[0.15cm] \hline \addlinespace[0.05cm]
Number of New Non-Catholic Churches&      -0.258         &      -0.251         &      -0.362         &      -0.349         \\
                    &     (0.150)         &     (0.141)         &     (0.150)\sym{**}  &     (0.146)\sym{**}  \\
                    &[-0.679 0.211]      &[-0.638 0.162]       &[-0.787 0.620]\sym{*}      &[-0.804 0.561]\sym{*}        \\
R-sq                &       0.020         &       0.026         &       0.024         &       0.028         \\
Observations        &        4529         &        4285         &        2440         &        2329         \\

 \addlinespace[0.15cm]\hline \addlinespace[0.15cm]
 Baseline Controls & Yes& Yes& Yes & Yes\\
Additional Controls & No & Yes& No & Yes\\
\addlinespace[0.15cm]\hline\hline
\multicolumn{5}{p{16cm}}{\scriptsize{\textbf{Note:} All columns report the correlation coefficients between a measure of the extent to which the respondent considers it justifiable for a man to beat his wife and the number of new non-Catholic churches established in a given Colombian region in a given year. The data come from the World Values Survey—specifically, waves 5, 6, and 7, which are the only waves in which this outcome variable is available. The sample is restricted to individuals who self-identify as either Catholic or non-Catholic Christians, representing approximately 80\% of the total sample. Regions, as defined by the World Values Survey, correspond to geographically contiguous groups of departments, with eight in total: Atlantic, Pacific, Orinoquía, Amazon, Western, Eastern-Central, and Central. The specific survey item associated with the outcome is: ``Please tell me for each of the following statements whether you think it can always be justified, never be justified, or something in between, using this card – For a man to beat his wife.'' Responses are recorded on a scale from 1 to 10, where 1 means ``Never justifiable'' and 10 means ``Always justifiable.'' All models include fixed effects for region and survey year, along with controls for gender, religious affiliation, age group, and regional population size.  Columns (3) and (4) further restrict the sample to individuals with above-median levels of religious attendance. Columns (2) and (4) additionally include controls for self-identified religiosity, marital status, self-reported income, and educational attainment. Robust standard errors (in parentheses) are clustered at the region level. Wild bootstrap confidence intervals are shown in square brackets. * denotes statistically significant estimates at 10\%, ** denotes significant at 5\% and *** denotes significant at 1\%.} }
\end{tabular}
}
\end{center}
\end{table}


\begin{table}[H]
\begin{center}
{
\renewcommand{\arraystretch}{0.8}
\setlength{\tabcolsep}{5pt}
\caption {Labor Market Outcomes and Non-Catholic Church Expansion: Department-Level Correlations}  
\label{tab_labormarketdeptos}
\vspace{-0.3cm}
\footnotesize
  \begin{tabular}{lccccccccc}
\hline\hline \addlinespace[0.15cm]
& (1)& (2)& (3)& (4) & (5)& (6)\\   
\addlinespace[0.1cm] \hline \addlinespace[0.1cm]
\multicolumn{1}{l}{\emph{\underline{Panel A}:}}                  & \multicolumn{6}{c}{Dependent Variable: Unemployment Rate} \\\addlinespace[0.12cm]\cmidrule[0.2pt](l){2-7} \addlinespace[0.05cm]
                        & \multicolumn{2}{c}{All} & \multicolumn{2}{c}{Women}& \multicolumn{2}{c}{Men}\\
\addlinespace[0.05cm] \cmidrule[0.2pt](l){2-3}\cmidrule[0.2pt](l){4-5}\cmidrule[0.2pt](l){6-7}
 \addlinespace[0.05cm]
Number of New Non-Catholic Churches&      -0.023\sym{***}&      -0.015\sym{**} &      -0.028\sym{***}&      -0.020\sym{**} &      -0.019\sym{***}&      -0.011\sym{**} \\
                    &     (0.006)         &     (0.006)         &     (0.010)         &     (0.009)         &     (0.005)         &     (0.005)         \\
R-sq                &       0.632         &       0.656         &       0.554         &       0.598         &       0.692         &       0.713         \\
Observations        &         479         &         408         &         479         &         408         &         479         &         408         \\

\addlinespace[0.05cm]\hline
\addlinespace[0.05cm]  \addlinespace[0.1cm]
\multicolumn{1}{l}{\emph{\underline{Panel B}:}}                  & \multicolumn{6}{c}{Dependent Variable: Labor Force Participation Rate} \\\addlinespace[0.12cm]\cmidrule[0.2pt](l){2-7} \addlinespace[0.05cm]
                        & \multicolumn{2}{c}{All} & \multicolumn{2}{c}{Women}& \multicolumn{2}{c}{Men}\\
\addlinespace[0.05cm] \cmidrule[0.2pt](l){2-3}\cmidrule[0.2pt](l){4-5}\cmidrule[0.2pt](l){6-7}
 \addlinespace[0.05cm]
Number of New Non-Catholic Churches&       0.026\sym{*}  &       0.011         &       0.030         &       0.016         &       0.021\sym{**} &       0.006         \\
                    &     (0.013)         &     (0.009)         &     (0.018)         &     (0.013)         &     (0.008)         &     (0.006)         \\
R-sq                &       0.739         &       0.799         &       0.766         &       0.819         &       0.713         &       0.759         \\
Observations        &         479         &         408         &         479         &         408         &         479         &         408         \\

\addlinespace[0.05cm]\hline
\addlinespace[0.05cm]  \addlinespace[0.1cm]
\multicolumn{1}{l}{\emph{\underline{Panel C}:}}                  & \multicolumn{6}{c}{Dependent Variable: Employment rate}\\\addlinespace[0.12cm]\cmidrule[0.2pt](l){2-7} \addlinespace[0.05cm]
                        & \multicolumn{2}{c}{All} & \multicolumn{2}{c}{Women}& \multicolumn{2}{c}{Men}\\
\addlinespace[0.05cm] \cmidrule[0.2pt](l){2-3}\cmidrule[0.2pt](l){4-5}\cmidrule[0.2pt](l){6-7}
 \addlinespace[0.05cm]
Number of New Non-Catholic Churches&       0.038\sym{***}&       0.020\sym{*}  &       0.041\sym{**} &       0.024\sym{*}  &       0.033\sym{***}&       0.014\sym{*}  \\
                    &     (0.013)         &     (0.010)         &     (0.017)         &     (0.012)         &     (0.009)         &     (0.008)         \\
R-sq                &       0.692         &       0.773         &       0.753         &       0.814         &       0.609         &       0.696         \\
Observations        &         479         &         408         &         479         &         408         &         479         &         408         \\

\hline \addlinespace[0.15cm]
Baseline controls & No & Yes& No& Yes & No & Yes\\ \addlinespace[0.15cm]
\hline\hline
\multicolumn{7}{p{16cm}}{\scriptsize{\textbf{Note:} All columns report the correlation coefficients between the corresponding labor market outcome and the number of new non-Catholic churches established in a given Colombian department in a given year. The labor market data come from the Gran Encuesta Integrada de Hogares (GEIH), conducted by Colombia’s National Administrative Department of Statistics (DANE), and cover the period from 2005 to 2019, with annual frequency. All models include department and year fixed effects. Columns (3), (5), and (6) additionally control for population size, GDP per capita, and the share of the population living in rural areas.. Robust standard errors (in parentheses) are clustered at the region level. * denotes statistically significant estimates at 10\%, ** denotes significant at 5\% and *** denotes significant at 1\%..} }
\end{tabular}
}
\end{center}
\end{table}


\begin{figure}[H]
\begin{center}
             \caption{Partial Correlation Between Domestic Violence and GDP per Capita at the Department Level}
        \label{fig_corrdomviolenceGDPdeptos}
        \vspace{-0.3cm}
\includegraphics[width=12cm,height=8cm]{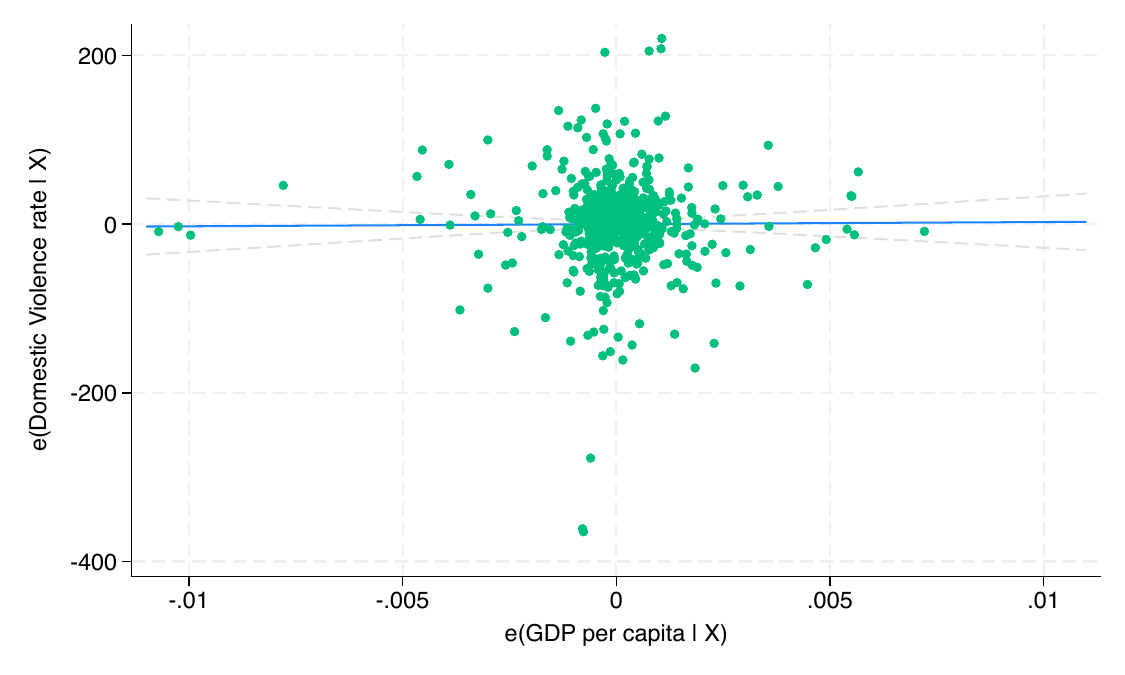}
     \begin{minipage}{12cm} \footnotesizes \textbf{Source:}  This figure presents a partial correlation plot illustrating the relationship between the domestic violence rate and GDP per capita across Colombian departments for the period 2005 to 2021. The analysis controls for total population, the proportion of the rural population, and the homicide rate, and includes fixed effects for both department and year. The estimated coefficient is 252.566, with a standard error of 1254.041.
\end{minipage}
\end{center}
\end{figure}


\begin{table}[H]
\begin{center}
{
\renewcommand{\arraystretch}{0.7}
\setlength{\tabcolsep}{15pt}
\caption {Effect of the First Non-Catholic Church on Domestic Violence:  Heterogeneous Effects by Number of Catholic Churches in 1995}  \label{twfe_domeviolenceMLrate_highlowcatholic_tab}
\vspace{-0.3cm}
\small
\centering  \begin{tabular}{lcccc}
\hline\hline \addlinespace[0.15cm]
  & \multicolumn{4}{c}{Dep. Var: Domestic Violence (per 100.000 inhabitants)} \\\addlinespace[0.1cm]\cmidrule[0.2pt](l){2-5}\addlinespace[0.05cm]
& (1)& (2)& (3)& (4) \\   \addlinespace[0.1cm] \cmidrule[0.2pt](l){2-3}\cmidrule[0.2pt](l){4-5}
            & \multicolumn{2}{c}{Above median \# of} & \multicolumn{2}{c}{Below median \# of}\\
                        & \multicolumn{2}{c}{Catholic churches in 1995} & \multicolumn{2}{c}{Catholic churches in 1995}\\\addlinespace[0.15cm]
\hline \addlinespace[0.15cm]
            \multicolumn{1}{l}{\emph{\underline{Panel A}:}}            & \multicolumn{4}{c}{\emph{Static}  Specification}\\
            \addlinespace[0.3cm]
Effect at t$\geq 0$ &     -25.963\sym{**} &     -31.757\sym{***}&       4.880         &       2.568         \\
                    &    (11.482)         &    (11.600)         &     (6.112)         &     (6.066)         \\
R-sq                &       0.676         &       0.687         &       0.817         &       0.823         \\
Observations        &        4000         &        3698         &        3420         &        3189         \\

\addlinespace[0.15cm]\hline\addlinespace[0.15cm]
            \multicolumn{1}{l}{\emph{\underline{Panel B}:}}            & \multicolumn{4}{c}{\emph{Dynamic} Specification}\\\addlinespace[0.3cm]
Effect at t=-3      &       3.702         &       2.494         &      -5.646         &      -7.890         \\
                    &    (10.677)         &    (10.977)         &     (6.454)         &     (6.496)         \\
Effect at t=-2      &       0.064         &      -3.820         &       0.350         &      -1.034         \\
                    &    (10.340)         &    (10.212)         &     (6.395)         &     (6.552)         \\
Effect at t=-1      &       0.000         &       0.000         &       0.000         &       0.000         \\
                    &         (.)         &         (.)         &         (.)         &         (.)         \\
Effect at t=0       &     -17.419\sym{*}  &     -21.461\sym{**} &       0.160         &      -1.410         \\
                    &     (9.964)         &    (10.655)         &     (5.504)         &     (5.782)         \\
Effect at t=+1      &     -17.698\sym{*}  &     -25.827\sym{**} &       9.449         &       6.978         \\
                    &     (9.493)         &    (10.393)         &     (6.200)         &     (6.573)         \\
Effect at t=+2      &     -14.166         &     -21.040\sym{*}  &       6.328         &       4.904         \\
                    &    (11.352)         &    (12.079)         &     (6.513)         &     (6.874)         \\
Effect at t=+3      &      -5.646         &      -5.362         &       4.598         &       2.380         \\
                    &    (12.716)         &    (11.797)         &     (7.326)         &     (7.806)         \\
Effect at t=+4      &      -8.756         &     -15.501         &      -1.635         &      -4.465         \\
                    &    (12.668)         &    (13.063)         &     (7.213)         &     (7.740)         \\
Effect at t=+5      &     -13.041         &     -22.755\sym{*}  &       4.837         &       0.557         \\
                    &    (10.744)         &    (11.988)         &     (7.248)         &     (8.015)         \\
Effect at t=+6      &     -16.163         &     -21.811\sym{*}  &      -1.904         &      -6.014         \\
                    &    (11.554)         &    (12.322)         &     (6.978)         &     (7.928)         \\
Effect at t=+7      &     -20.061\sym{*}  &     -27.633\sym{**} &      -3.849         &      -7.196         \\
                    &    (11.165)         &    (13.031)         &     (6.766)         &     (8.210)         \\
Effect at t=+8      &     -28.074\sym{**} &     -21.784         &      -1.514         &       2.362         \\
                    &    (13.441)         &    (15.386)         &     (8.266)         &    (10.483)         \\
R-sq                &       0.672         &       0.682         &       0.821         &       0.826         \\
Observations        &        3880         &        3586         &        3345         &        3120         \\

 \addlinespace[0.15cm]\hline \addlinespace[0.15cm]
Baseline controls & No & Yes& No & Yes\\
\addlinespace[0.15cm]\hline\hline
\multicolumn{5}{p{13.5cm}}{\scriptsize{\textbf{Note:} All columns in Panel A report the estimates from Eq. (\ref{didbaselines}) and all columns in Panel B report the estimates from Eq. (\ref{didbaseline}), in municipalities with an above median (columns (1) and (2)) and below median (columns (3) and (4)) number Catholic churches (per 100.000 inhabitants), using data from 1995.  All models include municipality and department $\times$ year fixed effects. The models with baseline controls (columns (2) and (4)) include the following (lagged) covariates:  log of total population, rurality index, and proportion with unsatisfied basic needs. Samples for regression models use data from 2005 to 2019.  The models in Panel B include 10 lags and 10 leads, normalized to the period prior to treatment. Robust standard errors (in parentheses) are clustered by municipality. * denotes statistically significant estimates at 10\%, ** denotes significant at 5\% and *** denotes significant at 1\%.} }
\end{tabular}
}
\end{center}
\end{table}


\begin{table}[H]
\begin{center}
{
\renewcommand{\arraystretch}{0.7}
\setlength{\tabcolsep}{15pt}
\caption {Effect of the First Non-Catholic Church on Domestic Violence: Heterogeneous Effects by Share of Population Aged 25 to 44}  \label{twfe_domeviolenceMLrate_highlowyoung_tab}
\vspace{-0.3cm}
\small
\centering  \begin{tabular}{lcccc}
\hline\hline \addlinespace[0.15cm]
  & \multicolumn{4}{c}{Dep. Var: Domestic Violence (per 100.000 inhabitants)} \\\addlinespace[0.1cm]\cmidrule[0.2pt](l){2-5}\addlinespace[0.05cm]
& (1)& (2)& (3)& (4) \\   \addlinespace[0.1cm] \cmidrule[0.2pt](l){2-3}\cmidrule[0.2pt](l){4-5}
            & \multicolumn{2}{c}{Above median} & \multicolumn{2}{c}{Below median}\\
                        & \multicolumn{2}{c}{population aged 25 to 44} & \multicolumn{2}{c}{population aged 25 to 44}\\\addlinespace[0.15cm]
\hline \addlinespace[0.15cm]
            \multicolumn{1}{l}{\emph{\underline{Panel A}:}}            & \multicolumn{4}{c}{\emph{Static} Specification}\\
            \addlinespace[0.3cm]
Effect at t$\geq 0$ &     -20.296\sym{**} &     -22.454\sym{**} &       1.289         &      -2.189         \\
                    &    (10.107)         &    (10.330)         &     (7.439)         &     (7.664)         \\
R-sq                &       0.767         &       0.769         &       0.702         &       0.709         \\
Observations        &        3700         &        3418         &        3720         &        3470         \\

\addlinespace[0.15cm]\hline\addlinespace[0.15cm]
            \multicolumn{1}{l}{\emph{\underline{Panel B}:}}            & \multicolumn{4}{c}{\emph{Dynamic}  Specification}\\\addlinespace[0.3cm]
Effect at t=-3      &       1.588         &      -0.368         &      -6.914         &      -6.577         \\
                    &     (9.781)         &    (10.121)         &    (10.069)         &    (10.082)         \\
Effect at t=-2      &      10.200         &       9.034         &      -9.206         &     -10.517         \\
                    &     (9.539)         &     (9.646)         &     (9.799)         &     (9.491)         \\
Effect at t=-1      &       0.000         &       0.000         &       0.000         &       0.000         \\
                    &         (.)         &         (.)         &         (.)         &         (.)         \\
Effect at t=0       &      -6.430         &      -7.404         &      -9.862         &     -10.672         \\
                    &     (8.855)         &     (9.287)         &     (9.329)         &     (9.735)         \\
Effect at t=+1      &      -2.415         &      -2.873         &      -6.920         &     -11.491         \\
                    &     (8.271)         &     (8.266)         &     (9.337)         &    (10.522)         \\
Effect at t=+2      &       3.603         &       1.149         &      -6.983         &      -9.069         \\
                    &     (9.940)         &    (10.754)         &     (9.371)         &     (9.725)         \\
Effect at t=+3      &       8.987         &       9.017         &      -8.160         &      -7.747         \\
                    &     (9.856)         &     (9.701)         &     (9.571)         &    (10.292)         \\
Effect at t=+4      &      -3.435         &      -6.021         &     -11.171         &     -15.916         \\
                    &     (9.328)         &     (9.580)         &     (9.923)         &    (10.464)         \\
Effect at t=+5      &      -5.546         &      -7.746         &     -10.494         &     -17.710\sym{*}  \\
                    &     (9.352)         &    (10.388)         &     (8.965)         &     (9.399)         \\
Effect at t=+6      &      -5.475         &     -11.168         &      -8.593         &     -11.569         \\
                    &    (10.226)         &    (10.743)         &     (8.598)         &     (9.314)         \\
Effect at t=+7      &      -5.486         &      -6.828         &      -7.078         &      -9.454         \\
                    &    (10.346)         &    (12.233)         &     (9.912)         &    (12.064)         \\
Effect at t=+8      &     -14.010         &     -10.536         &      -4.254         &       2.270         \\
                    &    (10.021)         &    (10.083)         &    (12.235)         &    (13.714)         \\
R-sq                &       0.769         &       0.771         &       0.705         &       0.712         \\
Observations        &        3700         &        3419         &        3720         &        3470         \\

 \addlinespace[0.15cm]\hline \addlinespace[0.15cm]
Baseline controls & No & Yes& No & Yes\\
\addlinespace[0.15cm]\hline\hline
\multicolumn{5}{p{13.5cm}}{\scriptsize{\textbf{Note:} All columns in Panel A report the estimates from Eq. (\ref{didbaselines}) and all columns in Panel B report the estimates from Eq. (\ref{didbaseline}), in municipalities with an above median (columns (1) and (2)) and below median (columns (3) and (4)) share of population aged 25 to 44, using data from 2004.  All models include municipality and  department $\times$ year fixed effects. The models with baseline controls (columns (2) and (4)) include the following (lagged) covariates:  log of total population, rurality index, and proportion with unsatisfied basic needs. Samples for regression models use data from 2005 to 2019.  The models in Panel B include 10 lags and 10 leads, normalized to the period prior to treatment. Robust standard errors (in parentheses) are clustered by municipality. * denotes statistically significant estimates at 10\%, ** denotes significant at 5\% and *** denotes significant at 1\%} }
\end{tabular}
}
\end{center}
\end{table}


\begin{table}[H]
\begin{center}
{
\renewcommand{\arraystretch}{0.8}
\setlength{\tabcolsep}{0pt}
\caption {Effect of the First Non-Catholic Church on Domestic Violence: Heterogeneous Effects by Number of Catholic Churches, Population Size, and Share of Young Adults}  \label{twfe_domeviolenceMLrate_heteroeffects_tab_int}
\vspace{-0.3cm}
\small
\centering  \begin{tabular}{lcccccc}
\hline\hline \addlinespace[0.15cm]
  & \multicolumn{6}{c}{Dep. Var: Domestic Violence (per 100.000 inhabitants)} \\\addlinespace[0.1cm]\cmidrule[0.2pt](l){2-7}\addlinespace[0.05cm]
& (1)& (2)& (3)& (4) & (5)& (6) \\   \addlinespace[0.1cm] 
\hline \addlinespace[0.15cm]
Effect at t$\geq 0$ &      -1.699         &      -4.024         &      -2.574         &      -2.941         &       1.000         &      -0.747         \\
                    &     (5.943)         &     (6.158)         &     (5.486)         &     (5.529)         &     (6.875)         &     (7.172)         \\
(Effect at t$\geq 0$)$\times$(High $\#$ of Cath. Churches)&     -16.893\sym{*}  &     -20.069\sym{**} &                     &                     &                     &                     \\
                    &     (9.254)         &     (9.755)         &                     &                     &                     &                     \\
(Effect at t$\geq 0$)$\times$(Low Population Size)&                     &                     &     -14.191\sym{*}  &     -17.381\sym{**} &                     &                     \\
                    &                     &                     &     (8.180)         &     (8.722)         &                     &                     \\
(Effect at t$\geq 0$)$\times$(High Share of Young Adults)&                     &                     &                     &                     &     -23.856\sym{**} &     -23.861\sym{**} \\
                    &                     &                     &                     &                     &    (11.338)         &    (11.602)         \\
Net Effect          &     -18.6**         &     -24.1**         &      -16.8*         &     -20.3**         &     -22.9**         &     -24.6**         \\
                    &      (9.24)         &      (9.67)         &(8.800)         &(9.290)         &      (9.93)         &     (10.03)         \\
R-sq                &       0.716         &       0.722         &       0.716         &       0.720         &       0.716         &       0.720         \\
Observations        &        7465         &        6929         &        7465         &        6929         &        7465         &        6929         \\

 \addlinespace[0.15cm]\hline \addlinespace[0.15cm]
Baseline Controls & No & Yes& No & Yes& No & Yes\\
\addlinespace[0.15cm]\hline\hline
\multicolumn{7}{p{17cm}}{\scriptsize{\textbf{Note:}  All columns report estimates from Equation (\ref{didbaselines}), augmented with interaction terms between the treatment variable and dummy variables indicating whether a municipality had: (i) a high (above-median) per capita number of Catholic churches in 1995 (columns (1) and (2)); (ii) a high (above-median) population size in 2004 (columns (3) and (4)); or (iii) a high (above-median) share of individuals aged 25 to 44 (columns (5) and (6)). All models include municipality and  department $\times$ year fixed effects. Specifications with baseline controls (columns (2), (4), and (6)) additionally include the following lagged covariates: the rurality index and the proportion of the population with unsatisfied basic needs. Column (2) also controls for the logarithm of total population. The estimation sample spans the years 2005 to 2019. Robust standard errors, clustered at the municipality level, are reported in parentheses.* denotes statistically significant estimates at 10\%, ** denotes significant at 5\% and *** denotes significant at 1\%} }
\end{tabular}
}
\end{center}
\end{table}


\begin{table}[H]
\begin{center}
{
\renewcommand{\arraystretch}{0.8}
\setlength{\tabcolsep}{5pt}
\caption {Effect of the First Non-Catholic Church on Domestic Violence: Heterogeneous Effects by Ethnic Fractionalization and Number of Civil Society Organizations}  
\label{twfe_domeviolenceMLrate_highlowcivilethnicivil_tab_int}
\vspace{-0.3cm}
\small
\centering  \begin{tabular}{lcccccc}
\hline\hline \addlinespace[0.15cm]
  & \multicolumn{6}{c}{Dep. Var: Domestic Violence (per 100.000 inhabitants)} \\\addlinespace[0.1cm]\cmidrule[0.2pt](l){2-7}\addlinespace[0.05cm]
& (1)& (2)& (3)& (4)& (5)& (6)  \\   \addlinespace[0.1cm] \hline \addlinespace[0.1cm]
Effect at t$\geq 0$ &      -10.35         &      -11.45         &      -4.311         &      -7.782         &      -3.053         &      -2.693         \\
                    &     (7.116)         &     (7.227)         &     (7.624)         &     (7.742)         &     (6.026)         &     (6.088)         \\
(Effect at t$\geq 0$)$\times$(Low EF)&      -0.387         &      -5.311         &                     &                     &      -2.672         &      -10.71         \\
                    &     (9.778)         &     (10.10)         &                     &                     &     (11.36)         &     (11.89)         \\
(Effect at t$\geq 0$)$\times$(Few CSOs)&                     &                     &      -14.86         &      -14.68         &      -17.96         &      -21.15         \\
                    &                     &                     &     (11.00)         &     (10.96)         &     (13.20)         &     (13.38)         \\
(Effect at t$\geq 0$)$\times$(Low EF)$\times$(Few CSOs)&                     &                     &                     &                     &       6.131         &       13.20         \\
                    &                     &                     &                     &                     &     (21.02)         &     (21.10)         \\
Net Effect          &       -10.7         &      -16.8*         &     -19.2**         &     -22.5**         &       -17.6         &      -21.3*         \\
                    &      (8.81)         &      (9.18)         &(9.300)         &       (9.4)         &      (12.3)         &     (12.33)         \\
R-sq                &       0.715         &       0.722         &       0.716         &       0.722         &       0.716         &       0.722         \\
Observations        &        7465         &        6929         &        7465         &        6929         &        7465         &        6929         \\

Baseline Controls & No & Yes& No & Yes& No & Yes\\
\addlinespace[0.15cm]\hline\hline
\multicolumn{7}{p{16.5cm}}{\scriptsize{\textbf{Note:} All columns report estimates from Equation (\ref{didbaselines}), augmented with interaction terms between the treatment variable and dummy variables indicating whether a municipality has: (i) low (below-median) ethnic fractionalization (EF) (columns (1) and (2)); (ii) a small (below-median) number of civil society organizations (CSOs) (columns (3) and (4)); or (iii) both low ethnic fractionalization and few CSOs (columns (5) and (6)).  All models include municipality and department $\times$ year fixed effects, and the following (lagged) covariates:  log of total population, rurality index, and proportion with unsatisfied basic needs. Samples for regression models use data from 2005 to 2019. Robust standard errors (in parentheses) are clustered by municipality. * denotes statistically significant estimates at 10\%, ** denotes significant at 5\% and *** denotes significant at 1\%.} }
\end{tabular}
}
\end{center}
\end{table}


\begin{table}[H]
\begin{center}
{
\renewcommand{\arraystretch}{0.7}
\setlength{\tabcolsep}{15pt}
\caption {Effect of the First Non-Catholic Church on Domestic Violence:  Heterogeneous Effects by Ethnic Fractionalization}  \label{twfe_domeviolenceMLrate_highlowethnic_tab}
\vspace{-0.3cm}
\small
\centering  \begin{tabular}{lcccc}
\hline\hline \addlinespace[0.15cm]
  & \multicolumn{4}{c}{Dep. Var: Domestic Violence (per 100.000 inhabitants)} \\\addlinespace[0.1cm]\cmidrule[0.2pt](l){2-5}\addlinespace[0.05cm]
& (1)& (2)& (3)& (4) \\   \addlinespace[0.1cm] \cmidrule[0.2pt](l){2-3}\cmidrule[0.2pt](l){4-5}
            & \multicolumn{2}{c}{Above median} & \multicolumn{2}{c}{Below median}\\
                        & \multicolumn{2}{c}{ethnic fractionalization} & \multicolumn{2}{c}{ethnic fractionalization}\\\addlinespace[0.15cm]
\hline \addlinespace[0.15cm]
            \multicolumn{1}{l}{\emph{\underline{Panel A}:}}            & \multicolumn{4}{c}{\emph{Static}   Specification}\\
            \addlinespace[0.3cm]
Effect at t$\geq 0$ &      -9.250         &     -11.636\sym{*}  &     -13.590         &     -19.507\sym{*}  \\
                    &     (6.167)         &     (6.357)         &    (10.768)         &    (11.233)         \\
R-sq                &       0.716         &       0.719         &       0.714         &       0.723         \\
Observations        &        3600         &        3357         &        3593         &        3316         \\

\addlinespace[0.15cm]\hline\addlinespace[0.15cm]
            \multicolumn{1}{l}{\emph{\underline{Panel B}:}}            & \multicolumn{4}{c}{\emph{Dynamic} Specification}\\\addlinespace[0.3cm]
Effect at t=-3      &       6.151         &       3.236         &      -3.450         &      -3.097         \\
                    &     (7.623)         &     (7.493)         &    (10.946)         &    (11.135)         \\
Effect at t=-2      &       8.645         &       6.681         &      -5.178         &      -7.844         \\
                    &     (6.765)         &     (6.824)         &    (11.032)         &    (10.777)         \\
Effect at t=-1      &       0.000         &       0.000         &       0.000         &       0.000         \\
                    &         (.)         &         (.)         &         (.)         &         (.)         \\
Effect at t=0       &      -0.536         &      -3.587         &     -14.898         &     -17.751\sym{*}  \\
                    &     (5.932)         &     (6.351)         &    (10.038)         &    (10.159)         \\
Effect at t=+1      &       7.097         &       4.062         &     -15.750         &     -22.296\sym{**} \\
                    &     (5.754)         &     (6.064)         &    (10.011)         &    (10.885)         \\
Effect at t=+2      &       5.436         &       1.377         &      -8.611         &     -12.918         \\
                    &     (6.376)         &     (6.718)         &    (11.033)         &    (11.549)         \\
Effect at t=+3      &      11.481\sym{*}  &       9.070         &      -6.898         &      -5.923         \\
                    &     (6.756)         &     (7.376)         &    (12.137)         &    (11.612)         \\
Effect at t=+4      &       5.339         &       2.853         &      -9.780         &     -16.762         \\
                    &     (6.907)         &     (7.584)         &    (11.683)         &    (11.896)         \\
Effect at t=+5      &      11.602         &       9.679         &     -12.504         &     -22.787\sym{*}  \\
                    &     (7.201)         &     (7.987)         &     (9.959)         &    (11.572)         \\
Effect at t=+6      &       6.837         &       5.540         &     -14.185         &     -25.327\sym{**} \\
                    &     (7.484)         &     (8.269)         &    (10.091)         &    (11.283)         \\
Effect at t=+7      &       5.845         &       3.491         &     -20.869\sym{**} &     -28.540\sym{**} \\
                    &     (6.360)         &     (7.609)         &    (10.220)         &    (12.384)         \\
Effect at t=+8      &       2.608         &       1.061         &     -19.624         &      -9.335         \\
                    &     (7.190)         &     (9.129)         &    (13.944)         &    (17.430)         \\
R-sq                &       0.718         &       0.721         &       0.715         &       0.724         \\
Observations        &        3600         &        3357         &        3593         &        3316         \\

 \addlinespace[0.15cm]\hline \addlinespace[0.15cm]
Baseline controls & No & Yes& No & Yes\\
\addlinespace[0.15cm]\hline\hline
\multicolumn{5}{p{13.5cm}}{\scriptsize{\textbf{Note:} All columns in Panel A report the estimates from Eq. (\ref{didbaselines}) and all columns in Panel B report the estimates from Eq. (\ref{didbaseline}), in municipalities with an above median (columns (1) and (2)) and below median (columns (3) and (4)) index of ethnic fractionalization, using data from 1993.  All models include municipality and department $\times$ year fixed effects. The models with baseline controls (columns (2) and (4)) include the following (lagged) covariates:  log of total population, rurality index, and proportion with unsatisfied basic needs. Samples for regression models use data from 2005 to 2019.  The models in Panel B include 10 lags and 10 leads, normalized to the period prior to treatment. Robust standard errors (in parentheses) are clustered by municipality. * denotes statistically significant estimates at 10\%, ** denotes significant at 5\% and *** denotes significant at 1\%} }
\end{tabular}
}
\end{center}
\end{table}


\begin{table}[H]
\begin{center}
{
\renewcommand{\arraystretch}{0.7}
\setlength{\tabcolsep}{15pt}
\caption {Effect of the First Non-Catholic Church on Domestic Violence:  Heterogeneous Effects by Number of Civil Society Organizations}  \label{twfe_domeviolenceMLrate_highlowcivil_tab}
\vspace{-0.3cm}
\small
\centering  \begin{tabular}{lcccc}
\hline\hline \addlinespace[0.15cm]
  & \multicolumn{4}{c}{Dep. Var: Domestic Violence (per 100.000 inhabitants)} \\\addlinespace[0.1cm]\cmidrule[0.2pt](l){2-5}\addlinespace[0.05cm]
& (1)& (2)& (3)& (4) \\   \addlinespace[0.1cm] \cmidrule[0.2pt](l){2-3}\cmidrule[0.2pt](l){4-5}
            & \multicolumn{2}{c}{Above median \# of} & \multicolumn{2}{c}{Below median \# of}\\
                        & \multicolumn{2}{c}{civil society organizations} & \multicolumn{2}{c}{civil society organizations}\\\addlinespace[0.15cm]
\hline \addlinespace[0.15cm]
            \multicolumn{1}{l}{\emph{\underline{Panel A}:}}            & \multicolumn{4}{c}{\emph{Static}  Specification}\\
            \addlinespace[0.3cm]
Effect at t$\geq 0$ &     -10.435         &     -11.190         &     -15.506         &     -20.124\sym{*}  \\
                    &     (7.851)         &     (7.909)         &    (10.656)         &    (11.270)         \\
R-sq                &       0.657         &       0.661         &       0.762         &       0.769         \\
Observations        &        3865         &        3579         &        3540         &        3295         \\

\addlinespace[0.15cm]\hline\addlinespace[0.15cm]
            \multicolumn{1}{l}{\emph{\underline{Panel B}:}}            & \multicolumn{4}{c}{\emph{Dynamic} Specification}\\\addlinespace[0.3cm]
Effect at t=-3      &      -4.229         &      53.560         &      -8.739         &     -12.146         \\
                    &     (8.335)         &    (54.479)         &    (13.367)         &    (13.328)         \\
Effect at t=-2      &      -1.390         &      62.123         &      -3.274         &      -9.999         \\
                    &     (7.080)         &    (45.446)         &    (14.525)         &    (13.622)         \\
Effect at t=-1      &       0.000         &       0.000         &       0.000         &       0.000         \\
                    &         (.)         &         (.)         &         (.)         &         (.)         \\
Effect at t=0       &      -7.562         &     -16.877         &     -22.249\sym{*}  &     -24.786\sym{**} \\
                    &     (8.524)         &    (23.386)         &    (11.679)         &    (11.595)         \\
Effect at t=+1      &      -8.058         &      -8.639         &     -12.376         &     -16.810         \\
                    &     (8.823)         &    (15.410)         &    (11.136)         &    (11.028)         \\
Effect at t=+2      &      -6.456         &      80.668         &      -8.128         &     -13.568         \\
                    &     (8.582)         &    (57.010)         &    (12.614)         &    (12.782)         \\
Effect at t=+3      &     -10.087         &     126.479         &       2.017         &      -3.278         \\
                    &     (9.344)         &    (85.167)         &    (11.715)         &    (11.741)         \\
Effect at t=+4      &      -8.889         &     126.116         &     -15.488         &     -26.287\sym{**} \\
                    &     (9.133)         &    (92.833)         &    (11.604)         &    (11.185)         \\
Effect at t=+5      &     -12.615         &     137.713         &     -11.667         &     -21.662\sym{*}  \\
                    &     (7.837)         &    (93.516)         &    (11.253)         &    (11.841)         \\
Effect at t=+6      &     -13.103\sym{*}  &      48.414         &     -11.187         &     -20.170         \\
                    &     (7.832)         &    (48.840)         &    (11.413)         &    (12.303)         \\
Effect at t=+7      &     -11.669         &      97.572         &     -11.362         &     -19.517         \\
                    &     (8.457)         &    (67.623)         &    (11.820)         &    (13.904)         \\
Effect at t=+8      &      -8.513         &      83.271         &     -23.146\sym{*}  &     -14.640         \\
                    &    (10.510)         &    (71.805)         &    (11.839)         &    (11.616)         \\
R-sq                &       0.658         &       0.884         &       0.764         &       0.771         \\
Observations        &        3865         &         225         &        3540         &        3295         \\

 \addlinespace[0.15cm]\hline \addlinespace[0.15cm]
Baseline controls & No & Yes& No & Yes\\
\addlinespace[0.15cm]\hline\hline
\multicolumn{5}{p{13.5cm}}{\scriptsize{\textbf{Note:} All columns in Panel A report the estimates from Eq. (\ref{didbaselines}) and all columns in Panel B report the estimates from Eq. (\ref{didbaseline}), in municipalities with an above median (columns (1) and (2)) and below median (columns (3) and (4)) number of civil society organizations (per 100.000 inhabitants), using data from 1995.  All models include municipality and   department $\times$ year fixed effects. The models with baseline controls (columns (2) and (4)) include the following (lagged) covariates:  log of total population, rurality index, and proportion with unsatisfied basic needs. Samples for regression models use data from 2005 to 2019.  The models in Panel B include 10 lags and 10 leads, normalized to the period prior to treatment. Robust standard errors (in parentheses) are clustered by municipality. * denotes statistically significant estimates at 10\%, ** denotes significant at 5\% and *** denotes significant at 1\%.} }
\end{tabular}
}
\end{center}
\end{table}


\newpage

\begin{table}[H]
\begin{center}
{
\renewcommand{\arraystretch}{0.7}
\setlength{\tabcolsep}{4pt}
\caption {Effect of the First Non-Catholic Church on Domestic Violence:  Heterogeneous Effects by Presence Coca Crops in 2000 and of Cases of Forced Recruitment in 1996}  \label{twfe_domeviolenceMLrate_coca_tab}
\vspace{-0.3cm}
\small
\centering  \begin{tabular}{lcccccccc}
\hline\hline \addlinespace[0.15cm]
  & \multicolumn{8}{c}{Dep. Var: Domestic Violence (per 100.000 inhabitants)} \\\addlinespace[0.1cm]\cmidrule[0.2pt](l){2-9}\addlinespace[0.05cm]
& (1)& (2)& (3)& (4)& (5)& (6)& (7)& (8) \\   \addlinespace[0.1cm] \cmidrule[0.2pt](l){2-3}\cmidrule[0.2pt](l){4-5}\cmidrule[0.2pt](l){6-7}\cmidrule[0.2pt](l){8-9}
            & \multicolumn{2}{c}{Presence of} & \multicolumn{2}{c}{No presence of}  & \multicolumn{2}{c}{Presence of} & \multicolumn{2}{c}{No presence of}\\\
                        & \multicolumn{2}{c}{coca crops} & \multicolumn{2}{c}{coca crops} & \multicolumn{2}{c}{forced recruitment} & \multicolumn{2}{c}{forced recruitment}\\\addlinespace[0.15cm]
\hline \addlinespace[0.15cm]
            \multicolumn{1}{l}{\emph{\underline{Panel A}:}}            & \multicolumn{8}{c}{\emph{Static}  Specification}\\
            \addlinespace[0.3cm]
Effect at t$\geq 0$ &     -14.296         &     -18.026         &     -10.289         &     -13.353\sym{*}  &      -4.176         &     -10.193         &     -11.931\sym{*}  &     -15.519\sym{**} \\
                    &    (10.328)         &    (11.412)         &     (7.403)         &     (7.468)         &    (11.801)         &    (11.957)         &     (7.078)         &     (7.239)         \\
R-sq                &       0.722         &       0.723         &       0.717         &       0.725         &       0.843         &       0.849         &       0.715         &       0.721         \\
Observations        &        1155         &        1077         &        6265         &        5811         &         780         &         727         &        6580         &        6105         \\

\addlinespace[0.15cm]\hline\addlinespace[0.15cm]
            \multicolumn{1}{l}{\emph{\underline{Panel B}:}}            & \multicolumn{8}{c}{\emph{Dynamic} Specification}\\\addlinespace[0.3cm]
Effect at t=-3      &      12.650         &      13.269         &      -5.965         &      -8.500         &      -6.256         &      -2.513         &      -3.359         &      -5.166         \\
                    &    (14.071)         &    (14.963)         &     (8.261)         &     (8.350)         &    (11.176)         &    (11.852)         &     (7.733)         &     (7.916)         \\
Effect at t=-2      &      26.225         &      24.004         &      -3.796         &      -6.899         &      21.297         &      24.326         &      -1.512         &      -4.613         \\
                    &    (17.727)         &    (17.343)         &     (7.857)         &     (7.651)         &    (18.859)         &    (19.897)         &     (7.624)         &     (7.424)         \\
Effect at t=-1      &       0.000         &       0.000         &       0.000         &       0.000         &       0.000         &       0.000         &       0.000         &       0.000         \\
                    &         (.)         &         (.)         &         (.)         &         (.)         &         (.)         &         (.)         &         (.)         &         (.)         \\
Effect at t=0       &       7.417         &       3.991         &     -11.514         &     -14.100\sym{*}  &      -0.131         &       2.484         &     -10.011         &     -12.941\sym{*}  \\
                    &    (11.613)         &    (11.887)         &     (7.577)         &     (7.836)         &    (14.457)         &    (14.293)         &     (7.230)         &     (7.488)         \\
Effect at t=+1      &      17.076         &      13.592         &      -9.179         &     -13.273\sym{*}  &      10.083         &       9.932         &      -7.567         &     -12.095         \\
                    &    (14.308)         &    (15.182)         &     (7.399)         &     (7.836)         &    (18.492)         &    (19.847)         &     (7.198)         &     (7.633)         \\
Effect at t=+2      &       9.944         &       8.622         &      -5.227         &      -8.657         &       4.899         &       6.133         &      -4.860         &      -8.461         \\
                    &    (13.075)         &    (14.626)         &     (7.947)         &     (8.276)         &    (15.673)         &    (17.041)         &     (7.595)         &     (7.879)         \\
Effect at t=+3      &      12.785         &       9.584         &      -2.835         &      -3.787         &      12.199         &      13.419         &      -3.066         &      -4.669         \\
                    &    (13.380)         &    (13.898)         &     (8.029)         &     (8.189)         &    (16.570)         &    (17.848)         &     (7.643)         &     (7.715)         \\
Effect at t=+4      &      -4.399         &      -6.802         &      -9.188         &     -14.727\sym{*}  &      20.706         &      17.160         &     -11.439         &     -17.172\sym{**} \\
                    &    (14.562)         &    (15.650)         &     (7.964)         &     (8.036)         &    (19.971)         &    (21.498)         &     (7.442)         &     (7.469)         \\
Effect at t=+5      &      -2.401         &      -9.513         &     -10.474         &     -16.899\sym{**} &      35.953\sym{*}  &      23.036         &     -14.384\sym{*}  &     -20.468\sym{***}\\
                    &    (12.758)         &    (14.477)         &     (7.953)         &     (8.355)         &    (20.169)         &    (25.668)         &     (7.323)         &     (7.747)         \\
Effect at t=+6      &      -5.019         &      -7.978         &     -10.224         &     -15.184\sym{*}  &      10.870         &       1.985         &     -11.302         &     -15.547\sym{*}  \\
                    &    (12.680)         &    (14.313)         &     (8.175)         &     (8.632)         &    (18.858)         &    (23.415)         &     (7.577)         &     (7.951)         \\
Effect at t=+7      &      -4.907         &     -13.803         &     -10.010         &     -13.116         &     -12.609         &     -39.729         &      -8.888         &     -11.261         \\
                    &    (13.219)         &    (15.778)         &     (8.316)         &     (9.764)         &    (17.500)         &    (25.706)         &     (7.761)         &     (9.034)         \\
Effect at t=+8      &     -18.551         &     -20.871         &     -12.734         &      -5.834         &     -11.373         &       3.478         &     -10.792         &      -5.296         \\
                    &    (14.877)         &    (17.115)         &     (9.744)         &    (10.329)         &    (20.367)         &    (21.405)         &     (9.270)         &     (9.662)         \\
R-sq                &       0.728         &       0.729         &       0.718         &       0.726         &       0.850         &       0.856         &       0.716         &       0.722         \\
Observations        &        1155         &        1077         &        6265         &        5811         &         780         &         727         &        6580         &        6105         \\

 \addlinespace[0.15cm]\hline \addlinespace[0.15cm]
Baseline controls & No & Yes& No & Yes& No & Yes& No & Yes\\
\addlinespace[0.15cm]\hline\hline
\multicolumn{9}{p{16.5cm}}{\scriptsize{\textbf{Note:} All columns in Panel A report estimates from Eq. (\ref{didbaselines}), and all columns in Panel B report estimates from Eq. (\ref{didbaseline}). The estimates are presented for municipalities with or without the presence of coca crops in 2000 (columns (1) to (4)) and with or without cases of forced recruitment in 1996 (columns (5) to (8)).  All models include municipality and  department $\times$ year fixed effects. The models with baseline controls (columns (2), (4), (6) and (8)) include the following (lagged) covariates:  log of total population, rurality index, and proportion with unsatisfied basic needs. Samples for regression models use data from 2005 to 2019.  The models in Panel B include 10 lags and 10 leads, normalized to the period prior to treatment. Robust standard errors (in parentheses) are clustered by municipality. * denotes statistically significant estimates at 10\%, ** denotes significant at 5\% and *** denotes significant at 1\%.} }
\end{tabular}
}
\end{center}
\end{table}


\begin{table}[H]
\begin{center}
{
\renewcommand{\arraystretch}{0.9}
\setlength{\tabcolsep}{3pt}
\caption {Correlation Between Domestic Violence and the Number of New Non-Catholic Churches and Civil Society Organizations (2005–2021)}  \label{tab_corrorgcivsoc}
\vspace{-0.3cm}
\footnotesize
\centering  \begin{tabular}{lcccccc}
\hline\hline \addlinespace[0.15cm]
  & \multicolumn{6}{c}{Dep. Var: Domestic Violence (per 100.000 inhabitants)} \\\addlinespace[0.1cm]\cmidrule[0.2pt](l){2-7}\addlinespace[0.05cm]
& (1)& (2) & (3)& (4)& (5)& (6) \\   \addlinespace[0.1cm] \hline \addlinespace[0.15cm]
& \multicolumn{2}{c}{Against any}& \multicolumn{2}{c}{Against}& \multicolumn{2}{c}{Against other} \\
& \multicolumn{2}{c}{household member}& \multicolumn{2}{c}{intimate partner}& \multicolumn{2}{c}{household members} \\
\cmidrule[0.2pt](l){2-3}\cmidrule[0.2pt](l){4-5}\cmidrule[0.2pt](l){6-7}\addlinespace[0.1cm]
Number of New Non-Catholic Churches&      -3.992         &      -5.261\sym{**} &      -2.774         &      -3.522\sym{*}  &      -1.219         &      -1.739\sym{**} \\
                    &     (2.729)         &     (2.567)         &     (1.924)         &     (1.829)         &     (0.841)         &     (0.773)         \\
Number of Civil Society Organizations&       0.128\sym{***}&       0.140\sym{***}&       0.089\sym{***}&       0.098\sym{***}&       0.039\sym{***}&       0.042\sym{***}\\
                    &     (0.037)         &     (0.034)         &     (0.025)         &     (0.024)         &     (0.011)         &     (0.011)         \\
R-sq                &       0.482         &       0.554         &       0.449         &       0.523         &       0.495         &       0.557         \\
Observations        &        1051         &        1034         &        1051         &        1034         &        1051         &        1034         \\

 \addlinespace[0.15cm]\hline \addlinespace[0.15cm]
Controls & No & Yes& No & Yes& No & Yes\\
\addlinespace[0.15cm]\hline\hline
\multicolumn{7}{p{15.1cm}}{\scriptsize{\textbf{Note:}  All columns report correlation coefficients between each domestic violence outcome and two key variables: (i) the number of new non-Catholic churches established in a given Colombian municipality between 2005 and 2021, and (ii) the number of civil society organizations operating in that municipality in 1995. All models include department fixed effects. The specifications with controls (columns (2), (4), and (6)) include the following covariates (averaged over the period 2005 to 2021): total population, proportion of the population with unsatisfied basic needs, municipal land area, altitude, distance to the nearest departmental capital, distance to Bogotá, homicide rate, and ethnic fragmentation. Robust standard errors (in parentheses) are clustered at the department level. * denotes statistically significant estimates at 10\%, ** denotes significant at 5\% and *** denotes significant at 1\%. }}
\end{tabular}
}
\end{center}
\end{table}


\begin{table}[H]
\begin{center}
{
\renewcommand{\arraystretch}{1}
\setlength{\tabcolsep}{3pt}
\caption {Effect of the First Non-Catholic Church on the Proportion of Population with Unsatisfied Basic Needs}  \label{poverty_tab}
\vspace{-0.3cm}
\small
\centering  \begin{tabular}{lcc}
\hline\hline \addlinespace[0.15cm]
    & \multicolumn{2}{c}{Dep. Variable:} \\
        & \multicolumn{2}{c}{Unsatisfied Basic Needs} \\\cmidrule[0.2pt](l){2-3}
& (1)& (2)    \\    \addlinespace[0.15cm] \hline \addlinespace[0.15cm]
First non-Catholic church in municipality at t-1&       0.551         &       0.848         \\
                    &     (0.645)         &     (0.680)         \\
Log of total population at t-1&                     &      -8.633\sym{*}  \\
                    &                     &     (5.039)         \\
Rurality index at t-1&                     &      16.899\sym{*}  \\
                    &                     &     (9.140)         \\
Homicide rate at t-1&                     &      -0.005         \\
                    &                     &     (0.006)         \\
R-sq                &       0.951         &       0.953         \\
Observations        &        1463         &        1386         \\

\addlinespace[0.15cm]\hline\hline
\multicolumn{3}{p{12cm}}{\footnotesize{\textbf{Note:} All models include municipality and department $\times$ year fixed effects.  Samples for regression models include data from 2005 to 2019.   Robust standard errors (in parentheses) are clustered by municipality. * denotes statistically significant estimates at 10\%, ** denotes significant at 5\% and *** denotes significant at 1\%.} }
\end{tabular}
}
\end{center}
\end{table}


\begin{table}[H]
\begin{center}
{
\renewcommand{\arraystretch}{0.7}
\setlength{\tabcolsep}{8pt}
\caption {Effect of the First Non-Catholic Church on Domestic Violence: Effect Before and After 2012}  \label{domv_before_after_2012}
\vspace{-0.3cm}
\small
\centering  \begin{tabular}{lcccccc}
\hline\hline \addlinespace[0.15cm]
  & \multicolumn{6}{c}{Dep. Var: Domestic Violence (per 100.000 inhabitants)} \\\addlinespace[0.1cm]\cmidrule[0.2pt](l){2-7}\addlinespace[0.05cm]
& (1)& (2) & (3)& (4)& (5)& (6) \\   \addlinespace[0.1cm] \hline \addlinespace[0.15cm]
& \multicolumn{2}{c}{Against any}& \multicolumn{2}{c}{Against}& \multicolumn{2}{c}{Against other} \\
& \multicolumn{2}{c}{household member}& \multicolumn{2}{c}{intimate partner}& \multicolumn{2}{c}{household members} \\
\cmidrule[0.2pt](l){2-3}\cmidrule[0.2pt](l){4-5}\cmidrule[0.2pt](l){6-7}\addlinespace[0.15cm]
            \multicolumn{1}{l}{\emph{\underline{Panel A}:}}            & \multicolumn{6}{c}{\emph{Before} 2012}\\
            \addlinespace[0.3cm]
Effect at t$\geq 0$ &     -19.017\sym{**} &     -18.727\sym{**} &     -10.540\sym{**} &     -10.242\sym{**} &      -8.477\sym{**} &      -8.485\sym{**} \\
                    &     (8.545)         &     (8.428)         &     (5.046)         &     (4.932)         &     (4.026)         &     (4.019)         \\
R-sq                &       0.757         &       0.757         &       0.775         &       0.774         &       0.657         &       0.657         \\
Observations        &        3979         &        3939         &        3979         &        3939         &        3979         &        3939         \\

\addlinespace[0.15cm]\hline\addlinespace[0.2cm]
            \multicolumn{1}{l}{\emph{\underline{Panel B}:}}            & \multicolumn{6}{c}{\emph{After} 2012}\\\addlinespace[0.3cm]
Effect at t$\geq 0$ &      -5.422         &     -11.827         &      -3.302         &      -6.844         &      -2.120         &      -4.982         \\
                    &     (9.542)         &    (11.885)         &     (6.735)         &     (8.551)         &     (3.148)         &     (3.723)         \\
R-sq                &       0.773         &       0.780         &       0.764         &       0.771         &       0.709         &       0.720         \\
Observations        &        3486         &        2988         &        3486         &        2988         &        3486         &        2988         \\

 \addlinespace[0.15cm]\hline \addlinespace[0.15cm]
Baseline controls & No & Yes& No & Yes& No & Yes\\
\addlinespace[0.15cm]\hline\hline
\multicolumn{7}{p{15cm}}{\scriptsize{\textbf{Note:} All columns  report the estimates  from Eq. (\ref{didbaselines}). All models include municipality and department $\times$ year fixed effects.  The models with baseline controls (columns (2), (4) and (6)) include the following (lagged) covariates:  log of total population, rurality index, and proportion with unsatisfied basic needs. Samples for regression models use data from 2005 to 2019. Robust standard errors (in parentheses) are clustered by municipality. * denotes statistically significant estimates at 10\%, ** denotes significant at 5\% and *** denotes significant at 1\%.} }
\end{tabular}
}
\end{center}
\end{table}


\newpage

\bibliographystyle{ecca}
\bibliography{bibdomviol}

\end{document}